\documentstyle[12pt]{article}
\begin{document}
\setlength{\oddsidemargin}{-0.1cm}
\setlength{\topmargin}{-1.cm}

\title{Linear Scaling Electronic Structure Methods}
\author{
Stefan Goedecker,
\\Max-Planck Institute for Solid State Research, 
\\Stuttgart, Germany
\\goedeck@prr.mpi-stuttgart.mpg.de }
\maketitle

Methods exhibiting linear scaling with respect to the size of the system, so called 
O(N) methods, are an essential tool for the calculation of the electronic 
structure of large systems containing 
many atoms. They are based on algorithms which take advantage of the decay properties 
of the density matrix. In this article the physical decay properties of the 
density matrix will first be studied for both metals and insulators.
Several strategies to construct O(N) algorithms will then be presented and 
critically examined. Some issues which are relevant 
only for self-consistent O(N) methods, such as the calculation of the 
Hartree potential and mixing issues, will also be discussed. 
Finally some typical applications of O(N) methods are briefly described. 

\pagebreak

\tableofcontents

\pagebreak

\section{Introduction}
The exact quantum mechanical equations for many-electron systems are 
highly intricate. Any attempt to solve these equations analytically for 
real systems is doomed to fail.  
Numerical methods such as Configuration Interaction based methods 
(McWeeny 1989, Fulde 1995) 
or Quantum Monte Carlo methods
(Hammond 1994, Nightingale 1998) 
can in principle solve these many-electron equations but because 
of the extremely high numerical effort required, their applicability is 
rather limited in practice. 

The bulk of all practical applications is therefore done within various 
independent-electron approximations such as the Hartree-Fock method 
(A. Szabo and N. Ostlund 1982), Density Functional methods (R. Parr and W. Yang 1989) or
Tight Binding methods (C. Goringe {\it et al.,} 1997a, Majewski and Vogl 1986). 
A comparison of the strength of different methods together with a selection of 
some interesting applications is given by Wimmer (1996).
Even these approximate quantum mechanical equations are still fairly complicated 
and in general not solvable by analytical methods. 
Finding efficient algorithms to solve the many-electron problem numerically within any 
of these approximations is imperative for the applicability of 
quantum mechanics to physics as well as to chemistry and 
materials science. 
Due to efforts in the past satisfactory algorithms are now available and 
computational electronic structure methods are making very 
important contributions to our understanding of matter at the microscopic level. 
The 1998 Nobel prize for W. Kohn and J. Pople is a landmark sign of the importance 
of this approach. 
Since computational electronic structure 
methods are used over a very wide spectrum of applications is it hard to summarize their use.
A hint of their versatility can be obtained by looking at the fraction of theoretical 
articles in several solid state and chemistry journals where computational electronic 
structure methods are used. As can be seen from Table 1 this fraction varies 
between 11 and 59 \% , being 27 \% in the two largest journals 
in solid state physics (PRB) and chemical physics (JCP). 

 \vspace{1cm}
 \noindent
 Table 1: The importance of computational electronic structure methods 
 as measured by the number of publications. Listed are the total number of 
 publications, the number of theoretical publications and the number of publications 
 using computational electronic structure methods in the period from January 1997 to 
 September 1988. To determine whether a paper belongs into this last category, 
 it has to contain either of the key words ``electronic structure calculation", 
 ``ab initio calculation", ``tight binding calculation" or ``density functional 
 calculation". I thank Dr Wanitschek for providing me with these data. 
 \newline
 \noindent
 \begin{tabular}{|c||c|c|c|} \hline \hline
  Journal                             & total      & theoretical & computational \\ \hline
 Physical Review B (Condensed Matter)         & 7924       &  4629      & 1263  \\ 
 Journal of chemical physics                  & 4066       &  2764      &  754  \\ 
 Theochem                                     &  905       &   856      &  502  \\ 
 Chemical physics letters                     & 2319       &  1209      &  490  \\ 
 International journal of quantum chemistry   &  609       &   582      &  271  \\ 
 Physical review letters                      & 4558       &  2834      &  250  \\ 
 Journal of physics: Condensed matter         & 1774       &   927      &  220  \\ \hline \hline
 \end{tabular}
 \vspace{1cm}

Due to the constant increase 
in computer power and due to algorithmic improvements the importance of 
computational methods is growing further. Whereas computational methods 
nowadays mainly supplement experimentally obtained information, they are expected 
to increasingly supersede this information.

This article will concentrate 
on recently developed methods that allow us to calculate the total energy within various 
independent-electron methods for large systems. Practically all physical 
observables can be obtained from the total energy, for instance in the form of derivatives 
with respect to certain external parameters. 
The reason why large systems containing many
atoms are accessible with these algorithms is their linear scaling 
with respect to the number of atoms. In principle linear scaling should also be 
obtainable for true many-electron methods. For the MP2 method such an 
algorithms has indeed recently been reported (Ayala and Scuseria, 1998).

Traditional electronic structure algorithms calculate 
eigenstates associated with discrete energy levels. The reason for this is probably 
historical since the prediction of these experimentally  
observed levels was the first big success of quantum mechanics. 
The disadvantage of this approach is that 
it leads to a diagonalization problem which has a cubic 
scaling in the computational effort. 
Direct diagonalization (Press {\it et al.,} 1986), which 
was the standard approach in the early days of the computational 
electronic structure era, has a cubic scaling with 
respect to the size of the Hamiltonian matrix, i.e. with respect to the 
number of basis functions $M_b$. Iterative diagonalization schemes (Saad 1996), 
preconditioned conjugate gradient minimizations 
(Teter {\it et al.,} 1989,  \v{S}tich {\it et al.,} 1989, Payne, {\it et al.,} 1992 ) 
and the Car-Parrinello method (Car and Parrinello 1985) for 
molecular dynamics simulations were a big algorithmic progress 
because of their improved scaling behavior. Their 
scaling was not any more proportional to the cube of the 
the number of basis functions but grew only like $M_b \log(M_b)$ if plane 
waves were used as a basis set. Nevertheless these methods still have a cubic scaling 
with respect to the number of atoms $N_{at}$, which comes from the orthogonality 
requirement of the wavefunctions. 
The reason why this orthogonalization step scales cubically can 
easily be seen. As the system grows, each wavefunction extends over a larger volume and 
has therefore to be represented by a larger basis set resulting in a longer vector.
At the same time there are more such wavefunctions and each wavefunction 
has to be orthogonalized to all the  others. Thus there are 3 factors that  
grow linearly, resulting in the postulated cubic behavior. 
The computer time $T_{CPU}$ required to do the calculation is thus given by 
\begin{equation} \label{cpu3}
T_{CPU} = c_3  N_{at}^3 \: ,
\end{equation}
where $c_3$ is a prefactor. It has to be pointed out that Equation~(\ref{cpu3}) gives 
only the asymptotic scaling behavior. Within Density Functional and Hartree Fock calculations 
there are other terms with a lower scaling which dominate for system sizes of less 
than a few hundred atoms due to their large prefactor. In the case of plane wave 
type calculations the Fast Fourier transformations necessary for the application of the 
potential to the wavefunctions consume most of the 
computational time for small systems, in the 
case of calculations using Gaussian type orbitals (Hehre 1996) 
it is the calculation of the Hartree potential.
This cubic scaling is a major bottleneck nowadays 
since in many problems of practical interest one has to do 
electronic structure calculations for systems containing many (a few hundred or more) 
atoms. Evidently, cubic scaling means that if one doubles the number of atoms in the 
systems the required computer time will increase by a factor of eight. By enlarging the 
system one therefore rapidly reaches the limits of the most powerful computers.
So called O(N) or low complexity algorithms are therefore a logical next step 
of algorithmic progress 
since they exhibit linear scaling with respect to the number of atoms
\begin{equation} \label{cpu1}
T_{CPU} = c_1  N_{at}  \: .
\end{equation}
These methods offer thus the potential to calculate very large 
systems. The prefactors $c_1$ and $c_3$ depend on the approximation used for the 
many-electron problem. For a Density Functional calculation with a large basis 
set the prefactors are of course much larger than for a Tight Binding calculation,
where the number of degrees of freedom per atom is much smaller. The prefactor $c_1$ 
depends also on what O(N) method is used, but in general the 
prefactor $c_1$ is always larger than $c_3$ assuming that the same 
independent-electron approximation
is used both in the traditional and O(N) version. There is therefore a so called 
cross over point. For system sizes smaller than the cross over point the traditional 
cubic scaling algorithms are faster, for larger systems the O(N) methods win. 
Tight Binding calculations are an ideal test emvironment for O(N) 
algorithms. Because of their rather small memory and CPU requirements 
one can easily treat systems comprising of a very large number of atoms 
and venture into regions beyond the cross over point. 
Contrary to what one might naively think, the importance of O(N) algorithms 
will also increase as computers get faster. Whereas at present it is 
difficult to access the cross over region situated at some 100 atoms using the 
Density Functional framework, this will be easy with faster computers and 
O(N) algorithms will be the algorithms of choice.

Even though O(N) algorithms contain many aspects of mathematics and computer science 
they have nevertheless deep roots in physics. Obtaining linear scaling is not 
possible by purely mathematical tricks but it is based on the understanding of the 
concept of locality in quantum mechanics. 
Conversely, the need of constructing 
O(N) algorithms was also an incentive to investigate locality questions more deeply, and 
has thus lead to a better understanding of this very fundamental concept. 
An algorithmic description of electronic structure in local terms 
can give a justification of the well established concepts of bonds and lone 
electron pairs in empirical chemistry. 

Since O(N) algorithms are based on 
a certain subdivision of a big system into smaller subsystems, techniques developed 
in this context might also be helpful in reaching another important goal for 
treating large systems, namely 
combining electronic structure methods of different accuracy such as empirical 
Tight Binding and Density Functional theory in a single system.

\section{Locality in Quantum Mechanics}
\label{general} 
Locality in Quantum Mechanics means that the properties of 
a certain observation region comprising one or a few atoms 
are only weakly influenced by factors that are spatially 
far away form this observation region. This fundamental characteristic 
of insulators is well established within independent-electron 
theories (Heine 1980) and it can even be carried over into the  
many-electron framework (Kohn 1964). 

\begin{figure}[ht]       
     \begin{center}
      \setlength{\unitlength}{1cm}
       \begin{picture}( 8.,4.5)           % figure dimensions
        \put(-4.,-2.0){\includegraphics{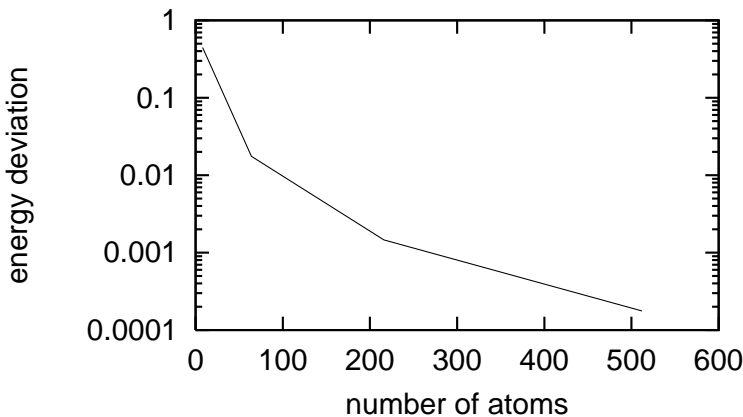}}   % VAX
       \end{picture}
       \caption{\label{sidecay} \it The deviation of the total energy per silicon atom 
                from its asymptotic bulk value 
                as a function of the size of the periodic volume in which it is
                embedded. The calculation was done with a Tight Binding scheme 
                using exact diagonalization  }
      \end{center}
     \end{figure}

Traditional chemistry is based on local concepts. Covalently bonded 
materials are described in terms of bonds and lone electron pairs.
It is standard textbook knowledge that the properties of a bond 
are mainly determined by its immediate neighborhood. 
The decisive factors are what type of atoms and how many of them 
(the coordination number) are surrounding it.
Second nearest neighbors and other more distant atoms 
have a very small influence. As an example let us look at the total energy 
of a hydrocarbon chain molecule $C_n H_{2n+2}$. In this case each 
$C H_2$ subunit is from an energetical point of view practically an 
independent unit. As one adds one $C H_2$ subunit, the energy increases 
by an amount which is nearly independent of the 
chain length. Already the insertion of a $C H_2$ subunit into the 
smallest chain $C_2 H_6$ gives an energy gain which agrees 
within $10^{-4}$ a.u. with the asymptotic value of the insertion 
energy for very long chains. This means that the electrons belonging 
to this inserted subunit already do not see any more the end of the 
chain for very short chain lengths. This example is 
a drastic illustration of a principle sometimes termed 
``nearsightedness'' (Kohn 1996). In other insulating materials 
the influence of the neighboring atoms decays slower. 
An example is shown in Figure~(\ref{sidecay}), where the 
total energy per silicon atom is plotted as a function of the size 
of its crystalline environment.

Even in metallic systems,
where the elementary bond concept is not any more valid, locality 
still exists. This is supported by the well known fact, 
that the total charge density 
in a metal is given with reasonable accuracy by the superposition of the 
atomic charge densities. Since atomic charge densities decay rapidly, this 
implies that the charge density at the midpoint of two 
neighboring atoms is mainly determined by the two closest atoms 
and very little by other more distant atoms. Another related 
example is given by V. Heine (Heine 1980) who points out, that 
the magnetic moment of an iron atom, which is embedded in an 
iron-aluminum alloy differs by less than 5 \% from the 
value for pure iron if the atoms are locally surrounded by only eight 
aluminum atoms. 

This locality is not at all reflected in 
standard electronic structure calculations which are based on  
eigenorbitals extending over the whole system, making both 
the interpretation of the results more difficult and requiring 
unnecessary computational effort. The simplistic 
bond concepts of empirical chemistry are certainly not adequate for 
electronic structure calculations aiming at high accuracy.
Nevertheless one might hope to incorporate some more general locality 
concepts into electronic structure calculation to make 
them both more intuitive and efficient. In the following we 
will therefore carefully examine  the range of interactions 
in quantum mechanical systems.

Self-consistent electronic structure methods require essentially two  steps.
The calculation of the potential from the electronic charge distribution 
and the  determination of the wavefunction for a given  potential.
In non-self-consistent calculations such as Tight Binding calculations, 
the first step is not needed. 

The calculation of the potential 
consists usually of two  parts, the exchange correlation potential, 
and the Coulomb potential. The exchange correlation potential 
is a purely local expression in Density Functional Theory and can 
therefore be calculated 
with linear scaling. In the Hartree Fock scheme one might first think that 
the exchange part is non-local, but a more profound examination reveals 
(section~\ref{coulomb}) that it is local even in this case.
The Coulomb potential on the other  hand is very 
long range and needs proper treatment. 
A naive evaluation of the potential $U$ arising from a charge 
distribution $\rho$ by subdividing space into subvolumes $\Delta V$ and summing over 
these subvolumes, 
$$
U( {\bf r}_i ) = \sum_j \frac{\rho({\bf r}_j)}{|{\bf r}_i-{\bf r}_j|} \Delta V \: ,
$$
would result in a quadratic scaling since both indices $i$ and $j$ 
have too run over all grid points in the system.
The Coulomb problem actually  arises not only in the context of electronic 
structure calculations but also in classical calculations of 
coulombic and gravitational systems such as galaxies of stars. Much 
effort has therefore been invested in this computational problem 
and several algorithms are known which solve the problem with 
linear scaling. These methods will be described in section~\ref{coulomb}.

The more interesting and more difficult part is to assess the role 
of locality for a given external potential. The appropriate quantity 
to study this property is the density matrix. The one-particle density 
matrix $F$ completely specifies our quantum mechanical system within 
the independent electron approximation and all quantities of interest 
can easily be calculated from it. The central quantities in any electronic 
structure calculation, the kinetic energy $E_{kin}$, the potential energy $E_{pot}$ 
and the electronic charge density $\rho$ are given by 
\begin{eqnarray}
E_{kin} & = & - \frac{1}{2} \int \left. \nabla^2_{\bf r} F({\bf r},{\bf r}') \right|_{{\bf r}={\bf r}'} d{\bf r}' \\
E_{pot} & = &  \int  F({\bf r}',{\bf r}') U({\bf r}') d{\bf r}' \\
\rho({\bf r}) & = &  F({\bf r},{\bf r}) \:,
\end{eqnarray}
where $U({\bf r}')$ is the potential. A related quantity which will frequently be 
used throughout the article is the band structure energy $E_{BS}$ 
defined as 
\begin{equation} \label{ebsdef}
E_{BS} = E_{kin} + E_{pot}
\end{equation}
and the grand potential
\begin{equation} \label{omdef}
\Omega = E_{BS} - \mu N_{el} \: ,
\end{equation}
where $\mu$ is the chemical potential and $N_{el}$ the number of electrons.
Subtracting $\mu N_{el}$ from $E_{BS}$ leaves $\Omega$ invariant under a constant potential 
offset. If one applies the shift ( $U({\bf r}) \rightarrow U({\bf r}) + const$) the 
potential energy will increase by $N_{el} \: const$. In order to conserve the 
total number of electrons, $\mu$ also has to be shifted ($\mu  \rightarrow  \mu + const$) 
and thus $\Omega$ remains constant. 

Discretizing the Hamiltonian $H$ which is the sum of the kinetic and 
potential energy as well as $F$ with respect to a finite orthogonal basis 
$\phi_i({\bf r})$, $i=1, ... , M_b$ 
one obtains
\begin{eqnarray}
H_{i,j} & = & \int \phi^*_i({\bf r}) 
              \left( - \frac{1}{2} \nabla^2_{\bf r} + U({\bf r}) \right) \phi_j({\bf r}) d{\bf r}  \\
F_{i,j} & = & \int \int \phi^*_i({\bf r}) F({\bf r},{\bf r}') \phi_j({\bf r}') d{\bf r} d{\bf r}' 
\label{fijdef}
\end{eqnarray}
and the expressions for the central quantities become
\begin{eqnarray} 
E_{BS} & = & Tr [F H] \label{ebstrace} \\
\Omega & = & Tr [F \: (H - \mu I) ] \label{omegatrace} 
\end{eqnarray}
\begin{equation} \label{rhotrace}
\rho({\bf r}) = \sum_{i,j} F_{i,j} \: \phi_i({\bf r}) \phi_j({\bf r}) \: ,
\end{equation}
where $Tr$ denotes the trace. It follows from Equation~(\ref{rhotrace}) that 
the total number of electrons $N_{el}$ in the system is given by
\begin{equation} \label{nel}
N_{el} = Tr [F] \: .
\end{equation}

Evaluating the traces using the eigenfunctions $\Psi_n$ 
of the Hamiltonian one obtains immediately the well known 
expressions for $N_{el}$, $E_{BS}$, $\Omega$ and 
$\rho$ within the context of conventional calculations which are 
based on diagonalization. Denoting the eigenvalues associated with the 
eigenfunctions $\Psi_n$ by $\epsilon_n$ one obtains
\begin{eqnarray} 
N_{el} & = & \sum_n f(\epsilon_n)  \label{diagnel} \\
E_{BS} & = & \sum_n f(\epsilon_n) \: \epsilon_n   \label{diagebs} \\
\Omega & = & \sum_n (f(\epsilon_n)-\mu) \: \epsilon_n  = 
             \sum_n f(\epsilon_n) \: \epsilon_n - \mu N_{el} \label{diaomega} \\
\rho({\bf r}) & = & \sum_n f(\epsilon_n) \: \Psi^*_n({\bf r}) \Psi_n({\bf r}) \label{diagrho}  \: .
\end{eqnarray}
The function $f$ is the the Fermi distribution 
\begin{equation}
f(\epsilon) =  \frac{1}{1+\exp (\frac{\epsilon-\mu}{k_B T})}  \: ,
\end{equation}
where $k_B$ is Boltzmann's constant and $T$ the temperature.
If we talk about temperature in this article, we always mean the 
electronic temperature since we are not considering the 
motion of the ionic degrees of freedom which might be associated 
with a different ionic temperature. 
In the expressions~(\ref{diagnel}),~(\ref{diagebs}),~(\ref{diaomega}), and~(\ref{diagrho}), as 
well as in the remainder of the whole article, we 
will use the convention, that all the subscripts indexing eigenvalues 
and eigenfunctions are combined orbital and spin indices, i.e. that we can put 
at most one electron in each orbital. This will eliminate bothering factors 
of 2. The usually relevant case of an unpolarized spin restricted 
system can always easily be obtained by cutting into half all sums 
over these indices and multiplying by 2.

In terms of the Hamiltonian $H$ the 
density matrix is defined as the following  matrix functional
\begin{equation} \label{foe}
F =  f(H) \:.
\end{equation}
Since $F$ is a matrix function of $H$ it has  the  same eigenfunctions $\Psi_n$ as H 
\begin{eqnarray}
 H \Psi_n & = & \epsilon_n \Psi_n   \label{hevcs} \\
 F \Psi_n & = & f(\epsilon_n) \Psi_n  \label{fevcs} \: .
\end{eqnarray}
The density matrix can consequently be written as
\begin{eqnarray} \label{fdens}
 F({\bf r},{\bf r}') = \sum_n f(\epsilon_n) \Psi^*_n({\bf r}) \Psi_n({\bf r}')  \: ,
\end{eqnarray}
where $n$ runs over all the eigenstates of the Hamiltonian.
From the functional form of the Fermi distribution it follows that 
the eigenvalues $f(\epsilon_n)$ are always in the interval [0:1].
At zero temperature the density matrix of an insulating system containing $N_{el}$ electrons 
will have $N_{el}$ eigenvalues of value one, all others being zero.
Thus the density matrix does not have full rank, but only rank $N_{el}$.
Hence we can write it as 
\begin{equation} \label{denst0} 
 F({\bf r},{\bf r}') = \sum_{n=occ} \Psi^*_n({\bf r}) \Psi_n({\bf r}') \: ,
\end{equation}
where $n$ runs now only over the $N_{el}$ occupied states.
It is easy to see that $F({\bf r},{\bf r}')$ is a projection operator in this case
\begin{equation} \label{denst0proj} 
 \int  F({\bf r},{\bf r}'') F({\bf r}'',{\bf r}') d{\bf r}''  = F({\bf r},{\bf r}') \: .
\end{equation}

\begin{figure}[ht]       
     \begin{center}
      \setlength{\unitlength}{1cm}
       \begin{picture}( 8.,9.9)           % figure dimensions
%        \put(-4.,-5.25){\special{psfile=h2opic_c.ps}}   % VAX
        \put(-4.,-5.25){\includegraphics{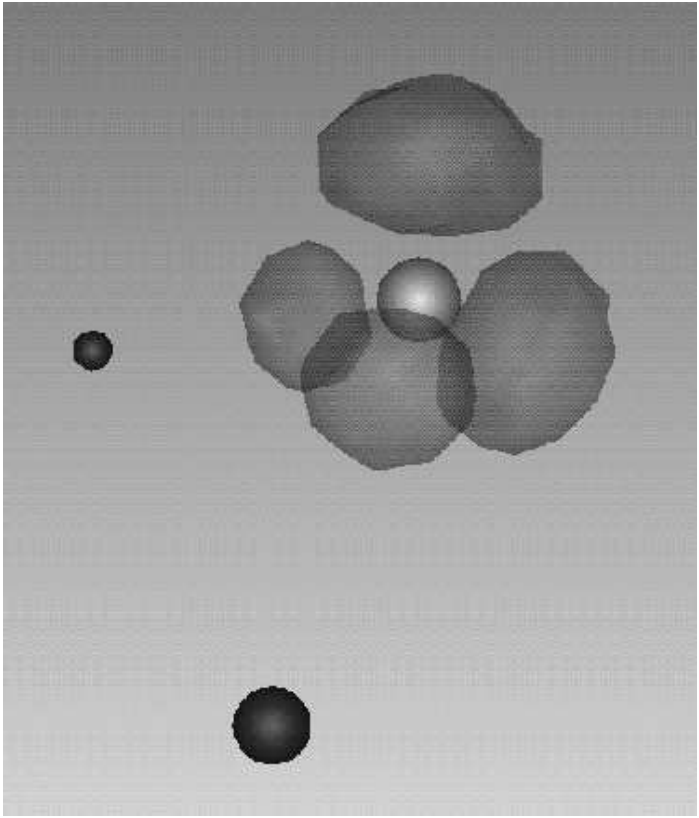}}   % VAX
       \end{picture}
       \caption{\label{water} \it A set of four Wannier orbitals (indicated by clouds) 
                for the water molecule.
                The oxygen nucleus in the center is shown as a bright ball 
                and the two hydrogen nuclei as nearly black balls.
                One sees two Wannier functions along the lines connecting the central 
                oxygen with the two hydrogen atoms representing bonds 
                as well as two lone electron pairs. The centers of the four Wannier functions 
                form a nearly tetragonal structure.}
      \end{center}
%       \caption{\label{water} \it A set of four Wannier orbitals for the water molecule.
%                The oxygen nucleus in the center is shown as a yellow ball 
%                and the two hydrogen nuclei as red balls.
%                One sees two Wannier functions along the lines connecting the central 
%                oxygen with the two hydrogen atoms representing bonds 
%                as well as two lone electron pairs. The centers of the four Wannier functions 
%                represented by by dark blue clouds form a nearly tetragonal structure.}
     \end{figure}

A new set of $N_{el}$ eigenfunctions $\Psi^{new}_n({\bf r})$ can be obtained 
by any unitary transformation of all the $N_{el}$ degenerate eigenfunctions 
$\Psi_n({\bf r})$ associated with eigenvalues one, 
\begin{equation} 
 \Psi^{new}_n({\bf r})  = \sum_{m=occ} U_{n,m} \Psi_m({\bf r})  \: ,
\end{equation}
where $U$ is a unitary $N_{el}$ by $N_{el}$ matrix. In the case of a crystalline periodic 
solids such a transformation can be used to generate 
the localized Wannier functions (Blount) from the extended eigenfunctions $\Psi_n$. 
We will refer to any set of orthogonal exponentially localized orbitals which 
can be used to represent the density matrix according to 
Equation~(\ref{denst0}) as Wannier functions. How to construct 
an optimally localized set of Wannier functions by the minimization 
of the  total spread $\sum_n <r^2>_n -<{\bf r}>^2_n$ in a 
crystalline periodic solid has recently been 
shown by Marzari and Vanderbilt (1997). It has been well known in the chemistry 
community (Chalvet 1976) that sets of maximally localized orbitals give excellent insight 
into the bonding properties of systems. 
In addition to the spread criterion used by Marzari {\it et al.} there are still other 
criteria in common use in the chemistry community. They are all in a certain sense arbitrary, 
but usually lead to the same interpretation of the bonding properties.
Figure~\ref{water} shows the four Wannier 
functions for the water molecule.

The density matrix $F({\bf r},{\bf r}')$ is a diagonally dominant operator, whose 
off-diagonal elements decay with increasing distance from the diagonal.
The exact decay behavior depends on the material. We will derive the 
decay properties within the theoretical framework of the 
description of periodic crystalline solids. For a periodic solid the 
density matrix is given by
\begin{eqnarray} \label{ftrans}
F({\bf r},{\bf r}') & = & \sum_n \frac{V}{(2 \pi)^3} \int_{BZ} d{\bf k} \: 
          f(\epsilon_n({\bf k})) \: \Psi^*_{n,{\bf k}}({\bf r}) \Psi_{n,{\bf k}}({\bf r}') \\
         & = & \sum_{n} \frac{V}{(2 \pi)^3} \int_{BZ} \:  d{\bf k} \: 
          f(\epsilon_n({\bf k})) \: u^*_{n,{\bf k}}({\bf r}) u_{n,{\bf k}}({\bf r}') 
          e^{i{\bf k}({\bf r}'-{\bf r})}  \nonumber \: ,
\end{eqnarray}
where $\Psi_{n,{\bf k}}({\bf r}) = u_{n,{\bf k}}({\bf r}) e^{i{\bf k}({\bf r})} $ 
are the Bloch functions associated with the wave vector ${\bf k}$ and band index $n$. 
The integral is taken over the Brillouin zone (BZ) and 
$V$ is the volume of the real space primitive cell. 

The Wannier functions $W_n$ of the $n$-th band in an insulating crystal 
are defined in the usual way
\begin{equation}
W_n({\bf r}-{\bf R}) = \frac{V}{(2 \pi)^3} \int_{BZ} d{\bf k} \: 
            e^{- i {\bf k} {\bf R}} \: \Psi_{n,{\bf k}}({\bf r}) \: .
\end{equation}
The Wannier functions are not uniquely defined.  
One can construct a different set of Bloch functions by multiplying 
them with a phase factor, 
$\Psi_{n,{\bf k}}({\bf r}) \leftarrow e^{i \omega({\bf k})} \Psi_{n,{\bf k}}({\bf r})$,  
where $\omega({\bf k})$ is an arbitrary function. This will obviously modify 
the Wannier functions. Further ambiguities arise in the case of degenerate bands (Blount). 
Because of these ambiguities in the construction of the 
Wannier functions it is advantageous to work with the density matrix 
where any phase factors cancel (Equation~(\ref{ftrans})) and where degeneracies 
do not cause any problems since one sums over all the occupied bands.

We will first discuss the decay properties of the density matrix in metallic systems. 
In this discussion we will assume that metals behave essentially like 
jellium and that exact results for jellium can be carried over to real metals. 

The decay properties of the density matrix  of a metallic system at zero temperature 
are well known (March). Because the integral in Equation~(\ref{ftrans}) contains 
a discontinuity in the metallic case, 
the density matrix decays only algebraically with 
respect to the distance between ${\bf r}$ and ${\bf r}'$. 
The decay is given by
\begin{equation} \label{march}
F({\bf r},{\bf r}') \propto k_F
                    \frac{\cos (k_F |{\bf r}-{\bf r}'|)}{|{\bf r}-{\bf r}'|^2} \: ,
\end{equation}
where the Fermi wave vector $k_F$ is related to the valence electron density by
$ \frac{N_{el}}{V} = \frac{k_F^3}{3 \pi^2}$ in a non-spin-polarized system.

Introducing a finite electronic temperature $T$ in a metal leads 
to a drastic change in this decay behavior. Instead of an algebraic decay one has 
a much faster exponential decay.
As shown independently by Goedecker (1998a) and Ismail-Beigi and Arias (1998), 
the decay at low temperatures is then given by
\begin{equation} 
F({\bf r},{\bf r}') \propto k_F
              \frac{\cos (k_F |{\bf r}-{\bf r}'|)}{|{\bf r}-{\bf r}'|^2}  \: 
               \exp \left( - c \frac{k_B T}{k_F} |{\bf r}-{\bf r}'| \right)  \: ,
\end{equation}
where $c$ is a constant on the order of 1. 
We thus find oscillatory behavior with an exponentially damped amplitude.
The decay rate depends linearly on temperature and the oscillatory part 
is described by the wave vector $k_F$. 
The related correlation function at finite temperature exhibits the same 
temperature dependence of the decay rate with respect to temperature
(Landau and Lifshitz, 1980).
In an insulator finite temperature plays no role as long as the thermal energy $k_B T$ 
is much smaller than the gap, which is usually fulfilled.

Let us next discuss the important case of an insulator with a band gap $\epsilon_{gap}$ at zero 
temperature. We will first present some numerical results, then we will 
put forward some arguments to explain the qualitative features of the density matrix 
and finally discuss in a more quantitative way the factors which determine
the exact decay rate.

Numerical calculations of the density matrix or the related Wannier functions 
show an oscillatory behavior with a decaying amplitude. There is exactly 
one node per primitive cell and logarithmic plots 
of the amplitude clearly reveal an exponential decay.
In the case of alkanes the decay of the density matrix calculated by the 
Hartree-Fock method has been studied 
and plotted on a logarithmic scale by Maslem {\it et al.} (1997). 
Interestingly, the decay depends also on 
the basis set used. Small low quality basis sets lead to a larger band gap and 
consequently to a faster decay of the density matrix. In the case of silicon, 
treated by Density Functional theory,
logarithmic plots revealing the exponential decay of the Wannier functions have also 
been done both for grid based basis sets (Goedecker unpublished) and 
atomic basis sets (Stephan 1998). 
Within the Tight Binding method the decay of the density matrix has also been 
studied numerically for crystalline and liquid carbon 
systems by Goedecker (1995) and for fullerenes by Itoh {\it et al.} (1996).

Let us now make plausible the exponential decay of the density matrix. 
The demonstration is based on the fact, that one can express
the Fourier components $\epsilon_n({\bf R})$ of the
band energy $\epsilon_n({\bf k})$ through the Wannier functions $W_n({\bf r})$
\begin{equation} \label{enkfour}
\epsilon_n({\bf R}) = 
\frac{V}{(2 \pi)^3} \int_{BZ} \epsilon_n({\bf k}) e^{-i{\bf k}{\bf R}} \: d{\bf k} = 
\frac{(2 \pi)^3}{V} \int_{space} W^*_n({\bf r}') H W_n({\bf r}'-{\bf R}) \: d{\bf r}' \: ,
\end{equation}
where ${\bf R}$ is a Bravais lattice vector. 
Now it is known, that the band energy $\epsilon_n({\bf k})$ is an analytic
function (Blount). This is actually not surprising. The first and
second derivatives of the band-structure have physical meaning since they
are related to the electron velocity and effective mass. 
So it is to be expected that higher derivatives exist as well.
Since the Fourier transform of an analytic function decays faster than 
algebraically (See Appendix) there exists a decay constant $\gamma$ and 
a normalization constant $C$ such that 
\begin{equation} \label{decins}
C e^{-\gamma R}  \geq  \epsilon_n({\bf R})  
  =  \frac{1}{V}  \int_{space} W^*_n({\bf r}') H W_n({\bf r}'-{\bf R}) \: d{\bf r}'  
\end{equation}
It is reasonable to expect that $H W_n({\bf r})$ will behave similarly as $W_n({\bf r})$.
In particular we expect $W_n({\bf r})$ to be small whenever $H W_n({\bf r})$ is small.  
So we will just drop $H$ in Equation~(\ref{decins}). In addition we will define this modified 
integral not only for lattice vectors ${\bf R}$ but for arbitrary vectors ${\bf r}$ to obtain.
\begin{equation} \label{decinsul}
C e^{-\gamma r}  \geq  
    \frac{1}{V}  \int_{space} W^*_n({\bf r}') W_n({\bf r}'-{\bf r}) \: d{\bf r}'  
\end{equation}
If Equation~(\ref{decinsul}) holds, then one can use the mean value theorem to show that 
\begin{eqnarray} 
C e^{-\gamma r}  & \geq & 
         \frac{1}{V}  \int_{space} 
         W^*_n({\bf r}') W_n({\bf r}'-{\bf r}) \: d{\bf r}' \nonumber \\
   & = & \frac{1}{V}  \sum_{{\bf R}'} \int_{cell}
        W^*_n({\bf r}'-{\bf R}') W_n({\bf r}'-{\bf R}'-{\bf r}) \: d{\bf r}' \nonumber \\ 
   & = &  \sum_{{\bf R}'}
        W^*_n({\bf s}({\bf r})-{\bf R}') W_n({\bf s}({\bf r})-{\bf R}'-{\bf r}) \: d{\bf r}' \nonumber  \\
   & = & F({\bf s}({\bf r}),{\bf s}({\bf r})-{\bf r}) \label{transins}
\end{eqnarray}
where the mean value ${\bf s}({\bf r})$ is a vector within the primitive cell. Assuming that 
the density matrix has the same order of magnitude within each cell one can neglect 
the dependence of ${\bf s}$ on ${\bf r}$ to obtain the final result
\begin{equation} 
C e^{-\gamma r}   \geq  F({\bf s},{\bf s}-{\bf r}) 
\end{equation}

\begin{figure}[ht]       
     \begin{center}
      \setlength{\unitlength}{1cm}
       \begin{picture}( 8.,6.5)           % figure dimensions
        \put(-1.5,-1.5){\includegraphics{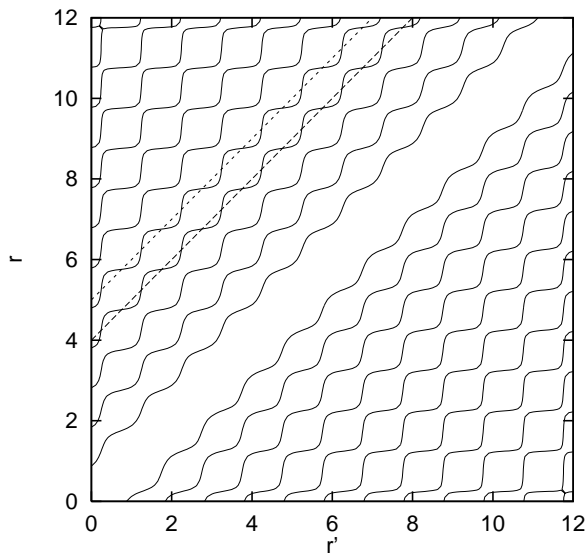}}   % VAX
       \end{picture}
       \caption{\label{nodes} \it The nodal structure  of the density matrix 
                 for a 1-dimensional model 
                 insulator with a bandwidth of 4 a.u. and a band gap of 2 a.u.. 
                 The length of the primitive cell is 1.
                 The nodes predicted by Equation~(\ref{nodpre}) are at the 
                 intersections with diagonal lines, two of which  are shown 
                 by the dashed lines.}
      \end{center}
     \end{figure}

The numerically observed nodal structure of the density matrix can be 
motivated in a very similar way.
Because of the orthogonality of the Wannier functions we have 
\begin{equation} 
0 = \int_{space} W^*_n({\bf r}') W_n({\bf r}'-{\bf R}) \: d{\bf r}'
\end{equation}
for any non-zero lattice vector ${\bf R}$. Doing the same sequence of transformation as 
in Equation~(\ref{transins}) one obtains
\begin{equation} \label{nodpre}
0 = F({\bf s}({\bf R}),{\bf s}({\bf R})-{\bf R})
\end{equation}
So there has to be one node in each cell. The numerically calculated nodal structure 
for a 1-dimensional model insulator is shown in Figure~\ref{nodes}. 

The next step is to examine in a more quantitative way which factors determine 
the rate of this exponential decay for  
an insulator with a band gap $\epsilon_{gap}$ at zero temperature.

Cloizeaux (1964) proved the exponential decay behavior of the zero 
temperature density matrix, which is a projection operator. Considering 
the extension of the band energy $\epsilon_n({\bf k})$ into the 
complex ${\bf k}$ plane he found that 
the minimal distance of the branch points of $\epsilon_n({\bf k})$ 
from the real axis determines the decay behavior.
For the Wannier functions, which are closely related to the density matrix by 
Equation~(\ref{denst0}), Kohn (1959) proved the 
same decay behavior in the case of a one-dimensional model crystal.
In a later publication Kohn (1993) claims that this distance to the real ${\bf k}$ axis 
should be related to the square root of the gap. Even though he did not present 
a derivation of this result, it was widely accepted to be generally valid.
Ismail-Beigi and Arias (1998) have however shown that Kohn's claim is not generally
valid. They demonstrated that in the Tight Binding limit the square 
root behavior can be found under certain circumstances, but that 
different behaviors 
can be found as well. In the weak binding limit, where the band-structure
can be obtained by perturbation theory from the band structure of the
free electron gas, they showed that 
the dependence is actually linear. 

\begin{equation} \label{insulator} 
F({\bf r},{\bf r}') \propto \exp( - \gamma |{\bf r}-{\bf r}'| )
             \hspace{1cm} where \:   
           \gamma = c \: \epsilon_{gap} \: a 
\end{equation}
The lattice constant is denoted by $a$, and $c$
is an unknown constant of the order of 1.

The dependence of the decay rate on the size of the band gap is a rather surprising 
relation. After all it follows from Equation~(\ref{ftrans}) that only the properties 
of the occupied bands enter into the calculation of the density matrix, whereas 
the size of the gap is not directly related to the occupied states. 
In the following we will give an intuitive explanation of the factors 
determining the decay rate. This explanation will again be based on Equation~(\ref{enkfour}) 
relating  the bandstructure to the decay properties of the density matrix. 
As is known from complex analysis, the distance 
of the singularities from the real axis is comparable to the length over which 
one has very strong variations along the real axis of a complex function.
Now, the long range decay properties of a Fourier transform 
are exactly determined by the length 
$\Delta k$ of such a region of strongest variation (See Appendix). 
One thus regains Cloizeaux's result that the decay rate is proportional to 
the distance of singularities from the real axis. 
Let us now explain the behavior found in the weak binding limit by Ismail-Beigi and Arias.
In the weak binding limit the effective mass establishes the connection 
between the gap and the important features of the occupied bands. 
The effective mass for the $n$-th band at the point ${\bf k}_0$ is defined as (Kittel 1963) 
\begin{equation}
\frac{1}{m} = 1 + \frac{2}{3} \sum_{m \neq n} 
    \frac{ | \int \Psi^*_{n,{\bf k}_0}({\bf r}) \nabla \Psi_{m,{\bf k}_0}({\bf r}) d{\bf r} |^2 }
         {\epsilon_n({\bf k}_0)-\epsilon_m({\bf k}_0)}
\end{equation}
Since we are only interested in order of magnitudes, 
we have here averaged over the diagonal elements of the effective mass tensor 
in order to obtain a effective mass which is a scalar quantity.
In the case of the weak binding limit, a gap will open up at the boundaries of the Brioullin 
zone and this gap will be small. The effective mass is therefore small 
and proportional to $a ^2 \epsilon_{gap}$, where we have assumed that the 
dipole matrix elements 
$ \int \Psi^*_{n,{\bf k}_0}({\bf r}) \nabla \Psi_{i,{\bf k}_0}({\bf r}) d{\bf r}$
are on the order of $\frac{1}{a}$. The band-structure near the boundaries 
of the Brioullin zone is then given by 
\begin{equation} \label{bseff} 
 \frac{1}{2 m} (\Delta k )^2 \propto \frac{1}{a^2 \epsilon_{gap}} (\Delta k )^2 
\end{equation}
where $\Delta k$ is the distance from the boundary, neglecting directional 
effects. Since the effective mass is small, the curvature of the band-structure 
is large in this region. Hence 
this region is just the region with the strongest variation. As is well known 
(Ashcroft and Mermin 1976), 
the perturbation theory arguments leading to Equation~(\ref{bseff}) are valid 
within an energy range of the order of $\epsilon_{gap}$. It then follows from 
Equation~(\ref{bseff}) that the corresponding range of 
$\Delta k$ is $\epsilon_{gap} \: a$, confirming the linear 
decay of the density matrix with respect to the size of the gap, i.e. 
$\gamma = c \: \epsilon_{gap} \: a $. 

Let us next show how a square root like behavior $\gamma = c \:  \sqrt{ \epsilon_{gap} }$ 
can arise for real crystals with a big gap. In this case the effective 
mass is of the order of one at all stationary points ${\bf k}_0$ in the Brioullin zone. 
Assuming that it is then of the order of one over the whole Brioullin zone, the 
region of largest variation is just the Brillouin zone itself. The decay constant is 
therefore simply related to the lattice constant $a$. 
\begin{equation} \label{insulator2} 
           \gamma =  c \frac{1}{a} 
\end{equation}
In order to get the square root dependence of the decay constant 
$\gamma$, one has to assume that
\begin{equation} \label{harrison} 
\epsilon_{gap} = C_{gap} \frac{1}{a^2}
\end{equation}
where $C_{gap}$ is a constant which is not or only weakly dependent on the material. 
Such a behavior has indeed been observed for certain classes of materials, 
where the tight binding limit is the most appropriate one, such as 
ionic crystals (Harrison 1980), but with a non-negligible variation of $C_{gap}$ 
across different materials.
A square root behavior of $\gamma$ can therefore be expected if one varies the 
lattice constant for a certain material, but the decay constants for different 
materials that happen to have the same gap are not necessarily comparable. 

\begin{figure}[ht]       
     \begin{center}
      \setlength{\unitlength}{1cm}
       \begin{picture}( 8.,5.0)           % figure dimensions
        \put(-1.5,-3.){\includegraphics{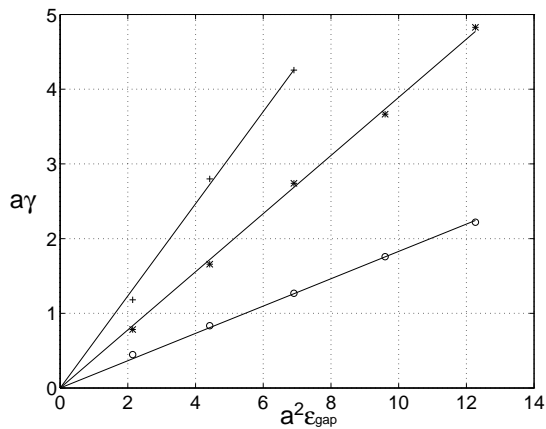}}   % VAX
       \end{picture}
       \caption{\label{beigi} \it The dependence of the decay constant $\gamma$ 
       on the gap. Plotted are the two dimensionless quantities $a \: \gamma$ versus 
       $a^2 \epsilon_{gap}$. The variation of the gap was obtained for a 
       three-dimensional cubic model crystal by varying the strength of the potential. 
       Circles refer to the [100], stars to the [110] and pluses to the [111] direction.
       This figure is reproduced with kind permission of the authors 
       from Ismail-Beigi and Arias (1998)} 
      \end{center}
     \end{figure}

In practice the distinction between the Tight Binding and weak binding case 
may not always be clear. Unless 
the region of strongest variation is really a very small fraction of the whole 
Brioullin zone, all the prefactors which were neglected in these considerations 
might be important enough to blur out differences. The importance of these prefactors 
can also be seen from the fairly strong directional dependence of the decay rate.  
Ismail-Beigi and Arias (1998) found such a strong directional dependence in numerical tests 
to confirm the linear dependence of the decay constant on the size of the 
gap (Figure~\ref{beigi}). Stephan {\it et al.} (1998) found the same behavior 
during Tight Binding studies of carbon.  So a statement in an old paper by Kohn (1964),
namely that the decay length of the Wannier functions is of the order of the 
interatomic spacing, is for practical purposes probably in many cases the best 
available characterization of localization. 

\begin{figure}[p]       
     \begin{center}
      \setlength{\unitlength}{1cm}
       \begin{picture}( 8.,5.5)           % figure dimensions
        \put(-1.5,-1.){\includegraphics{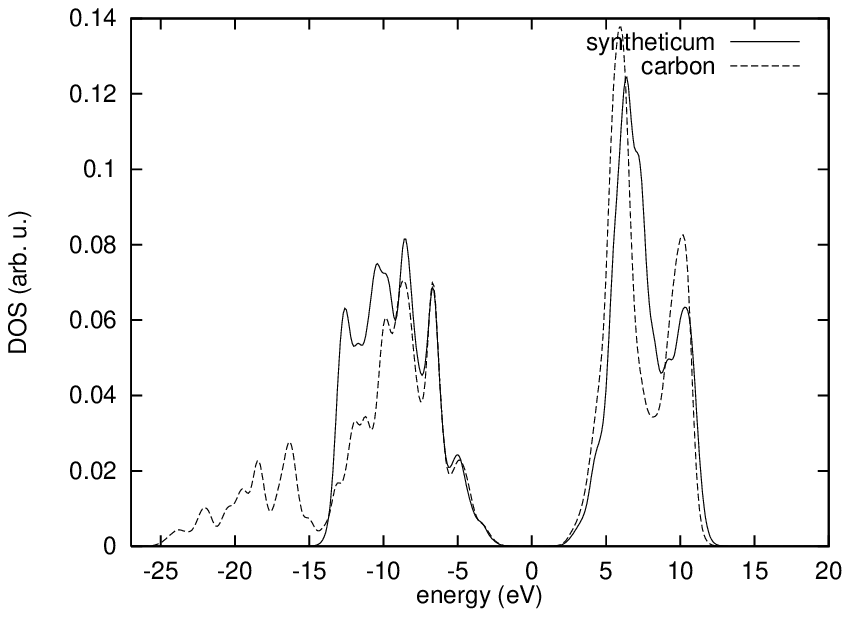}}   % VAX
       \end{picture}
       \caption{\label{syndos} \it Comparison of the density of states 
          of carbon and syntheticum. As one sees both have roughly the 
          same gap.  The lower part of the valence band is however drastically different. 
          The valance band of syntheticum is roughly only half as wide as the 
          one of carbon}
      \end{center}
     \end{figure}

As a numerical illustration of this surprising result, that only a small 
part of the Brillouin zone where one has the strongest variation 
determines the decay behavior of the Wannier functions, 
we compared the decay behavior of carbon in the diamond structure with 
"syntheticum" in the same structure. The artificial element "syntheticum" was 
computer generated within the Tight Binding context in such a way that 
its top part of the conduction band as well as the gap is nearly identical to real carbon, 
whereas the lower part of the valence band is drastically different as shown in Figure~\ref{syndos}.
More precisely, Carbon was characterized by the parameters of Goodwin (1991) and 
to obtain syntheticum $\epsilon_{s}$ was modified from -5.16331 to -1.16331 and 
$V_{s s \sigma}$ was modified from -4.43338 to -2.43338.

\begin{figure}[p]       
     \begin{center}
      \setlength{\unitlength}{1cm}
       \begin{picture}( 8.,5.5)           % figure dimensions
        \put(-1.5,-1.){\includegraphics{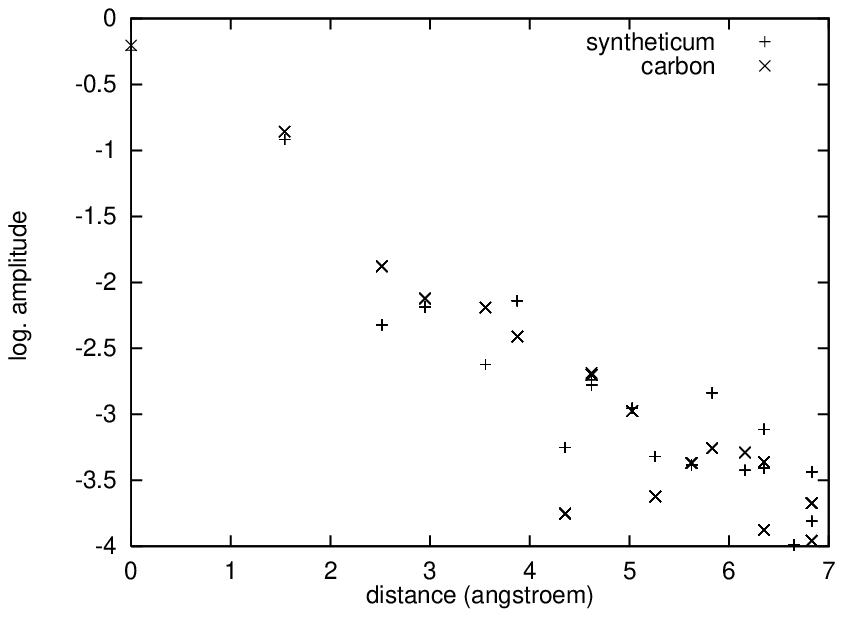}}   % VAX
       \end{picture}
       \caption{\label{syndec} \it Comparison of the decay behavior of the 
           density matrices for carbon and syntheticum. They are both very 
           similar. The moderate scattering comes from the fact that 
           the density matrix does not decay equally fast in all directions}
      \end{center}
     \end{figure}

Figure~\ref{syndec} shows the decay behavior of the density matrix. As one can see 
the decay behavior is very similar in both cases. We note that not only the 
gap is similar but also the effective mass since the density of states 
at the top of the valence band has the same behavior in both materials. 

% This example does however not completely 
% rule out that other factors in addition to the gap might influence the decay. 
% For instance the effective mass for both carbon and syntheticum is practically the same 
% as can be seen from the nearly identical density of states at the top of the valence band. 

All the above arguments apply to simple and mainly periodic materials.
Advanced electronic structure calculations however frequently study materials which are 
not in this class. The localization properties of such materials have not yet been 
studied systematically and so there is some incertitude about 
which orbitals are localized and to what extent (Kohn 1995). If the localization properties 
are unknown one should better not impose any localization constraints. In this case 
some of the discussed O(N) techniques still give a quadratic scaling, which 
also allows us to gain computational efficiency compared to the traditional 
cubically scaling algorithms.

\section{Basic strategies for O(N) scaling}
\label{four}
Most O(N) algorithms are built around the density matrix or its 
representation in terms of Wannier functions and take advantage of its 
decay properties. To obtain linear scaling one has to cut off the exponentially decaying 
quantities when they are small enough. This introduces the concept of a localization region.
Only inside this localization region the quantity is calculated, outside it is 
assumed to vanish. For simplicity the localization region is usually taken to be 
a sphere, even though the optimal shape might be different (Stephan 1998).
In the Tight Binding context the boundary of the localization region can either be defined 
by a geometric distance criterion or in terms of the number of "hops", i.e. 
the number of steps one has to do along bonds connecting neighboring atoms to reach this 
boundary (Voter {\it et al.,} 1996). Different localization regions generally have significant overlaps.
The localization regions thus do not form a partition of the computational volume and 
one atom in general belongs to several localization regions.

In a numerical calculation the density operator $F(r,r')$ is discretized with respect 
to a basis. The basis set has to be chosen such that the matrix elements 
$F_{i,j}$ reflect the decay properties of the operator $F(r,r')$. This will 
obviously only be the case if the basis set consists of localized functions, such 
as atom centered Gaussian type basis functions. Sets of orthonormal basis functions 
usually facilitate the calculations. Unfortunately all currently used localized basis 
sets are non-orthogonal. In the context of the orthogonal 
Tight Binding scheme (Goringe {\it et al.,} 1997a, Majewski and Vogl 1986) one 
just assumes the existence of an 
basis set which is both atom centered and orthogonal. 
Since only the parameterized 
Hamiltonian matrix elements enter in the calculation, there is no need to explicitly ever 
construct such a basis set. In the following sections, we will follow this 
practice and assume in all relevant parts that we are dealing with such 
a localized orthogonal basis set. The non-orthogonal case will be discussed 
in section~\ref{nonorthog}. Whenever we refer from now on to a localization 
region, we actually mean the subset of all basis functions which are contained 
within this spatial localization region.

Obviously the size of the localization region needed to obtain 
a certain accuracy depends on the decay properties of the density matrix as 
well as on the selected accuracy threshold.
It also depends on the quantity one wants to study. Generally, the total energy as 
well as derived quantities such as the geometric equilibrium configurations are 
surprisingly insensitive to finite localization regions, because these quantities 
are not strongly influenced by the exponentially small tails which are cut off 
by the introduction of a localization region. This insensitivity also holds true, 
even though to a much lesser extent, for metals. As we have seen above the introduction 
of a finite temperature leads to an exponential decay of the density matrix 
which in turn justifies truncation. 
In a metal, the difference between the finite and the zero temperature 
total energy $\Delta E$ is proportional to the 
square of the temperature, $\Delta E \propto T^2$, (Ashcroft and Mermin 1976) and thus 
rather small. There are however quantities which 
are very sensitive to finite localization regions. 
In the modern theory of polarization in solids (King-Smith and Vanderbilt 1989), 
the polarization can be expressed 
in terms of the centers of the Wannier functions 
$\int W({\bf r }) {\bf r } W({\bf r }) d{\bf r }$. 
Using this formula (Fernandez {\it et al.,} 1997) one has a strong influence of the 
tails of the Wannier functions because they get strongly weighted by the factor 
of ${\bf r }$ in the integral. Since the tails are much more influenced by the 
boundary of the localization region than the central part, this quantity is more sensitive 
to the size of the localization region. 

There are even quantities which are not at all directly accessible by 
a solution which is given in terms of density matrices or Wannier functions. The Fermi 
surface in a metal which can be calculated via the eigenvalues of the band 
structure $\epsilon_n({\bf k})$ is such an example. 

It is also clear that one can gain significant computational efficiency only if the 
size of the system is larger than the size of the localization region.
When this criterion is fulfilled depends not only on the decay properties 
of the density matrix of the system but also on its dimensionality. 
In the case of a linear chain molecule with a large band gap, it might be enough 
to have a localization region containing just two neighboring atoms on each side.
So the localization region would just contain 5 atoms and for systems larger than 5 atoms 
one might potentially gain computational efficiency by using an O(N) method. 
If one has a 3 dimensional system 
with a comparable gap, then a spherical localization region extending out to the second 
neighbors would contain some 60 atoms and the crossover point would already be much larger. 
For a system with a small gap such as silicon or for metallic systems the 
crossover point is even larger.

There are essentially six basic approaches to achieve linear scaling. 
\begin{itemize}
\item
The Fermi Operator Expansion (FOE) is based on Equation~(\ref{foe}). 
In this approach one 
finds a computable functional form of $F$ as a function of $H$ to build up the 
density matrix. Two possible representations 
based on a Chebychev expansion and a rational expansion will be discussed. 
\item
The Fermi Operator Projection (FOP) is closely related to the FOE method.
The computable form of $F$ is however not used to construct the entire 
density matrix but to find the space spanned by the occupied states, 
i.e. the space corresponding to the eigenfunctions
associated with the unit eigenvalues of the Density matrix 
at zero temperature. These eigenfunctions can be 
considered as Wannier functions in the generalized sense defined before. 
\item
In the Divide and Conquer (DC) method for the density matrix 
the relevant parts of the density matrix are patched together 
from pieces that were calculated for smaller subsystems.
\item
In the Density Matrix 
Minimization (DMM) approach, one finds the  density matrix 
by a minimization of an energy expression based on the density matrix. 
\item
In the Orbital Minimization approach (OM), one finds a set of Wannier functions
by minimization of an energy expression. 
\item 
The Optimal Basis Density Matrix Minimization scheme (OBDMM) contains aspects 
of both the OM and DMM methods. In addition to finding a density matrix with respect 
to the basis, one also finds an optimal basis by additional 
minimization steps. The number of basis functions has to be at least equal 
to the number of electrons in the system, but can be bigger as well. 
\end{itemize}

A major difference between these methods is whether they calculate the full 
density matrix or only its representation in terms of Wannier functions. 
The later approach applies only to insulators while the former is in 
also applicable to systems with fractional occupation numbers 
(i.e. $f(\epsilon_n)$ is not either 1 or 0) such as 
metals or systems at finite electronic temperature. 

In the following each of these six approaches will be presented in detail.

\subsection{The Fermi Operator Expansion}
The FOE (Goedecker and Colombo 1994, Goedecker and Teter 1995) 
is the most straightforward approach for the calculation of the density matrix.
The basic idea in this approach is to find a representation of the matrix 
function~(\ref{foe}) which can be evaluated on a computer. 
Several such representations are possible. We will discuss a Chebychev and 
a rational representation.

\subsubsection{The Chebychev Fermi Operator Expansion}
\begin{figure}[ht]       
     \begin{center}
      \setlength{\unitlength}{1cm}
       \begin{picture}( 8.,5.0)           % figure dimensions
        \put(-1.5,-0.5){\includegraphics{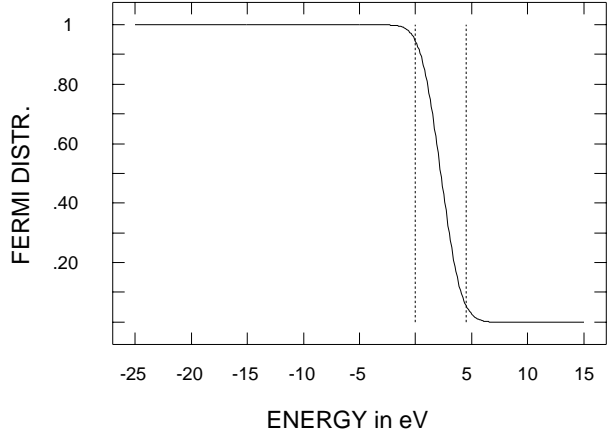}}   % VAX
       \end{picture}
       \caption{\label{foedist} \it The Fermi distribution as obtained by a Chebychev 
                fit of degree 40 in the case of a diamond structure. 
                The bandgap is in between the two vertical lines.}
      \end{center}
     \end{figure}

One of the most basic 
operations a computer can do are matrix times vector multiplications. The 
simplest representation of the density matrix would therefore be a polynomial 
representation 
$$ F \approx p(H) = c_0 I + c_1 H + c_2 H^2 + \: ...  \: + c_{n_{pl}} H^{n_{pl}} \: .$$
where $I$ is the identity matrix.
Unfortunately polynomials of high degree become numerically unstable. This 
instability can however be avoided by introducing a Chebychev polynomial representation, 
which is a widely used numerical method (Press {\it et al.,} 1986)
\begin{equation}
p(H) = \frac{c_0}{2} I + \sum_{j=1}^{n_{pl}} c_j T_j(H) \: .
\end{equation}
Since the Chebychev polynomials are defined only within the interval [-1:1], 
we will assume in the following that the eigenvalue spectrum of $H$ 
falls within this interval. This can always be easily achieved by scaling and 
shifting of the original Hamiltonian. 
The Chebyshev matrix polynomials $T_j(H)$ satisfy the recursion relations
\begin{eqnarray}
T_0(H) & = & I  \\
T_1(H) & = & H  \\
T_{j+1}(H) & = & 2\: H \: T_j(H) - T_{j-1}(H) \: .
\end{eqnarray}
The expansion coefficients of the Chebychev expansion can easily be determined. 
The eigenfunction representation (Equation~(\ref{fevcs})) of $F$ is, 
\begin{equation}
< \Psi_n|F|\Psi_m>  = f(\epsilon_n) \: \delta_{n,m} \: .
\label{eq:fdia1}
\end{equation}
Evaluating the polynomial expansion in the same eigenfunction representation we obtain
\begin{equation}
< \Psi_n|p(H)|\Psi_m>  = p(\epsilon_n) \: \delta_{n,m} \: ,
\label{eq:fdia2}
\end{equation}
where
\begin{equation}
p(\epsilon) = \frac{c_0}{2}  + \sum_{j=1}^{n_{pl}} c_j T_j(\epsilon) \: .
\end{equation}
Comparing Equation~(\ref{eq:fdia1}) and Equation~(\ref{eq:fdia2})
we see that the polynomial $p(\epsilon)$ has to
approximate the Fermi distribution in the energy interval [-1:1] where
the scaled and shifted Hamiltonian has its eigenvalues. 
How to find the Chebychev expansion coefficients 
for a scalar function is described in standard textbooks on numerical 
analysis (Press {\it et al.,} 1986). Actually it is not necessary to take the exact Fermi distribution. 
In practically all situations one is interested in the limit of zero 
temperature. Hence any function which approaches a step functions 
in the limit of zero temperature can be used. In the case of simulations 
of insulators for instance it is advantageous to take the 
function $f(\epsilon) = \frac{1}{2} (1-{\rm erf}(\frac{\epsilon-\mu}{\Delta \epsilon}))$
shown in Figure~\ref{foedist} 
since it decays faster to 0 respectively 1 away from the chemical potential.
The term Fermi distribution will in the following always be used in this broader sense.
The energy resolution $\Delta \epsilon$ is chosen to be a certain fraction of the size 
of the gap (Goedecker and Teter 1995). In the case of metals, $\Delta \epsilon$ is 
chosen by considerations of numerical convenience. Large values of $\Delta \epsilon$ 
will give lower accuracy results. However as pointed out before, the convergence 
of the total energy with respect to $\Delta \epsilon$ is quadratic and so 
highly accurate total energies can be obtained with rather high values 
of $\Delta \epsilon$ (Goedecker and Teter 1995). Small values of $\Delta \epsilon$ 
make the calculation numerically expensive. The detailed scaling behavior of the 
numerical effort in the limit of vanishing gaps is analyzed in section~\ref{compare}, 
where it is found that actually the increase in the size of the localization region 
is the limiting factor in all methods. 

Even if one wants to study electronic properties in the limit of zero electronic temperature
it is important that one nevertheless uses a finite temperature 
Fermi distribution for the Chebychev fit. Using the zero temperature 
step function introduces so-called Gibbs oscillations in the fit and spoils 
the Chebychev fit over the whole interval. 

How to eliminate these Gibbs oscillations in the zero temperature case by 
the so called kernel polynomial method (Voter {\it et al.,} 1996, Silver {\it et al.,}1996) 
can be used as a starting point for an alternative derivation of the 
FOE method. The basic idea is to expand 
a delta function as a polynomial using damping factors to suppress large 
oscillations. This representation of an approximate delta function 
can then be integrated to obtain a smooth representation of the 
Fermi distribution. Used this way the kernel polynomial method is thus just 
another way to derive the expansion coefficients for the 
Chebychev expansion (Kress {\it et al.,} 1998).
In addition the kernel polynomial method can also be used 
to smear out the density of states rather than the 
zero temperature Fermi distribution resulting in a method with practically identical 
computational requirements but some slightly different properties. One useful 
property is that the smeared density of states energy is an approximate lower bound to the 
energy, whereas the smeared Fermi energy is an approximate upper bound (Voter {\it et al.,} 1996).

Coming back to the original motivation for a polynomial representation, 
let us now show how the density matrix can be constructed using only matrix times 
vector multiplications. Let us denote by $t^j_l$ the l-th column of the Chebychev matrix 
$T_j$. Now each column of these Chebychev matrices satisfies the same recursion relations
\begin{eqnarray}
|t^0_l> & = & |e_l>  \label{recurs} \\
|t^1_l> & = & H|e_l>    \nonumber \\
|t^{j+1}_l> & = & 2\: H \: |t^j_l> - |t^{j-1}_l> \nonumber \: .
\end{eqnarray}
where $e_l$ is a unit vector that has zeroes everywhere except at the $l$-th entry.
So Equation~(\ref{recurs}) demonstrates that we indeed need only matrix vector multiplications.
Once we have generated the $l$-th columns of all the Chebychev matrices, we can obtain the 
$l$-th column $f_l$ of the density matrix just by forming linear combinations
\begin{equation}
|f_l> = \frac{c_0}{2} |t^0_l> + \sum_{j=1}^{n_{pl}} c_j |t^j_l> \: .
\end{equation}
As we have described the method so far it has a quadratic scaling instead of the 
linear scaling we finally want to achieve. If we have 
$M_b$ basis functions, the density 
matrix is a $M_b \times M_b$ matrix and we have to calculate $M_b$ full columns. 
For the calculation 
of each column, we have to do $n_{pl}$ matrix times vector multiplications, each of which 
costs $M_b n_H$ operations assuming the matrix $H$ is a sparse matrix with $n_H$ off-diagonal 
elements per row/column. 
So the total computational cost is $M_b^2 \: n_{pl} \: n_H$. 
The degree of the polynomial $n_{pl}$ 
and the width $n_H$ of the Hamiltonian are independent of the size of the system, 
whereas $M_b$ is proportional to the number of atoms in the system. The overall 
scaling with respect to the number of atoms is therefore quadratic. 

In order to do the correct shifting and scaling of the original Hamilton to map 
its eigenvalue spectrum on the interval [-1:1] we have to know its 
lowest and highest eigenvalues $\epsilon_{min}$ and $\epsilon_{max}$. 
In addition we have to know the chemical potential $\mu$. 
There are auxiliary matrix functions of $H$ that can help 
us to determine these quantities.  These functions of $H$ can be build up in the 
same way as the density matrix. Since the recursive build up 
of the Chebychev matrices is the most costly part, the additional cost for evaluating 
other functions is negligible. To determine whether we have a vanishing density 
of states beyond an energy $\epsilon_{up}$ we can for instance construct a Chebychev 
fit $p_{up}(\epsilon)$ to a function which is zero (to within a certain tolerance) for energies 
below $\epsilon_{up}$, but blows up for energies larger than $\epsilon_{up}$.
If $Tr[p_{up}(H)]$ does not vanish we have a non-vanishing density of states beyond 
$\epsilon_{up}$. A similar procedure can be applied to determine a lower 
bound for the density of states. The determination of the chemical potential in 
an insulator can be done along the same lines as well (Bates and Scuseria 1998).
Without any significant extra cost one can build up several Fermi distributions 
with different chemical potentials until one finds the correct chemical potential 
leading to charge neutrality. In a metallic system the search for the chemical 
potential can be accelerated since it is possible to predict with high accuracy 
how the number of electrons changes in response to a change in the chemical potential.
From Equation~(\ref{nel}) it follows
\begin{equation} \label{dnel}
\frac{\partial N_{el}}{\partial \mu}  = - Tr [p'(H)] \: ,
\end{equation}
where $p'$ is the derivative of the Chebychev polynomial $p$ that approximates the 
Fermi distribution. The Chebychev expansion coefficients of $p'$ can be calculated from 
the coefficients for $p$ (Press {\it et al.,} 1986). Using  the finite difference 
approximation of Equation~(\ref{dnel}),  
\begin{equation} \label{fddnel}
\Delta \mu = \frac{ \Delta N_{el}}{Tr [p'(H)]} \: ,
\end{equation}
it is possible to find the correction $\Delta \mu$ to the chemical potential which will 
nearly exactly eliminate an excess of $\Delta N_{el}$ electrons 
due to an incorrect initial chemical potential. 
The correct chemical potential in a metallic system can thus be found 
with very high accuracy with a few iterations.

The desired linear scaling can be obtained by introducing a localization region for each column, 
outside of which the elements are negligibly small. For the $k$-th column, this localization 
region will be centered on the $k$-th basis function. 
If we use atom centered basis functions, 
then the localization region will consequently be centered on the atom to which 
this $k$-th basis function belongs.
We have then to calculate only that part of each column which corresponds to this 
localization region. This means that we can use a truncated Hamiltonian  $H(k)$ 
which retains only the matrix elements corresponding to the basis 
functions contained within the localization region $k$. 
Denoting the number of basis functions in this region by $M_{loc}$ (which might 
actually depend on the localization region $k$ being considered), the overall 
computational cost is then $M_b M_{loc} \: n_{pl} \: n_H$ and thus scales linearly.
Let us stress, that the size of the localization region is independent of 
the degree of the polynomial. If one uses for instance a polynomial of 
degree $n_{pl}=50$, the recursion in Equation~(\ref{recurs}) will extend over the 
50 nearest neighbor shells without localization constraint for a Hamiltonian coupling  
only nearest neighbors. The localization region however is typically much smaller 
comprising just a few nearest neighbor shells.
Imposing a localization region introduces some subtleties. For instance the 
eigenvalues of the truncated density matrix are not anymore exactly given 
by $p(\epsilon_n)$ and $F$ is not any more strictly symmetric.
More importantly, strictly speaking we can no longer use the Trace notation, 
since we use different local Hamiltonians $H(k)$ to build up the different columns 
of the density matrix. The band-structure energy $E_{BS}$ has now to be written as 
\begin{equation} \label{etrun}
E_{BS} = \sum_k \sum_j [p(H(k))]_{k,j} [H(k)]_{j,k} \: .
\end{equation}

Another important quantity are the forces. The force acting on the $\alpha$-th 
atom at position $R_{\alpha}$ is obtained by differentiating the total energy 
with respect to these positions. The total energy consists of the band structure 
part and possibly other contributions. We will only discuss the non-trivial 
part of the force arising from the differentiation of the band structure energy $E_{BS}$.  
For simplicity let us assume that we have a simple polynomial expansion and 
not a Chebychev expansion. Let us also assume that we calculate the full 
density matrix, i.e. that we do not truncate $H$ by introducing a localization region. 
We then obtain
\begin{equation} 
\frac{d E_{BS}}{d R_{\alpha}}  =  
      \frac{ d}{d R_{\alpha}} Tr \left[ H \sum_{\nu} c_{\nu} H^{\nu}  \right] = 
      \sum_{\nu} c_{\nu} Tr \left[ \frac{\partial H^{\nu+1}}{\partial R_{\alpha}} \right] \: .
\end{equation}
Let us consider for instance the term for which $\nu = 2$
\begin{equation} \label{term3full}
      \frac{d Tr[H^{3}]}{d R_{\alpha}} = 
    Tr \left[   H H \frac{\partial H}{\partial R_{\alpha}} \right] + 
    Tr \left[     H \frac{\partial H}{\partial R_{\alpha}} H \right] + 
    Tr \left[       \frac{\partial H}{\partial R_{\alpha}} H H  \right] 
  = 3 Tr \left[   H H \frac{\partial H}{\partial R_{\alpha}}   \right] \: ,
\end{equation}
 where we used that $Tr[A B] = Tr[B A]$. The final result for the force, which 
also holds in the case of a Chebychev expansion, is thus 
\begin{equation} \label{foeforce}
\frac{d E_{BS}}{d R_{\alpha}}  =  
       Tr \left[ \left( p(H) + H p'(H) \right) \frac{\partial H}{\partial R_{\alpha}} \right] \: .
\end{equation}
In the case of an insulator, the second term in the brackets $H p'(H)$ is very small 
compared to the first term $p(H)$ at small but finite temperatures 
and it vanishes in the limit of zero temperature.
The reason for this is that the eigenvalues of the matrix $p'(H)$ 
are $p'(\epsilon_n)$. Since at zero temperature 
$p'(\epsilon)$ is nonzero only at the chemical potential which is in
the middle of the gap, all eigenvalues are zero and the matrix is
identically zero. Nevertheless it is recommendable to retain this term 
in numerical calculations because it leads to forces consistent with the total 
energy.  

In the case where we calculate only part of the density matrix, i.e. where  
we have a truncated Hamiltonian $H(k)$ going with the energy expression~(\ref{etrun})  
we cannot use the properties of the 
trace to simplify the force expression as we did in Equation~(\ref{term3full}).
The equation corresponding to Equation~(\ref{term3full}) therefore reads

\begin{eqnarray} \label{term3trun}
   \sum_{k,j1,j2,k}
& & [H(k)]_{k,j1} [H(k)]_{j1,j2} 
     \left( \frac{\partial H(k)}{\partial R_{\alpha}} \right)_{j2,k} +  \\
& & [H(k)] _{k,j1} 
    \left( \frac{\partial H(k)}{\partial R_{\alpha}}\right)_{j1,j2} [H(k)]_{j2,k} + \nonumber \\
& & \left( \frac{\partial H(k)}{\partial R_{\alpha}} \right)_{k,j1} [H(k)]_{j1,j2} [H(k)]_{j2,k}      \nonumber \: .
\end{eqnarray}
Similar results hold for all the other terms with different values of $\nu$. 
In the case of a Chebychev expansion the situation is completely analogous,
just the formulas are more complicated. The force formula has been worked out in this case 
by Voter {\it et al.} (1996) and is given by
\begin{equation} \label{forcevoter}
      \frac{d T_j(H)}{d R_{\alpha}} = 
      \frac{d T_{j-2}(H)}{d R_{\alpha}} +
      \sum_{i=0}^{j-1} (1+k_i)(1+k_{j-1-i}) T_i(H) 
                 \frac{\partial H}{\partial R_{\alpha}} 
                 T_{j-1-i}(H) \: ,
\end{equation}
where $k_j = 0$ if $j \leq 0$ and $k_j = 1$ otherwise.
In the typical Tight Binding context 
$\frac{\partial H}{\partial R_{\alpha}}$ is a very sparse matrix. If it contains 
$n_{D}$ non-zero elements, we need of the order of $n_{pl}^2 \: n_{D} M_b$ operations
to evaluate all the forces according to Equation~(\ref{forcevoter}). 
The error incurred by using the approximate formula 
~(\ref{foeforce}) based on the trace is usually negligible if the localization 
region is large enough. Since the approximate formula can be evaluated with 
order $n_{pl} \: n_{D} M_b$ operations, it might actually be preferable to do so.
In a molecular dynamics simulation, the largest deviations in the conservation of the 
total energy come from events where atoms enter or leave localization regions 
and this kind of error is not taken into account by either force formula.

All the above force formulas were derived for the case where we have a constant chemical 
potential and where the polynomial representing the Fermi distribution does thus 
not change. Frequently one wants however to do simulations for a fixed number of electrons  
rather than for a fixed chemical potential. In this case one has to readjust the 
chemical potential for each new atomic configuration. The chemical potential is thus 
a function of all the atomic positions $\mu = \mu(R_{\alpha})$, but the explicit functional 
form of this dependence is not known. The force formula can however also be adapted 
to this case (B. Roberts and P. Clancy 1998). 
Ignoring the above warnings and using again trace 
notation for simplicity we have
\begin{eqnarray}
E_{BS} & = & Tr[H \: p(H-\mu I)] \\
N_{el} & = & Tr[p(H-\mu I)] 
\end{eqnarray}
and consequently
\begin{eqnarray}
\frac{d E_{BS}}{d R_{\alpha}} & =  &
    Tr \left[ ( H \: p' + p) \frac{\partial H}{\partial R_{\alpha}} \right] - 
    Tr [ ( H \: p' ) ] \frac{\partial \mu}{\partial R_{\alpha}}  \label{foededr} \\
\frac{d N_{el}}{d R_{\alpha}} & =  &
    Tr \left[ p' \frac{\partial H}{\partial R_{\alpha}} \right] - 
    Tr [ p' ]  \frac{\partial \mu}{\partial R_{\alpha}}  
                          \label{foedndr} \: .
\end{eqnarray}
Since $\frac{d N_{el}}{d R_{\alpha}}$ has to be equal to zero, we can solve 
Equation~(\ref{foedndr}) for $\frac{\partial \mu}{\partial R_{\alpha}}$ and 
insert it into equation~(\ref{foededr}) to obtain the force under the constraint of 
a constant number of electrons.

Let us finally derive a force formula for the case where 
a local charge neutrality condition is enforced (Kress {\it et al.,} 1998). 
The motivation for this approach is that
in non-self-consistent Tight Binding calculations one frequently finds an unphysically 
large transfer of charge between atoms. In a self-consistent calculation the 
electrostatic potential, built up by a charge transfer, is counteracting a further 
charge flow and thus limits charge transfer to reasonably small values.
Some Tight Binding schemes (Horsfield {\it et al.,} 1996a) enforce a so-called 
local charge neutrality 
condition requiring that the total charge associated with an atom in a molecule 
or solid be equal to the charge of the isolated atom. This is done by determining 
a potential offset $u_{\alpha}$ for each atom $\alpha$ in the system which 
will ensure this neutrality. The total Hamiltonian $H$ of the system is then given 
by $H_0+U$ where $H_0$ is the Hamiltonian without any potential bias and $U$ a 
diagonal matrix containing the atomic potential offsets $u_{\alpha}$.
The band structure energy is given by
\begin{equation} \label{foe273}
E_{BS} = Tr[(H_0 + U) \: p(H_0 + U)] -  \sum_{\alpha} Q_{\alpha} u_{\alpha} \: ,
\end{equation}
where the term containing the atomic valence charges $Q_{\alpha}$ has been subtracted 
to make the expression invariant under the application of a uniform
potential bias to all atoms in the system.
Expressed in terms of the density matrix 
the local charge neutrality condition becomes 
\begin{equation} \label{foe499}
\sum_l p(H)_{\alpha,l;\alpha,l} = Q_{\alpha} \: .
\end{equation}
In Equation~(\ref{foe499}) we have labeled the basis functions by a composite index 
where $\alpha$ indicates on which atom the basis function is centered and where 
$l$ describes the character of the atom centered basis function. If we have carbon 
atoms, for which $Q_{\alpha}=4$, $l$ would for instance denote the 4 orbitals 
$2s$, $2px$, $2py$, $2pz$.
Using Equation~(\ref{foe499}), Equation~(\ref{foe273}) then simplifies to
\begin{equation} \label{foe473}
E_{BS} = Tr[H_0  \: p(H_0 + U)]  \: .
\end{equation}
Taking the derivative we get
\begin{equation} \label{foe573}
 \frac{d E_{BS}}{d R_{\alpha}}  = 
  \sum_{\beta} \frac{\partial E_{BS}}{\partial u_{\beta}} 
               \frac{\partial u_{\beta}}{\partial R_{\alpha}} +
                   \frac{\partial E_{BS}}{\partial R_{\alpha}}  \: ,
\end{equation}
where  
\begin{equation} \label{foe673}
  \frac{\partial E_{BS}}{\partial u_{\beta}} = 
        Tr \left[ H_0 \: p'(H) \frac{\partial H}{\partial u_{\beta}} \right] \: .
\end{equation}
As discussed above, the matrix $p'(H)$ is close to zero in an insulator 
at sufficiently low temperature and can often be neglected. 
The forces are therefore approximately given by
\begin{equation} \label{foe773}
 \frac{d E_{BS} }{d R_{\alpha} }  = 
        Tr \left[ H_0 \: p(H) \frac{\partial H}{\partial R_{\alpha}} \right] \: .
\end{equation}
It has to be pointed out that to get sufficiently high accuracy the degree of 
the polynomial has to be higher than in the case of Tight Binding case without 
local charge neutrality.

The degree $n_{pl}$ of the polynomial needed to represent the Fermi distribution is 
proportional to 
\begin{equation} \label{npldegree}
\frac{\epsilon_{max}-\epsilon_{min}}{\Delta \epsilon} \: .
\end{equation}
This follows from the fact, that the $n$th order Chebychev polynomial has 
$n$ roots and so a resolution that is roughly proportional to $1/n$.
For the usual Tight Binding Hamiltonians the ratio in Equation~(\ref{npldegree}) 
is not very large and for silicon and carbon systems without gap states 
polynomials of degree 50 are 
sufficient. In contexts other than Tight Binding this ratio can however 
be fairly large and polynomial representation would become very inefficient.

\subsubsection{The Rational Fermi Operator Expansion}
\begin{figure}[ht]       
     \begin{center}
      \setlength{\unitlength}{1cm}
       \begin{picture}( 8.,3.5)           % figure dimensions
        \put(-.5,-.5){\includegraphics{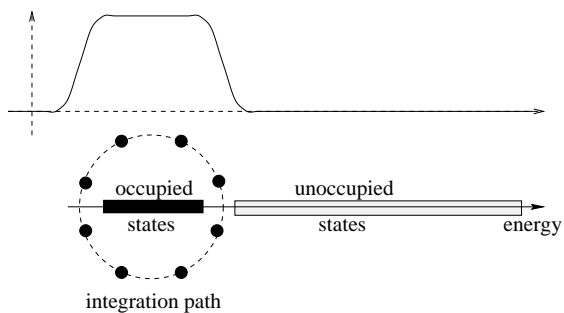}}   % VAX
       \end{picture}
       \caption{\label{figcontint} \it A discretization of the contour 
         integral in the complex energy plane 
         of Equation~(\ref{zercont}). The resulting Fermi 
         distribution is shown on top.}
      \end{center}
     \end{figure}

A rational representation of the density matrix (Goedecker 1995) is in this case more efficient
\begin{equation} \label{foepade}
 F = \sum_{\nu} \frac{w_{\nu}}{H-z_{\nu}}  \: .
\end{equation}
As is well known, the function $f(\epsilon)$ given by 
\begin{equation} \label{zercont}
f(\epsilon) = \frac{1}{2 \pi i} \oint \frac{dz}{\epsilon-z}
\end{equation}
is equal to 1 if $\epsilon$ is within the volume encircled by the 
contour integration path and zero otherwise. If the integration path 
contains the occupied states as shown in Figure~(\ref{figcontint}) it 
can therefore be used as a zero temperature Fermi distribution. 
The exact finite temperature Fermi distribution $f(\epsilon_n({\bf k}))$ can be obtained by 
replacing the path in this contour integral 
by a sequence of paths which do not intersect the real axis 
(Goedecker 1993, Gagel 1998, Nicholson and Zhang 1997).
Actually, as already mentioned above, it is usually not necessary to have the 
exact Fermi distribution. The electronic temperature is just determined by the 
slope (and possibly some higher derivatives) of the distribution at the Fermi energy. 
We will also refer to such generalized distributions as Fermi distributions. 
A distribution of this type can be obtained 
by discretizing the zero temperature contour integral from Equation~(\ref{zercont}) 
as shown in Figure~\ref{figcontint}. 

In principle any other set of 
$z_{\nu}$'s and $w_{\nu}$'s can be used as long as it satisfies
\begin{equation} 
 f(\epsilon) \approx \sum_{\nu=1}^{n_{pd}} \frac{w_{\nu}}{\epsilon-z_{\nu}}  \: ,
\end{equation}
where $n_{pd}$ is the degree of the rational approximation.
How can we now evaluate Equation~(\ref{foepade}) on a computer?
Denoting $\frac{I}{H-z_{\nu}}$ by $F_{\nu}$ we have 
\begin{eqnarray}
 (H-z_{\nu}) F_{\nu} & = & I \label{padeeq} \\
  F & = & \sum_{\nu} w_{\nu} F_{\nu}  \label{padesm} \: .
\end{eqnarray}
So we have first to invert all the matrices $H-z_{\nu}$ and then to form linear 
combinations of them. The inversion is equivalent to the solution of $M_b$ linear 
systems of equations. This can be effectuated using iterative techniques so that in the 
end everything can again be done by matrix times vector multiplications.
A rational approximation can represent the sharp variation near the chemical potential of a 
low temperature Fermi distribution in a more efficient way than a Chebychev approximation.
Whereas in the Chebychev case the degree of the polynomial is given by
Equation~(\ref{npldegree}) the degree of the rational approximation $n_{pd}$ is 
given by
\begin{equation}  \label{npddegree} 
n_{pd} = \frac{\mu-\epsilon_{min}}{\Delta \epsilon} \: .
\end{equation}
This $n_{pd}$ in contrast to $n_{pl}$ does not
depend on the largest eigenvalue $\epsilon_{max}$.
Once $n_{pd}$ is of the order of magnitude given by Equation~(\ref{npddegree}) 
one has exponential convergence to the zero temperature Fermi distribution.
In the case where the 
integration points and weights are obtained by discretizing the contour integral of 
Figure~\ref{figcontint} this exponential behavior is immediately comprehensible since 
an equally spaced integration scheme gives exponential convergence for periodic 
functions (Sloan and Joe, 1994). 
Since $n_{pd}$ is usually reasonably small, the success of the 
method will hinge upon whether it is possible to solve the linear system of equations 
associated with each integration point with a small number of iterations.
The number of iterations in an iterative method such as a conjugate gradient 
scheme (Press {\it et al.,} 1986) is related to whether it is possible to find 
a good preconditioning scheme. In the case of plane wave calculations  a good 
preconditioner can be obtained from the diagonal elements 
and of the order of 10 iterations are required. 
In other schemes using Gaussians for instance it is not quite clear whether 
good preconditioners can be found. When the Hamiltonian depends on the 
atomic positions $R_{\alpha}$, Equations~(\ref{padeeq}) and ~(\ref{padesm}) can 
be differentiated to obtain the derivative $\frac{d F}{d R_{\alpha}}$, 
which is needed for the calculation of the forces.

\subsection{The Fermi Operator Projection Method}
The FOE method is used to calculate the full 
density matrix. This can be inefficient if the number of basis functions 
per atom is very large.
As was mentioned before, the density matrix at zero temperature does not have full 
rank.  In the case of an insulator it can be constructed from $N_{el}$ 
Wannier functions~(\ref{denst0}).  
If one has a numerical representation of the zero temperature density operator, 
which is actually a projection operator, that eliminates all components belonging to 
eigenvalues above the Fermi level, one can apply it to a set 
of trial Wannier functions $\tilde{V}_n$, $n=1,...,N_{el}$ to generate a set of 
orbitals which span the space of the Wannier functions. 
The numerical representation of the density operator can again either be a 
Chebychev or rational one. We will first discuss the rational case (Goedecker 1995).

\begin{figure}[ht]       
     \begin{center}
      \setlength{\unitlength}{1cm}
       \begin{picture}( 8.,4.5)           % figure dimensions
        \put(-4.,-2.0){\includegraphics{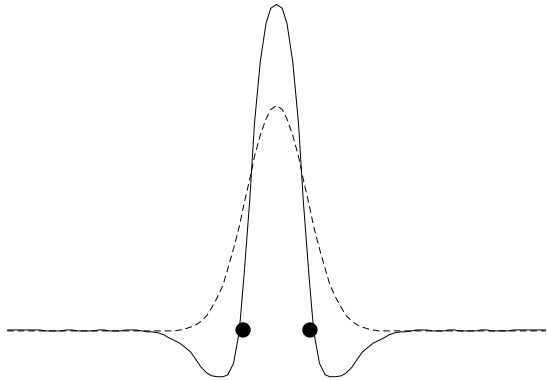}}   % VAX
       \end{picture}
       \caption{\label{project} \it The effect of applying the density operator, 
                which is a projection operator in the eigenvalue space, 
                to a Gaussian (dashed line) centered in the middle of a 
                bond between two silicon atoms denoted by discs. The resulting 
                function $\tilde{W}$ is shown by the solid line. The 
                orthogonal Wannier function $W$ obtained by symmetric orthogonalization 
                is practically indistinguishable from $\tilde{W}$ on this scale. 
                The calculation was done using Density Functional Theory with 
                 pseudopotentials}
      \end{center}
     \end{figure}

To do the projection with a rational representation, a system of equations 
analogous to ~(\ref{padeeq}) and ~(\ref{padesm}) has to be solved for 
each trial Wannier function $\tilde{V}_n$ and at each integration point $\nu$
\begin{eqnarray}
 (H-z_{\nu}) \tilde{W}_{n,\nu} & = & \tilde{V}_n \label{wanneq} \\
  \tilde{W}_n & = & \sum_{\nu} w_{\nu} \tilde{W}_{n,\nu}  \label{wannsm} \: .
\end{eqnarray}
Thus the saving comes from the fact that one has to solve this system of 
equations~(\ref{wanneq}) just for $N_{el}$ right hand sides, whereas one has 
$M_b$ right hand sides in Equation~(\ref{padeeq}).  
Obviously the solution of the Equation~(\ref{wanneq}) has to be 
done not within the whole computational volume but only within the 
localization region to obtain linear scaling. 
The functions $\tilde{W}_n$ will now span our subspace unless one of our trial 
functions $\tilde{V}_n$ was chosen in such a way that it has zero overlap 
with the space of the occupied orbitals, which is highly unlikely.
To obtain a set of valid Wannier functions $W_n$ one has still to orthogonalize 
the orbitals $\tilde{W}_n$. Since the $W_n$'s are localized 
the overlap matrix is a sparse matrix and can be calculated with linear scaling.
In the typical Density Functional context, the inversion of this matrix is a 
rather small part, even if it is done with cubic scaling. In a Tight Binding 
context it is much more important and a linear scaling method 
has been devised by Stephan {\it et al.} (1998) for the inversion. 
The construction of the Wannier functions by projection 
according to Equations ~(\ref{wanneq}) and ~(\ref{wannsm})
is illustrated in Figure~\ref{project} in the case of a silicon crystal. 
In this case one knows 
that the Wannier functions are bond centered and it is therefore natural to choose 
a set of bond centered functions as an initial guess. In this example 
we took simple Gaussians. As shown in Figure~\ref{project}, the projection 
modifies the details of the Gaussian but does not significantly change its 
localization properties. 

Chebychev based projection methods have been introduced in connection with other techniques 
by Sankey {\it et al.} (1994) and by Stephan {\it et al.} (1998). 

\subsection{The Divide and Conquer method}
The original 
formulation of the DC method (W. Yang 1991a, W. Yang 1991b, Zhao and Yang 1995)
was based on a subdivision of the electronic density. 
To calculate the density at a certain point an ordinary electronic structure 
calculation is done for a sub-volume containing  this point. Since the electronic 
density is only influenced by features in a rather small 
neighborhood, the density 
obtained in this way is a good approximation to the density one would obtain 
by doing a calculation in the whole volume occupied by the molecule or solid 
under consideration. 
The more recent formulation of this theory (W. Yang and T-S. Lee 1995)  
is however also based on the 
density matrix and we will discuss this version in more detail. The idea is 
to calculate certain regions of the density matrix by considering 
sub-volumes and then to generate the full density matrix by adding up these 
parts with the appropriate weights. How to calculate the density matrix 
for the sub-volumes is in principle unspecified but is usually done using 
ordinary electronic structure calculations based 
on exact diagonalization. Let us illustrate the principle 
for synthesizing the density matrix from its subparts in a pictorial way 
by considering a one-dimensional chain molecule. For simplicity let us 
also assume that we have atom centered basis functions.

\begin{figure}[ht]       
     \begin{center}
      \setlength{\unitlength}{1cm}
       \begin{picture}( 10.,4.7)           % figure dimensions
        \put(-2.,-0.0){\includegraphics{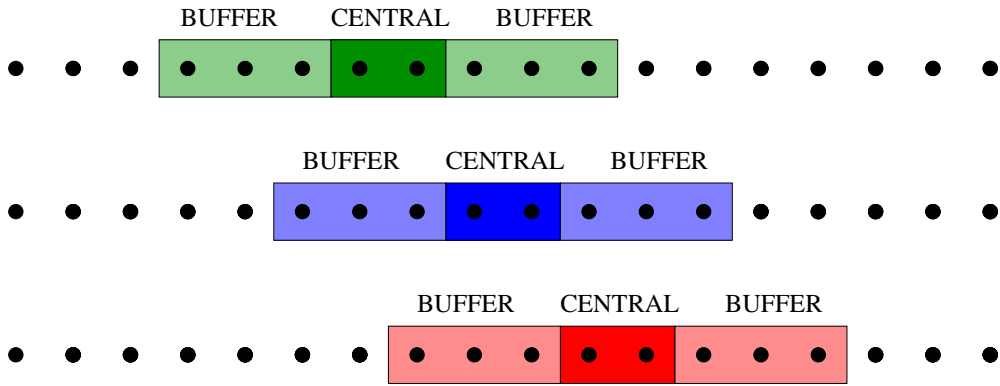}}   % VAX
       \end{picture}
       \caption{\label{dcregs} \it The different sub-volumes described in the text which are 
                necessary for the application of the DC method to a chain molecule.
                The atoms are denoted by black dots. }
      \end{center}
     \end{figure}

First one divides the whole computational volume into sub-volumes which we will 
call central regions. 
Three such central regions are displayed in Figure~(\ref{dcregs}) by 
dark shading.
% the dark green, blue and red colors. 
Around each central region one puts a buffer region 
denoted by light shading. 
% denoted by light green, blue and red color. 
The sum of these two regions corresponds 
to the localization region of the other O(N) methods.

\begin{figure}[ht]       
     \begin{center}
      \setlength{\unitlength}{1cm}
       \begin{picture}( 10.,4.0)           % figure dimensions
        \put(1.,-0.3){\includegraphics{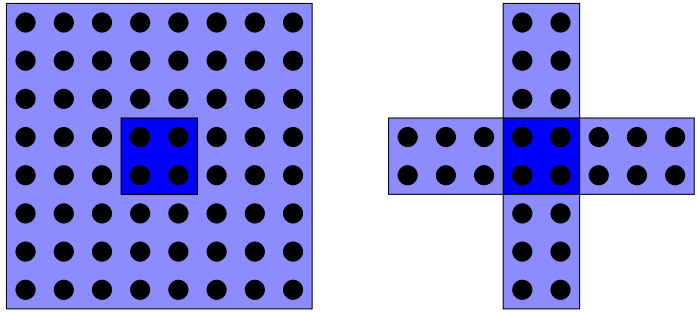}}   % VAX
       \end{picture}
       \caption{\label{dctrans} \it From the full density matrix calculated for 
                a certain localization region (left hand side) only the 
                cross shaped part (right hand side) is used }
      \end{center}
     \end{figure}

Once one has set up this subdivision, one does an electronic structure calculation 
within each localization region to obtain the density matrix belonging to this region.
These different calculations are only coupled by the requirement that the 
Fermi level is the same in all the localization regions. Typically a very small 
temperature is used to stabilize the search for the overall Fermi level.
From the density matrices obtained in this way, one cuts off all the corners, 
i.e. the regions where both matrix indices belong to basis functions in the 
buffer region. 
% In the case of the blue region this series of transformations is shown 
% in Figure~(\ref{dctrans}).
This transformation is shown for one localization region in Figure~(\ref{dctrans}).

\begin{figure}[ht]       
     \begin{center}
      \setlength{\unitlength}{1cm}
       \begin{picture}( 10.,5.5)           % figure dimensions
        \put(0.,-.3){\includegraphics{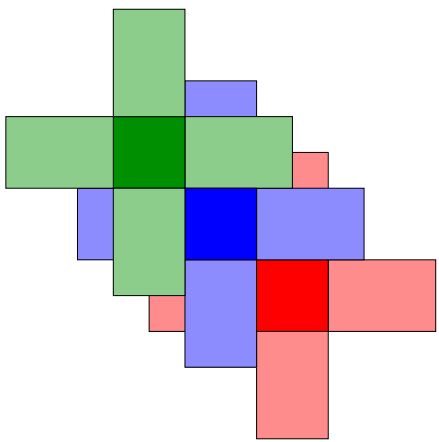}}   % VAX
       \end{picture}
       \caption{\label{dcmat} \it The full density matrix is obtained by adding up 
                the contributions from the different localization regions. 
                In this figure only 3 contributions from the localization 
                regions of Figure~\ref{dcregs} are shown.}
      \end{center}
     \end{figure}

Using these cross-shaped blocks one then finally synthesizes the density matrix
as shown in Figure~(\ref{dcmat}) by adding up the different contributions. The regions 
shown in dark 
% colors
shading 
which do not overlap with other regions have thereby weight one, 
whereas the overlapping regions indicted by light 
% colors 
shading 
have each weight one half such that the sum of the weights is one as well 
in the overlap regions.

The achievement of linear scaling  in the DC and OM methods is conceptually related. 
In both methods certain parts of the density matrix are calculated independently.
The main difference is that in the FOE method no calculated parts 
of the density matrix are discarded in the way
done in the DC method as depicted in Figure~\ref{dctrans}. The FOE method can thus 
be considered as some DC method where only the central part of the density
matrix which is not contaminated by the boundary of the localization region 
is calculated. The fact that in the DC method only a small part of the density matrix 
obtained by costly diagonalizations  
is used, while the largest part associated with the buffer region 
is thrown away results in a large prefactor (Equation~(\ref{cpu1})) for this method.
This is evidently a particularly serious disadvantage if large localization regions are 
needed as will be discussed in more detail in section~\ref{compare}. 

The calculation of the forces acting on the atoms within the DC method is also 
described by Yang and Lee (1995). Their force formula is based on the Hellmann-Feynman 
theorem (Feynman 1939) as well as some other terms such as Pulay forces 
(Pulay 1977) which arise from the use of atom centered basis sets and auxiliary 
charge densities. As has been discussed in the case of the FOE method the 
Hellman Feynman expression for the force~(\ref{foeforce}) 
is not exactly consistent with the total energy expression 
in a non-variational method, since it is based on the assumption that 
one is allowed to take traces. Even though the density matrix in the DC method is 
not calculated via a polynomial expansion, the analysis given for the FOE method also 
applies to the DC method since conceptually one can represent any 
matrix functional of $H$ as a polynomial Taylor expansion. 
The total energy will consequently not have its 
minimum exactly at the same place where the Hellman Feynman forces vanish if both quantities 
are calculated with the DC method. In the case of the FOE method there is 
a simple analytic expression for the calculation of the total energy,  
even in the case where localization constraints are imposed (Equation~(\ref{etrun})). 
One can therefore differentiate it without using the simplifications arising from the 
use of traces to obtain consistent forces. No such simple prescription 
can be written down for the DC method which would allow the calculation of consistent 
forces. Evidently this compatibility problem becomes negligible for large 
localization regions and there are certainly practical applications where 
small inconsistencies of forces and energies are tolerable.

\subsection{The Density Matrix Minimization approach}
\begin{figure}[ht]       
     \begin{center}
      \setlength{\unitlength}{1cm}
       \begin{picture}( 8.,5.0)           % figure dimensions
        \put(-2.,-1.0){\includegraphics{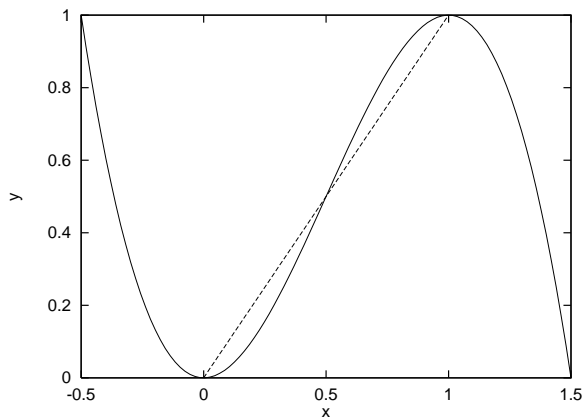}}   % VAX
       \end{picture}
       \caption{\label{weeny} \it The McWeeny (1960) purification function 
                $3 x^2 - 2 x^3$ }
      \end{center}
     \end{figure}

The DMM approach of Li, Nunes and Vanderbilt (1993) is another approach where the full 
density matrix is constructed. In contrast to the FOE method one obtains the 
density matrix $F$ in the limit of zero temperature, so no adjustable 
temperature parameter enters the calculation. The density matrix is obtained by 
minimizing the following functional for the grand potential $\Omega$ 
with respect to $F$
\begin{equation} \label{dmmfunc}
\Omega = Tr[ (3 F^2 - 2 F^3) (H - \mu I)] \: .
\end{equation} 
There is no constraint imposed during the minimization so 
all the matrix elements of $F$ are independent degrees of freedom.  
Nevertheless the final density matrix will obey the correct constraint of 
being a projector if no localization constraints are imposed. 
This is related to the fact that the 
matrix $3 F^2 - 2 F^3$ is a purified version of $F$ as can be seen 
from Figure~\ref{weeny}. If $F$ has eigenvalues close to zero or one then 
the purified matrix will have eigenvalues that are even closer to the same values.
It is also clear from Figure~\ref{weeny} that the eigenvalues of the 
purified matrix are contained in the 
interval [0;1] as long as the eigenvalues of $F$ are in the interval 
[ -1/2 ; 3/2 ]. 

The gradient of $\Omega$ as given by Equation~(\ref{dmmfunc}) with respect 
to $F$ is itself a matrix and it is given by
\begin{equation} \label{dmmgrad}
\frac{\partial \Omega}{\partial F} =  3 ( F \: H' + H' \: F) - 
                    2 (F^2 \: H' + F \: H' \: F + H' \: F^2) \: ,
\end{equation}
where $H' = (H - \mu I)$.
In order to verify that Equation~(\ref{dmmfunc}) defines a valid functional we have to 
show two things. First, that the grand potential expression~(\ref{dmmfunc}) 
gives the correct result if we insert the exact density matrix $F$, and second, that 
the gradient~(\ref{dmmgrad}) vanishes in this case. From Equation~(\ref{denst0proj}) we see that 
the exact $F$ is a projection operator, i.e that $F^2 = F$. Therefore $(3 F^2 - 2 F^3) = F$ 
and the grand potential expression~(\ref{dmmfunc}) agrees indeed with the 
correct result~(\ref{omegatrace}). Using in addition the fact that $H'$ and the exact $F$ commute 
(as follows from Equations ~(\ref{hevcs}),~(\ref{fevcs}) ) it is also evident 
that the gradient in Equation~(\ref{dmmgrad}) vanishes. 
The gradient vanishes however not only for the ground 
state density matrix $F$ but also for any excited state density matrix.
In order to exclude the possibility of local minima, we have to verify, that these 
stationary points are no minima. This can easily be done 
(D. Vanderbilt, private communication) using the fact 
that the functional is a cubic polynomial with respect to all its 
degrees of freedom. Let us suppose that there are two minima. 
Inspecting the functional along the line connecting these 
two minima we would obviously again find these two minima, which is 
a contradiction because a cubic polynomial cannot have two minima.
Thus we have proved by contradiction that the DMM functional 
has only one single minimum.

There is a second thing which is worrying at first sight with this functional. If the density 
matrix for an insulator at zero temperature is of the correct form  
(i.e. if the occupation numbers $n_l$ are integers)  
the gradient~(\ref{dmmgrad}) will vanish independently of the 
value of the chemical potential. 
This ambiguity however disappears as soon as one has 
fractional occupation numbers. Let us consider an approximate density matrix of the form
\begin{equation} \label{fap}
F =  \sum_l n_l |\Psi_l> <\Psi_l|  \: .
\end{equation}
Then it is easy to see that 
\begin{eqnarray} 
 \Omega & = &  
     \sum_l  (\epsilon_l -\mu)  ( 3 n_l^2 - 2 n_l^3)  \label{dmmmmx} \\
\frac{\partial \Omega}{\partial F} & = &  
     \sum_l  6 (\epsilon_l -\mu)  n_l (1- n_l) |\Psi_l> <\Psi_l| \label{dmmmix}  \: .
\end{eqnarray}
The polynomial of Equation~(\ref{dmmmmx}) is the same as the one shown 
in Figure~\ref{weeny} and we see that 
components corresponding to eigenvalues larger than the chemical potential 
are damped until they vanish in the minimization process, whereas components 
corresponding to smaller eigenvalues are amplified until they reach the value one. 
Thus the chemical potential 
will determine the number of electrons to be found in the system as it should.
The above statements are actually only correct if all the $n_l$'s are 
contained in the interval [-1/2: 3/2]. If this is not the case then one can see 
from Figure~\ref{weeny}, that there can be runaway solutions, where 
some $n_l$ tend to $\pm \infty$. When we implemented the scheme we however 
never encountered in practice such a runaway case.

Having convinced ourselves, that the functional defined in Equation~(\ref{dmmfunc}) 
is well behaved, let us now estimate the number of iterations which are necessary 
to minimize it. As is well known, the error reduction per iteration step 
depends on the condition number $\kappa$ which is the ratio 
of the largest curvature $a_{max}$ to the smallest curvature $a_{min}$ at the minimum.
These curvatures could be determined exactly by calculating the Hessian matrix 
at the minimum. Let us instead only 
derive an estimate of these curvatures by calculating the curvature along 
some representative directions. To do this 
let us now consider a ground state density matrix where some fraction $x$ of an 
excited state is mixed in

\begin{equation} 
F(x) =  \sum_{n=1}^{N_{el}} \Psi^*_n({\bf r}) \Psi_n({\bf r})  
     - x \Psi^*_I({\bf r}) \Psi_I({\bf r}) + x \Psi^*_J({\bf r}) \Psi_J({\bf r}) \: .
\end{equation}
The index $I$ is a member of the $N_{el}$ eigenstates below $\mu$ 
and the index $J$ refers to a state above $\mu$.
The expectation value of the OM functional for this density matrix is given by 
\begin{eqnarray} 
\Omega(x) & = & Tr[ (3 F(x)^2 - 2 F(x)^3) (H - \mu I)] \\
 & = & \sum_{n=1}^{N_{el}} \epsilon_n + \left( 3x^2-2x^3 \right) (\epsilon_J-\epsilon_I) 
 \nonumber
\end{eqnarray}
and its second derivative by 
\begin{equation} \label{dmmgscurve}
 \left. \frac{\partial^2 \Omega(x)}{\partial x^2} \right|_{x=0} = 
    6 (  \epsilon_J -  \epsilon_I ) \: .
\end{equation}
The largest curvature will roughly be $\epsilon_{max} - \epsilon_{min}$ 
and the smallest curvature of the order of the HOMO-LUMO separation 
$\epsilon_{gap} = \epsilon_{N_{el}+1} - \epsilon_{N_{el}}$
The condition number is thus given by
\begin{equation} \label{dmmkappa}
\kappa = \frac{a_{max}}{a_{min}} \approx \frac{\epsilon_{max} - \epsilon_{min}}{\epsilon_{gap}} \: .
\end{equation}

In the conjugate gradient method, which is the most efficient method to minimize 
the DMM functional, the error $e_k$ decreases as follows (Saad)
\begin{equation}
    e_k \propto \left( \frac{ \sqrt{\kappa}-1 } 
                            { \sqrt{\kappa}+1 } \right)^k \: .
\end{equation}
The error $e_k$ is defined in this context as the length of the vector which is the 
difference between the exact and approximate solution at the k-th iteration step. 
Under realistic  conditions  $\kappa$ is large and the number of iterations $n_{it}$
to achieve a certain accuracy is therefore proportional to
\begin{equation} \label{cgnit} 
    n_{it} \propto \sqrt{ \kappa } 
                 = \sqrt{ \frac{\epsilon_{max} - \epsilon_{min}}{\epsilon_{gap}} } \: .
\end{equation}
This is an important result since it indicates that in an insulator 
the number of iterations is independent of system size. This result 
is also confirmed by numerical tests.

The use of a conjugate gradient scheme requires line minimizations along these 
conjugate directions. For arbitrary functional forms this has to be done 
by numerical techniques such as bisection (Press {\it et al.,} 1986). In the case of the 
DMM functional we have however a cubic form along each direction.
The four coefficients determining the cubic form can be calculated with 
four evaluations of the functional. Once these 4 coefficients are known the 
minimum along this direction can easily be found.

Doing a series of  minimization steps as outlined above will in general result in 
a density matrix which does not lead to the correct number of electrons. Thus one has to do 
some outer loops where one searches for the correct value of the chemical potential.
For better efficiency, these two iterations loops can however be merged into one 
loop where one alternatingly minimizes the energy and adjusts the chemical 
potential (S-Y. Qui {\it et al.,} 1994).

The forces on the atoms are given by
\begin{equation} \label{dmmfrc} 
\frac{d \Omega}{d R_{\alpha}} =  
\frac{\partial \Omega}{\partial F}  \frac{\partial F}{\partial R_{\alpha}} +   
\frac{\partial \Omega}{\partial H}  \frac{\partial H}{\partial R_{\alpha}}  \: . 
\end{equation}
Since the method is variational, $\frac{\partial \Omega}{\partial F}$ 
vanishes at the solution and the force formula  simplifies to 
\begin{equation} \label{dmmforce} 
\frac{d \Omega}{d R_{\alpha}} =  
\frac{\partial \Omega}{\partial H}  \frac{\partial H}{\partial R_{\alpha}} = 
Tr \left[ (3 F^2 - 2 F^3) \frac{\partial H}{\partial R_{\alpha}}  \right] 
\end{equation}
which can easily be evaluated.

\begin{figure}[ht]       
     \begin{center}
      \setlength{\unitlength}{1cm}
       \begin{picture}( 8.,8.0)           % figure dimensions
        \put(-4.,-1.5){\includegraphics{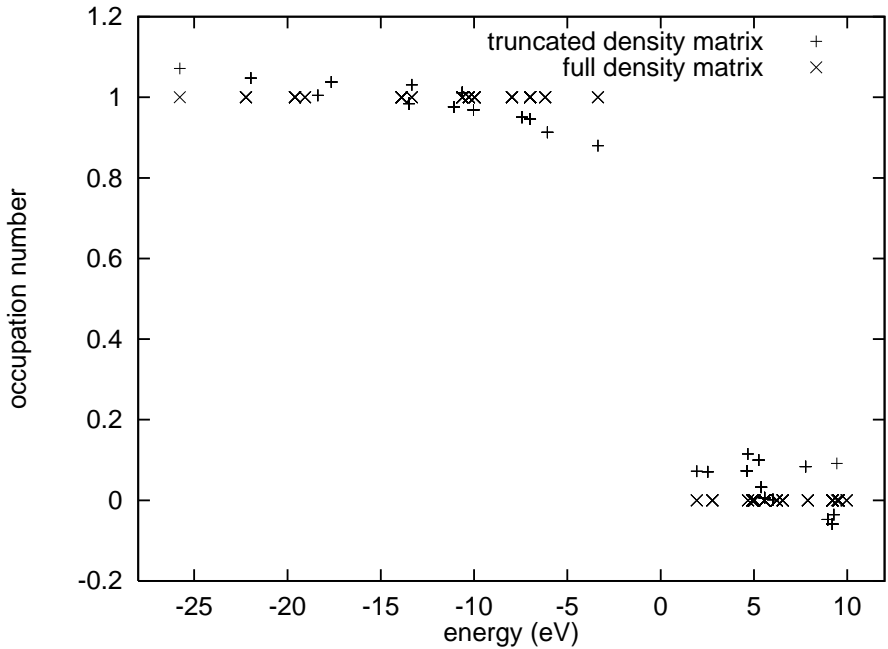}}   % VAX
       \end{picture}
       \caption{\label{dmmfig} \it An analysis of the eigenvectors of the 
                full and truncated density matrix. In the case of the full 
                density matrix the eigenvectors were chosen to be simultaneously 
                eigenvectors of both $F$ and $H$, and the eigenvalues with 
                respect to $F$ (occupation numbers) are plotted versus the eigenvalues with 
                respect to $H$. In the case of the truncated density matrix, the 
                eigenvectors cannot anymore simultaneously diagonalize $F$ and $H$. 
                Therefore the eigenvalues with respect to $F$ are plotted versus 
                their energy expectation values with respect to $H$. 
                Note that in the energy expression~(\ref{dmmfunc}) the purified 
                density matrix $3 F^2 -2 F^3$ enters instead of $F$.  The occupation numbers 
                of the purified version are closer to zero or one.}
      \end{center}
     \end{figure}

The introduction of a localization region leads again to some subtleties.
Whereas in the unconstrained case the eigenvalues of the final density matrix 
$F$ will all be either zero or one, this is not any more the case when a 
localization region is introduced. So the truncated $F$ is not any 
more a projection matrix but it is given by
\begin{equation} 
F =  \sum_{m=1}^{M_b} n_m \Psi^*_m({\bf r}) \Psi_m({\bf r})   \: ,
\end{equation}
where now $\Psi_m$ are the eigenfunctions of the truncated $F$ and the 
occupation numbers $n_m$ their 
eigenvalues. In a certain sense the localization constraint 
introduces a finite electronic temperature. This is actually not surprising after 
the discussion of the relation between the temperature and the 
localization properties in section~\ref{general}. Figure~\ref{dmmfig} shows the 
energy expectation values of the eigenvectors of $F$ versus the occupation numbers, 
for the case of a crystalline Si cell of 64 atoms, 
where the localization region extends up to the second nearest neighbors.
As one sees, the energy expectation values $<\Psi_m|H|\Psi_m>$ 
of the eigenvectors of $F$ are very close to the exact eigenvalues of $H$. 

This close correspondence of the eigenvectors of $F$ 
to the eigenvectors of $H$ explains why the number of iterations 
needed to find the minimum 
does not increase as one introduces localization constraints.
Equation~(\ref{dmmgscurve})
remains approximately valid if the occupation numbers 
for the occupied states are close to 1 and if the occupation 
numbers for the unoccupied states are very small as well as 
if the energy expectation values $<\Psi_m|H|\Psi_m>$ 
are close to to the exact eigenvalues of the Hamiltonian.
These conditions are fulfilled as discussed above. Hence the 
condition number for the minimization process does not change 
appreciably in the truncated case.  

All the arguments used to prove the absence of local minima remain valid in the 
truncated case as well. The force formula Equation~(\ref{dmmforce}) remains equally valid.

An alternative derivation of this algorithm has been given by Daw (1993).
He considers a differential equation which describes the evolution of a 
density matrix when the electronic temperature 
is cooled down from infinity to zero. The change of the density matrix during 
this process is equal to the gradient of Equation~(\ref{dmmgrad}).

\subsection{The Orbital Minimization approach}
\label{ommethod} 
The OM method (Mauri {\it et al.,} 1993, Ordejon {\it et al.,} 1993, Mauri and Galli 1994, 
Ordejon {\it et al.,} 1995, Kim {\it et al.,} 1995) also 
calculates the grand potential in the limit of zero 
temperature. In contrast to the two previous methods, it 
does not calculate the density matrix directly but expresses it via the 
Wannier functions according to Equation~(\ref{denst0}). These Wannier 
functions are obtained by minimizing the following unconstrained functional.

\begin{equation} \label{omfunc} 
 \Omega = 2 \sum_n \sum_{i,j} c^n_i H'_{i,j} c^n_j 
          - \sum_{n,m} \sum_{i,j} c^n_i H'_{i,j} c^m_j \sum_l c^n_l c^m_l \: ,
\end{equation}
where $c^n_i$ is the expansion coefficient of the $n$-th Wannier 
orbital with respect to the $i$-th basis function and $H'_{i,j}$ 
are the matrix elements of the shifted Hamiltonian $H-\mu I$ with 
respect to the basis functions. In the original 
formulation (Mauri {\it et al.,} 1993, Ordejon {\it et al.,} 1993, Mauri and Galli 1994, 
Ordejon {\it et al.,} 1995) only $N_{el}$ orbitals were included in the 
orbital sums in Equation~(\ref{omfunc}) (i.e. $n = 1 ... N_{el}$ , 
$m = 1 ... N_{el}$). In the formulation of Kim (1995) more than $N_{el}$ 
orbitals are included in the sums. The functional of Equation~(\ref{omfunc}) can be derived by 
considering the ordinary band structure energy expression
\begin{equation} \label{convfunc} 
 E_{BS} = \sum_n \sum_{i,j} c^n_i H'_{i,j} c^n_j 
\end{equation}
and by incorporating the
orthogonality constraint by a Taylor expansion of the inverse of the overlap matrix $O$ 
between the occupied orbitals
\begin{equation} \label{omover} 
O_{n,m} = \sum_l c^n_l c^m_l
\end{equation}
up to first order.
A family of related functionals can be obtained by Taylor expansions 
to higher order (Mauri and Galli 1994, Galli 1996). 
Since these functionals do not offer any significant 
advantage and are not used in calculations we will not discuss them. 
The gradient of the functional of Equation~(\ref{omfunc}) is given by
\begin{equation} \label{omgrad} 
 \frac{ \partial \Omega}{\partial c^n_k} = 4 \sum_j  H'_{k,j} c^n_j 
          - 2 \sum_m \sum_j  H'_{k,j} c^m_j \sum_l c^n_l c^m_l
          - 2 \sum_m c^m_k \sum_{i,j}  c^n_i H'_{i,j} c^m_j  \: .
\end{equation}

Let us first discuss this functional in the case where no localization 
constraint is imposed on the orbitals. It is easy to see that the 
functional~(\ref{omfunc}) is invariant under unitary  transformations of the 
occupied (i.e. $N_{el}$ lowest) orbitals. So we can derive our results in terms 
of eigenorbitals rather than Wannier orbitals. The coefficients 
$c^n_i$ are then the expansion coefficients of the eigenorbitals.
Using the fact that in this case $ \sum_l c^n_l c^m_l = \delta_{n,m}$ 
and that $\sum_{i,j} c^n_i H'_{i,j} c^m_j = \delta_{n,m} (\epsilon_n -\mu) $ we obtain
\begin{eqnarray} 
 \Omega & = & 2 \sum_n \sum_{i,j} c^n_i H'_{i,j} c^n_j 
          - \sum_{n,m} \sum_{i,j} c^n_i H'_{i,j} c^m_j \delta_{n,m} \nonumber \\
        & = &  \sum_n \sum_{i,j} c^n_i H'_{i,j} c^n_j  \nonumber \\
        & = & \sum_n \epsilon_n - \mu N_{el} \nonumber 
\end{eqnarray}
which is the standard expression~(\ref{omegatrace}) for the grand potential.
Similarly the gradient expression can be simplified obtaining 
\begin{eqnarray} 
 \frac{ \partial \Omega}{\partial c^n_k} & = & 4 \sum_j  H'_{k,j} c^n_j 
          - 2 \sum_m \sum_j  H'_{k,j} c^m_j \delta_{n,m}
          - 2   c^n_k \sum_m  \delta_{n,m} (\epsilon_m -\mu)  \\
          & = & 2 \sum_j  H'_{k,j} c^n_j - 2  c^n_k (\epsilon_n -\mu) = 0 \nonumber \: .
\end{eqnarray}
So the functional has indeed a vanishing gradient at the ground state and it gives 
the correct ground state energy. As was the case for the DMM functional the gradient 
vanishes not only for the set of ground state orbitals but also for 
any set of excited states. So we have to verify that these stationary points are 
not local minima but saddle points. We do this by picking a certain direction 
along which the curvature is negative.
In the OM case an excited state is 
described by a set of $N_{el}$ orbitals $\Psi_n$ where  at least one index $n=I$ 
corresponding to an occupied orbital $I$ is replaced by an unoccupied orbital $J$.
Let us now consider the variation of the grand potential $\Omega(x)$ 
under a transformations of the form $\Psi_J \rightarrow \cos (x) \Psi_J + \sin (x) \Psi_I$.  
One can show that the curvatures at these stationary points is given by
\begin{equation} \label{dmmcurve}
 \left. \frac{\partial^2 \Omega(x)}{\partial x^2} \right|_{x=0} =
   - 4 (  \epsilon_J -  \epsilon_I ) \: .
\end{equation}
Since the unoccupied eigenvalue $\epsilon_J$ is higher in energy than the 
occupied one $\epsilon_I$, the curvature is negative and we have indeed  a saddle point. 
In the same way we can also again show that the condition number is given 
by Equation~(\ref{dmmkappa}). 

\begin{figure}[ht]       
     \begin{center}
      \setlength{\unitlength}{1cm}
       \begin{picture}( 8.,5.5)           % figure dimensions
        \put(-2.,-1.5){\includegraphics{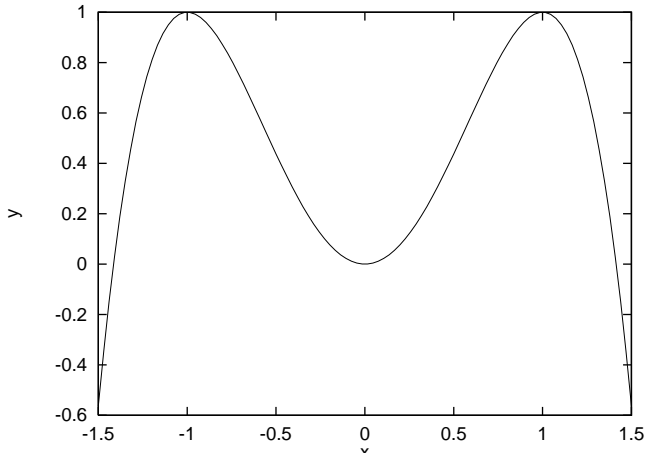}}   % VAX
       \end{picture}
       \caption{\label{ompoly} \it The function $(2-x^2) x^2 $ relevant in 
                Equation~(\ref{ompl}) }
      \end{center}
     \end{figure}

Also in analogy to the DMM functional one can show (Kim {\it et al.,} 1995)
that in the formulation of Kim, the chemical potential $\mu$ determines the 
number of electrons by amplifying 
components below $\mu$ and annihilating components above it. 
Considering a state consisting of a set of eigenfunctions $\Psi_n$ of $H$ where 
each eigenfunction is multiplied by a scaling factor $a_n$, the expectation value 
for the grand potential becomes
\begin{equation} \label{ompl} 
 \Omega = \sum_n (2-a_n^2) a_n^2 (\epsilon_n-\mu)  \: .
\end{equation}
The relevant function $(2-x^2) x^2 $ is shown in Figure~(\ref{ompoly}).
One can see that the minimum of Equation~(\ref{ompl}) is attained 
by $a_n=0$ if $\epsilon_n > \mu$ and by $a_n= \pm 1$ if $\epsilon_n < \mu$.
Again this is only true if $a_n$ is within a certain safety interval. 
Otherwise there can be runaway solutions. Infinitesimally close 
to the solution $\mu$ becomes ill defined in an insulator as it should. 

Whereas the DMM functional keeps all its good properties when one introduces a 
localization constraint, the OM functional looses most of them.
The localization constraint is introduced in the OM functional by allowing 
each Wannier orbital to deviate from zero only within its 
own localization region. These localization regions are usually atom centered 
and contain a few shells of neighboring atoms. 
The basic idea of the OM functional, namely of describing an electronic 
system by a set of Wannier functions with finite support is already problematic. 
Orthogonality and finite support are mutually exclusive properties and so the 
orbitals which one obtains in the minimization process are necessarily 
non-orthogonal. The true Wannier functions are however orthogonal. 
In addition, as we have seen in the DMM case, a density 
matrix which is truncated has full rank, i.e. none of its eigenvalues is 
exactly zero. Thus $N_{el}$ Wannier orbitals are not sufficient to 
represent the density matrix in this case. The generalized formulation of 
Kim (1995) where more than $N_{el}$ orbitals are used alleviates this 
problem, but does not completely fix it unless the number of orbitals is equal to 
the number of basis functions $M_b$. 

When implemented with localization constraints 
the OM functional exhibits the following problems:
\begin{itemize}
\item The functional has multiple minima (Ordejon {\it et al.,} 1995). Depending on the initial guess 
      one obtains thus different answers, some of which are physically 
      meaningless (Kim 1995). 
      As we have shown above in Equation~(\ref{dmmcurve}) 
      the functional has no multiple minima in the non-truncated case. 
      The analysis we used to show this was based on the eigenfunctions.
      Since the Wannier functions have no resemblance to the eigenfunctions 
      this analysis cannot be carried over into the localized regime. In the case of the 
      DMM method the absence of multiple minima could be proven using the fact 
      that the DMM functional is cubic. The OM functional~(\ref{omfunc}) is however 
      quartic with respect to its degrees of freedom and will thus in general 
      have multiple minima. The problem of the multiple minima is 
      attenuated by the formulation of Kim (1995) but it is not completely 
      eliminated since the functional still has quartic character.
      As a byproduct of the multiple minimum problem, the total energy cannot 
      be conserved in molecular dynamics simulations, which is an important 
      requirement. Here again, energy conservation is better in the Kim 
      formulation but still far from perfect (Kim 1995). 

      In practical applications of the OM method for electronic structure methods 
      great care is usually taken to construct input guesses which correspond 
      to the physical bonding properties of the molecule under consideration 
      (Itoh {\it et al.,} 1996). If the minima that is closest in distance is always 
      selected during the subsequent line minimizations then one will most likely 
      end up in a physically reasonable minima that reflects the bonding properties 
      of the input guess (Stephan private communication). 
      This is especially true if the localization regions are 
      large and if the topology of the total energy surface within a reasonably 
      large region around the physical minimum is not too different from 
      the one of the non-truncated case. Such a procedure 
      is of course not applicable in systems where the exact bonding properties 
      are unknown.  

\item The number of iterations is very large whenever any localization 
      constraint is imposed together with tight convergence criteria.  
      This is due to the deterioration of the condition number,
      a phenomenon which is easy to understand (Ordejon 1995).  
      Introducing a localization 
      region destroys the strict invariance of the band structure energy under unitary 
      transformations among the occupied orbitals. When the localization region is
      large, this invariance will still approximately exist and 
      one can find certain directions around the minimum where the energy varies
      extremely slowly and where the curvature is therefore much smaller than the
      smallest curvature $\epsilon_{gap}$ in the unconstrained case. Whereas directions 
      where the curvature is strictly zero do not affect the condition number, 
      these very small curvatures will have a negative effect on the condition 
      number (Equation~(\ref{dmmkappa})) and the required number of iterations is consequently 
      much larger in the constrained case than in the unconstrained case.
      Even though the condition number deteriorates with increasing localization 
      region, the detrimental effect on the number of iterations will disappear 
      at a certain point where the gradient due to these small
      curvatures becomes smaller than the numerical threshold determining the 
      convergence criterion of the minimization procedure. 

\item The optimal localization regions would be centered on the centers of the 
      Wannier functions. Since these centers are not known a priori, 
      atom centered localization regions are usually chosen. 
      In this case the Wannier functions do not generally 
      exhibit the correct symmetry (Ordejon 1995). 
      As a consequence molecular geometries 
      obtained from this functional can have broken symmetry as well. In a $C_{60}$ 
      molecule for instance there are only two equivalent sites. When treated with the OM 
      functional they are however all slightly different (Kim {\it et al.,} 1995). 

\item As follows from Equation~(\ref{ompl}) there can be runaway solutions.
      We have encountered this problem in test cases with random numbers 
      as input guess. If one would construct a more sophisticated input guess, 
      based on the bonding properties of the system, this would however probably not occur.
      In the DMM method the possibility of runaway solutions also exists but is never found 
      in practice even with the most trivial input guess.

\item If the method is used in the context of self-consistent calculations, where the 
      electronic charge density is used to calculate the Hartree and exchange 
      correlation potential, problems arise, since the total charge is not conserved 
      during the minimization iteration (Mauri and Galli 1994). 
\end{itemize}

To overcome the competing requirements of orthogonality and localization, 
a related approach has recently been proposed by Yang (1997) where the 
orbitals are allowed to be non-orthogonal.
This approach of Yang has up to now not been applied in connection 
with a localization constraint. 

\subsection{The Optimal Basis Density Matrix Minimization method}
Despite its many advantages in the Tight Binding context, the DMM method 
has the big disadvantage that it is very inefficient if one needs 
very large basis sets (i.e. many basis functions per atom). 
Large basis sets are typically required in grid 
based Density Functional calculations. In this case it just becomes impossible 
to calculate and store the full density matrix in the DMM method even though 
it is a sparse matrix. From this point of view the Wannier function based 
methods are advantageous since they do not require the full density matrix.
The basic idea of the OBDMM method (Hierse and Stechel 1994, Hernandez and Gillan 1995) 
is now to contract first the fundamental basis 
functions into a small number of new basis functions and then to set up 
the Hamiltonian and overlap matrix in this new small basis. A generalized 
version of the DMM method which can be applied to the non-orthogonal context
(a subject which will be discussed later in the article) is then used to 
solve the electronic structure problem in this basis. The essential point is 
that one tries to do the contraction in an optimal way. This is done by minimizing 
the total energy also with respect to the degrees of freedom determining the 
contracted basis functions $\Psi_n$. 
Formulated mathematically the density matrix is given by 
\begin{equation}  \label{gillan}
 F({\bf r},{\bf r}') = \sum_{i,j} \Psi^*_i({\bf r}) K_{i,j} \Psi_j({\bf r}') \: . 
\end{equation}
The matrix $K$ is a purified version of the the density matrix within 
the contracted basis $L$ and it is given by 
\begin{equation}  \label{gillan2}
K = 3 LOL -2 LOLOL \: , 
\end{equation}
where $O$ is the overlap matrix among the contracted orbitals. 
The main difference between the formulation of Hierse and Stechel and of 
Hernandez and Gillan is that in the first formulation the number of contracted 
basis functions $\Psi_i$ is equal to the number of electrons, whereas in the second 
approach it can be larger. In the formulation of Hernandez and Gillan the 
basis set can for instance be chosen to have the size of a minimal basis set.
The difference to standard minimal basis sets from quantum chemistry is 
that it is optimally adapted to its chemical environment since the contraction 
coefficients are not predetermined but found variationally. 
In practice the full density matrix is found by a double loop minimization 
procedure. In the inner loop one has the ordinary DMM procedure to find 
the density matrix for a given contracted basis set. In the outer loop 
one searches for the optimally contracted basis functions $\Psi_i$ for fixed $L$. 

Unfortunately the minimization of the contracted basis functions $\Psi_i$ is ill
conditioned (Gillan {\it et al.,} 1998) and the number of iterations is therefore at present 
very large. As already explained before ill-conditioning occurs if the 
curvatures in the minimum along different directions are widely different. 
Three causes for the ill-conditioning are reported by Gillan (1998).  
\begin{itemize}
\item Length scale ill-conditioning: \newline 
This problem is actually not related to the OBDMM algorithm itself but to the 
(uncontracted) basis functions which are taken to be so-called ''Blip" functions 
in the present implementation.
This kind of problem can be found in all iterative electronic structure algorithms 
if grid based basis functions such as finite elements are used. Its origin is 
easy to understand.  Let us imagine that we are searching for the 
lowest state of jellium using a localized basis set associated with an equally 
spaced grid. By symmetry the solution is a constant vector, i.e. all 
basis functions have the same amplitude in the solution vector. Let us now assume that 
we explore the energy surface around the minimum along several directions. 
Let us first ''go" into a direction where we add components in such a way that 
the sign of the amplitude of each neighboring basis function changes. 
This corresponds to a high frequency plane wave and since the kinetic energy 
of such a plane wave is big, the total energy will rapidly increase if 
we add a such a contribution to our solution vector. If on the other hand 
we add contributions that correspond to low frequency plane waves the energy 
will increase much more slowly. Since in grid based methods the basis functions 
are usually narrow and since one can thus construct high frequency 
functions the condition number can be very bad. As one can suspect from the 
above explanation the different curvatures can be estimated by doing a 
Fourier analysis. With this information one can then use preconditiong techniques 
to cure the length scale ill-conditioning problem. Such a scheme has 
been proposed by Bowler and Gillan (1998).  

\item Superposition ill-conditioning: \newline
This ill-conditioning problem is essentially identical to the ill-conditioning problem 
of the OM functional. If we have $N_{el}$ contracted basis functions and no localization 
constraints the total energy is invariant with respect to unitary transformations 
of these functions. The introduction of a localization constraint destroys 
this invariance but there is an approximate invariance left with manifests 
itself in very small curvatures in the minimum along certain directions.  

\item Redundancy ill-conditioning: \newline
This problem can only be found in the formulation of Hernandez and Gillan, 
where the number of contracted basis functions is larger than the number 
of electrons. In this case one spans a space that contains not only 
the occupied orbitals but also some unoccupied. As was shown 
before in the context of the DMM functional introducing a localization 
constraint will not assign zero occupation numbers, but only very 
small occupation numbers to components corresponding to the unoccupied 
states in the unconstrained case. 
Since these components corresponding to the unoccupied states have now 
very little weight they have little influence on the total energy 
and one has again certain directions where the total energy changes 
very slowly resulting in very small curvatures. 

\end{itemize}

Another open question is whether the OBDMM has local minima. The functional is 
a 6-th order polynomial with respect to the expansion coefficients of the 
contracted basis functions as can be seen from Equation~(\ref{gillan}) 
and (\ref{gillan2}). The two overlap matrices in Equation~(\ref{gillan2}) 
give each a quadratic term, the two contracted orbitals in Equation~(\ref{gillan}) 
a linear term. Minimization with respect to the 
contracted basis functions should therefore exhibit local minima. 
Local minima have however not been reported with this method so far. 
Perhaps the following DMM minimization step which is free of local minima saves the method 
from overall local minima.

\section{Comparison of the basic methods}
\label{compare}
It is certainly not possible to claim that a specific method is the best for all 
applications. Nevertheless the methods presented so far differ in many respects 
and one can therefore clearly judge under which limiting circumstances certain 
methods will fail or perform well. In the following the methods presented 
so far will therefore be compared under several important aspects. 
The comparison will be done in two categories. The first category applies to 
electronic structure methods requiring a small number of degrees of freedom per atom. 
The Tight Binding method belongs to the first category requiring a few 
basis functions per atom (or just a few degrees of freedom in the case of 
semiempirical Tight Binding). But we will also include the standard quantum chemistry 
methods into this first category, where one typically needs from a few up to a few dozen 
Gaussian type basis functions per atom. The second category contains methods which 
are grid based such as finite difference schemes (Chelikowsky {\it et al.,} 1994), or where the basis 
functions can be associated with grid points such as in finite element basis functions 
(White {\it et al.,} 1989)
or blip (Hernandez {\it et al.,} 1997) basis functions. In these methods one has typically 
many hundred degrees of freedom per atom. Even though the density matrix is 
a sparse matrix, O(N) methods which calculate the 
full density matrix can not be applied to the second category of electronic 
structure methods. The memory requirements alone are already prohibitive. As pointed out 
before we can expect that the localization region in a 3-dimensional structure 
comprises on the order of 100 atoms. The density matrix will exhibit 
significant sparseness only for larger system. Assuming that we just have 100 basis 
functions per atom the storage of the density matrix would require about 1 Gigabyte 
of memory which is the upper limit of current workstations.
The comparison in the large basis set class will therefore comprise only 
the methods which are Wannier function based namely FOP, OM and OBDMM. 
The comparison in the small basis set class will comprise 
FOE, DC, DMM and OM, excluding two methods 
which are explicitly targeted at large basis sets, namely FOP and OBDMM. 

\subsection{Small basis sets}

The comparison of the methods applicable to small basis sets is 
based on the following criteria:

\begin{itemize}
\item Scaling with respect to the size of the localization region: \newline
The size of the localization region is taken as the 
number of atoms contained within it. 
Only the FOE method has a linear scaling with respect to the size of the 
localization region. As one increases the size of the localization region 
the nonzero part of each column of the Chebychev matrices increases linearly 
implying also a linear increase in the basic matrix time vector multiplication part.
In the DMM method the CPU time increases quadratically since the numerical 
effort for the basic matrix times matrix multiplications grows quadratically 
with respect to the number of off-diagonal elements of the matrix. 
Neglecting ill-conditioning problems, the OM method exhibits quadratic scaling, since  
the numerical effort for the calculation of the overlap matrix 
(Equation~(\ref{omover})) between 
the Wannier orbitals increases quadratically. As the localization region 
grows there are more matrix elements and the calculation of each matrix 
element is more expensive since each orbital is described by a longer vector. 
Because the ill-conditioning problem becomes more severe for large localization regions 
the number of iterations increases in reality in a way which is difficult to model 
resulting  in an effective scaling that is stronger than linear.
The DC method has a cubic scaling with respect to the size of the 
localization region if each sub-volume is treated with diagonalization schemes.
From the comparison of the scaling behavior of all these methods one can thus 
conclude, that the FOE method will clearly perform best if large localization regions 
are needed. The FOE method is thus also the only method 
which can be faster than traditional cubically scaling algorithms if 
no localization constraints are imposed. In this case its overall scaling behavior is 
quadratic whereas all other methods have 
a cubic scaling with a prefactor which is significantly larger than the 
one for exact diagonalization. 

\item Scaling with respect to the accuracy: \newline
A detailed comparison of the polynomial FOE method and the DMM method under 
this aspect has recently been given by Baer and Head-Gordon (1997)
for systems of different dimensionality. Their analysis is based 
on the assumption that the decay constant $\gamma$ is given by 
the tight binding limit of Equation~(\ref{insulator}). 
They conclude, that in the one dimensional case the DMM has the best 
asymptotic behavior, but its prefactor is much larger than the one of the 
FOE method, so that the FOE method is more efficient in the relevant 
accuracy regime. In the two dimensional case they have the same asymptotic 
behavior, but the FOE method has again a much smaller prefactor. In the 
most relevant three dimensional case the FOE method has both the best 
asymptotic behavior and prefactor.
These results are plausible after the preceeding discussion of the scaling with 
respect to increasing localization region size. When one wants to improve the 
accuracy the most important factor is the enlargement of the localization region.
It is also clear that in higher dimensions the number of atoms within the 
localization region grows faster than in lower dimensions and that the scaling 
with respect to the number of atoms will thus become the decisive factor in 
3 dimensions. In lower dimensions the number of iterations has higher relative 
importance, favoring the DMM method. 
A comparison of the FOE and DMM method 
applied to quasi two-dimensional huge tight binding fullerenes 
by Bates and Scuseria (1998) is also 
in agreement with the above statements. They found 
that the FOE and DMM methods give nearly the same performance 
with a small advantage for the FOE method.

As discussed before, the scaling of the OM and DC methods is stronger than quadratic 
with respect to the size of the localization region. 
It is therefore clear that the required numerical effort for increased accuracy 
will grow even faster than in the DMM method. 

\item Scaling with respect to the size of the gap: \newline 
In the FOE method the degree $n_{pl}$ of the Chebychev polynomial increases 
linearly with decreasing gap (Equation~(\ref{npldegree})). At the same time the 
density matrix decays more slowly. It follows from Equation~(\ref{insulator}) that 
the linear extension of the 
localization region grows as $\epsilon_{gap}^{-1}$ in the applicable weak binding limit. 
The volume of the localization region 
and the number of atoms contained in it grow consequently as $\epsilon_{gap}^{-3}$. 
Taking into account the number of iterations (Equation~(\ref{npldegree})), 
the total numerical effort increases thus as $\epsilon_{gap}^{-4}$. 
In the DMM method the number of iterations also increases 
with decreasing gap but more slowly namely like $\epsilon_{gap}^{-1/2}$ 
as follows from Equation~(\ref{cgnit}). Taking into 
account the above discussion of the scaling properties of the DMM method with increasing 
localization region we obtain the overall scaling of $\epsilon_{gap}^{-13/2}$, which is 
higher than the scaling behavior of the FOE method. Obviously the scaling behavior 
of the OM and DC methods is worse. So in contrast to what one might first 
think the FOE methods performs best in this limit. 
In three-dimensional metallic systems, the FOE method is thus to be expected to 
be the only method which will work efficiently at good accuracies. 

\item Finding a first initial guess: \newline 
No initial guess is required in the FOE and DC methods (except perhaps for the 
potential in a selfconsistent calculation). In the DMM method an extremely simple 
and efficient input guess for the density matrix is just a diagonal matrix 
that sums up to the correct number of electrons. In the OM method this point is 
somewhat problematic. As mentioned above a Wannier function represents typically a 
bond or lone electron pair. So if one can draw the standard Lewis structure of a 
molecule, where bonds are denoted by lines and a lone electron pair by a pair 
of dots, one knows where the Wannier functions should be centered and the Lewis 
formula can be the basis for the initial guess. This procedure can not be used any 
more if the molecule is characterized by two or more Lewis structures that are 
resonating. Especially if the two Lewis structures correspond to an electron transfer 
over a distance which is larger than the range of the localization region, serious 
problems are to be expected with Wannier function based methods. In such a case it might 
actually not only be impossible to find an initial guess, but it might also be impossible 
at all to describe such a molecule by $N_{el}$ localized Wannier functions. 

\item Number of iterations in electronic structure calculations: \newline 
In the variational methods (OM, DMM) the number of iterations depends on the condition 
number of the energy expression. As pointed out the OM energy expression is 
ill conditioned under localization constraints and therefore the required number 
of iterations is very large. Even for modest accuracy several hundred iterations 
are required ( Mauri and Galli 1994, Ordejon {\it et al.,} 1995). In the DMM method on the 
other hand the number of iterations is equal to the number of iterations one 
needs in ordinary electronic structure calculations (non O(N)), namely 20 to 30 
(Nunes private communication). 
The quantity that corresponds to the number of iterations in the FOE method is the 
degree $n_{pl}$ of the polynomial. While it is difficult to compare the cost 
of one Chebychev recursion step with the cost of a DMM minimization step, 
such a comparison can be done in the case of the OM method. In each OM 
line minimization step one has to calculate the minimum of a quartic polynomial 
which requires at least 3 applications of $H$ to the wavefunction. 
One Chebychev recursion step requires one application of $H$.  

\item Number of iterations in molecular dynamics simulations: \newline
In molecular dynamics simulations as well as structural relaxations steps and 
selfconsistent mixing schemes the 
density matrix or respectively the Wannier functions 
from the previous step are a good input guess for the next step. Good initial input 
guesses are beneficial in all methods except in the polynomial FOE method and the DC method. 
It is difficult to quantify the possible savings of such a reuse. 
To preserve the quality of the solution of the preceeding step as an input guess 
in a Molecular Dynamics simulation, it can be necessary to make the time step smaller 
than the integration scheme would allow. How large the maximum time step can be 
depends of course also on the order and properties of the time integration scheme 
used to propagate the Molecular Dynamics simulation. Similar remarks apply to 
the case of structural relaxations. The decisive factor determining the number 
of iterations per molecular dynamics step is in this context 
again the condition number of the functional. With the DMM methods of the order 
of 2 to 3 steps are needed both for accurate molecular dynamics simulations (Qiu 1994) 
and for structural relaxations (Nunes private communication). 
The smallest number of iterations that was used in molecular dynamics simulations 
with the OM method was 10, but at the price of a very poor energy conservation. 

\item Cross over point for standard Tight Binding systems: \newline 
The FOE method has the lowest reported cross over point for the standard 
carbon test-system in the crystalline diamond structure. For the FOE method 
it is around 20 atoms (Goedecker 1995), 
and for the DMM it is estimated (Li {\it et al.,} 1993) to be around 90 atoms.  
No crossover points were ever given for the OM and DC method and presumably 
they are much higher.
All quoted cross over points for electronic structure calculations 
are for an accuracy of roughly 1 percent in the 
cohesive energies, but in the relevant publications not all computational details
are listed to ensure that these numbers are really comparable in all respects. 
The low crossover point of the FOE method can be understood in terms of its 
scaling behavior with respect to the size of the localization region discussed above. 
For small systems the size of the localization region equals the size of the 
whole system.  The FOE method therefore starts off 
with a quadratic scaling behavior, whereas all other methods start off with a 
cubic behavior. Consequently the cross over point for all other methods 
can only be for system sizes larger than the localization region, whereas 
the crossover point in the FOE method can already be at smaller system sizes if it 
is implemented efficiently.

In the context of molecular dynamics simulations the cross over points are different 
because some of the variational methods can benefit from good input guesses. 
For the FOE method the cross over point remains at 20 atoms, 
for the OM method Mauri and Galli quote 40 atoms, and for the 
DMM method Qui {\it et al.} (1994) quote 60 atoms. 
Again no crossover point is given for the DC method.
It has to be stressed however that 
the number quoted by Qui {\it et al.,} (1994) was for highly accurate molecular dynamics runs where 
the total energy was conserved, while in the benchmarks of Mauri and Galli 
no satisfactory energy conservation was obtained. 

\item Influence of the range of a sparse Hamiltonian matrix on the performance: \newline 
In the FOE method the numerical effort increases strictly linearly with respect 
to the number of nonzero elements per column which depends cubically on the range of the 
Hamiltonian matrix. In the case of the DMM method it can be shown (Li {\it et al.,} 1993) 
that one has to calculate intermediate product matrices only up to a range 
which is the sum of the range of the density matrix and the Hamiltonian 
matrix. As long as the range of the Hamiltonian is small compared to the 
range of the density matrix the number of operation increases therefore 
only very weakly with respect to an increasing Hamiltonian range. The DMM 
method therefore outperforms the FOE method under such circumstances 
(Daniels and Scuseria 1998).
Hamiltonian matrices of relatively large range are found 
in the context of Density Functional calculations using Gaussian basis sets.
For Tight Binding calculations, in contrast, the range of the Hamiltonian is usually 
small. The OM method shows the same behavior as the FOE method. The numerical effort 
increases linearly with respect to the 
number of nonzero elements per column of the Hamiltonian. 
In the DC method the numerical effort is independent of the bandwidth, but 
it is not expected that even in this case the DC method might outperform the FOE or 
DMM method. 

\item Scaling with respect to the size of the basis set: \newline
Let us now consider the case, where the number of atoms as well as all other 
relevant quantities, such as the size of the localization region, 
are fixed and where we only increase the number of basis 
functions per atom $m_b$. Both the size and the number of off-diagonal elements 
per column of the  
density matrix will then increase. We will also assume that the Hamiltonian is a 
sparse matrix with $m_b$ off diagonal elements per column. 
In the DMM method the numerical effort will 
consequently grow cubically with respect to $m_b$, since the number of operations 
needed for the multiplication of two sparse matrices of size $n$ with $m$ off-diagonal 
elements per column is proportional to $n \: m^2$.
The DC method scales cubically as well since it is based on diagonalization.
The FOE method equally scales 
cubically, since three factors are increasing. The number of columns in the 
density matrix, the number of coefficients in each column and the number of off-diagonal 
elements of the Hamiltonian matrix.
In addition to the arguments showing the unrealistically large memory 
requirements of these methods when used with large basis sets, we thus also find 
a cubic scaling which prohibits the use of these algorithms in this context. 

In the OM method both the application of the Hamiltonian to the orbitals as well as 
the calculation of the overlap between the orbitals scale quadratically with respect to $m_b$.

\item Memory requirements: \newline 
The DMM method requires the storage of the whole sparse density matrix.
If one takes advantage of the fact that the matrix is symmetric 
storage can actually be cut into half. 
The OM method requires only the storage of the truncated Wannier orbitals
and so the memory requirements are reduced by about 50 percent in the typical 
Tight Binding context compared to the case where one stores the 
all the nonzero elements of the density matrix without taking advantage of its 
symmetry. If the Kim formulation is used the gain can 
come down to less. In both the DC and FOE method only 
the subparts respectively the columns of the density matrix which 
are consecutively calculated need to be stored. The storage requirements are 
therefore greatly reduced compared to the DMM and OM methods, namely by a factor 
of roughly $N_{el}$.

\item Parallel implementation: \newline 
Parallel computers and clusters of workstations are standard tools in the 
high performance computing environment. The suitability of an algorithm for 
parallelization is therefore also an important aspect. It is of course 
always possible to parallelize any program, the question is just whether 
this can be done in a coarse or fine grained way, i.e. with a 
small or large communication to computation ratio.  Only a coarsed 
grained parallel program will run efficiently on clusters of workstations 
with relatively slow communications as well as on a very large number of 
processors of a massively parallel computer. Both the FOE and the 
DC algorithms are intrinsically parallel algorithms, meaning that the big computational 
problem is subdivided into smaller subproblems which can be solved practically 
independently. In the case of the FOE method (Goedecker and Hoisie 1997) 
the calculations of the 
different columns of the density matrix are practically 
independent (Equation~(\ref{recurs})). In the case 
of the DC method the calculations of the different patches of the density matrix 
are practically independent as well. 
Both methods can therefore be implemented in a coarse grained way. 
A parallel program based on the FOE method won the 1993 
Gordon Bell prize in parallel computing for its outstanding performance on a cluster of 
8 workstations, obtaining half of the peak speed of the whole configuration (Goedecker and 
Colombo 1994). Impressive speedups of up to 400 have been obtained 
with the FOE method on a 800 processor parallel machine (Kress {\it et al.,} 1998). 
Even though it is more difficult to implement the OM method in parallel 
two such implementations have been reported.
In the OM method two parallelization schemes are conceivable. In the first 
scheme (Canning {\it et al.,} 1996) one associates to each 
processor a certain number of localized orbitals.
This data structure is optimal for the application of the Hamiltonian to the 
orbitals, but requires communication for the calculation of the overlap between 
the orbitals. The second scheme (Itoh {\it et al.,} 1995) associates 
the coefficients of all the orbitals 
whose localization region has an intersection with a certain region of space 
to a certain processor. This data layout is optimal for the calculation of the 
overlap matrix, but requires communication for the application of the Hamiltonian.
The OBDMM method has also been parallelized (Goringe {\it et al.,} 1997b).
Since the OBDMM is more complex than the other methods 
that have been implemented on parallel machines three different parallelization 
paradigms are required.

\item Quality of forces: \newline 
In the case of the variational (DMM and OM) methods the force formula is particularly 
simple (Equation~\ref{dmmforce}) since only the Hellman Feynman term survives. It has to 
be stressed that this formula is however only exact if one has succeeded in 
reducing the gradient with respect to all variational quantities really to zero. 
If, in a simulation, the gradient is not reduced to zero within the required precision 
because too many iterations would be required, errors will creep into the 
calculated forces, making them inconsistent with the total energy. 
From this point of view the situation is easier in the FOE method. 
Since the FOE method is not an iterative method (in the sense that one iterates 
until a certain accuracy criterion is met), 
the force formula of Voter (Equation~(\ref{forcevoter})) will always give forces 
consistent with the total energy. As has already been discussed no 
consistent force formula exists for the DC method. 

Consistent forces are a prerequisite for the conservation of the total energy in 
Molecular Dynamics simulations. Even with consistent forces there are however 
other factors which can cause deviations from perfect total energy conservation 
in Molecular Dynamics simulations such as finite time steps and events where 
atoms enter or leave the localization region. 

\item Cases where the methods become inefficient: \newline 
Cases where different methods become inefficient have already implicitly been  
pointed out when discussing the previous criteria. Let us finally mention 
a special case where the FOE method is inefficient. As discussed above a 
small gap implies usually highly extended density matrices and the FOE method 
is highly competitive. There can however 
be exceptions to this rule. If by symmetry restrictions there is 
practically no coupling between two well localized states, which are 
close together, their energy levels can be split by a very small amount. If the Fermi 
level falls in between these two levels a very high degree polynomial is needed 
to separate these two levels in an occupied and an unoccupied one. 
This scenario can be found in the case of a vacancy in the silicon crystal.
A Jahn Teller distortion leads to a very small splitting between an 
occupied and an unoccupied gap level. Using a high electronic temperature 
and a low degree polynomial in the FOE method does not reproduce this 
Jahn Teller distortion. A detailed study of this effect is given by Voter {\it et al.,} (1996) 
showing that a polynomial of degree 200 is needed instead of the typical polynomial of degree 50 
needed for bulk silicon in the Tight Binding context. From an energetic point of 
view it is frequently not necessary to track such Jahn Teller distortions, since they  
lead to a rather small energy lowering. 
In molecular dynamics simulations of metallic 
systems this suppression of the opening of a gap can even 
be beneficial (Goedecker and Teter 1995) since it leads to 
a smoother density of states around the Fermi level.

\end{itemize}

In summary, we see that the performance 
depends critically on many parameters which can change from one application 
to another. Performance superiority claims based on test runs of a 
particular system have therefore to be taken with caution.

\subsection{Large basis sets}
Whereas the methods which are mainly applicable in the context of small 
basis sets showed important differences under the various comparison criteria,
the behavior of the FOP, OM and OBDMM methods are quite similar under most 
of these criteria. The comparison of the methods which are applicable 
to large basis sets will therefore be based only on a smaller set of important criteria.  

\begin{itemize}
\item Scaling with respect to the size of the basis set: \newline
As pointed out before the methods compared in this sections all 
have a reasonable scaling with respect to the size of the basis set 
allowing thus their use in the context of very large basis sets. 
In contrast to the discussion of the same point within the small basis set 
framework, the number of nonzero matrix elements of the Hamiltonian 
is typically independent of the resolution of the grid, i.e. of the number 
of basis functions. The most important part of the FOP, OM and OBDMM algorithms,
the application of the Hamiltonian matrix to a wave vector scales 
therefore linearly. At the same time all these algorithms require at 
some stage the calculation of an overlap matrix among the occupied 
orbitals. This part scales quadratically as discussed before. 

\item Finding a first initial guess: \newline 
As discussed in the comparison part dealing with small basis sets, 
it can be difficult to find an initial guess for Wannier function based methods.
This problem does not exist in the OBDMM method if the number of 
orbitals is larger than the number of electrons. In this case the orbitals 
are just basis functions and by analogy with the usual tight binding or 
LCAO basis sets it should always be possible to generate a physically 
motivated initial guess for these orbitals.

\item Required number of iterations: \newline 
As mentioned both the OM and OBDMM methods suffer from ill-conditioning problems 
and require therfore a frequently excessive number of iterations for the 
iterative minimization. No such ill-conditioning exists for the FOP method.

\item Cases where the methods become highly inefficient: \newline 
None of the 3 methods have ever been applied to metallic systems, and they 
are all expected to fail in this case. 

\end{itemize}

\section{Other O(N) methods}
The recursion method is a well established method which also exhibits O(N) scaling.
It is principally a method to calculate the density of states $D(\epsilon)$, but  once the 
density of states is known, the band structure energy can be calculated by 
integrating $\epsilon D(\epsilon)$ up to the Fermi level. 
The recursion method has been described 
extensively (Haydock 1980, Gibson {\it et al.,} 1993) and we will therefore not review it. 
Let us just point out that 
within the original formulation of the recursion method only the diagonal 
elements of the density matrix could be calculated. For the calculation of the 
forces one would however also need the off-diagonal elements. So the applicability 
of the recursion method is significantly reduced compared 
to the O(N) method described in section~\ref{four}, which all gave access 
to the forces. There have been several attempts to overcome this 
limitation (Aoki 1993, Horsfield {\it et al.,} 1996b, Horsfield {\it et al.,} 1996c, Bowler {\it et al.,} 1997). 
In contrast to the methods of section~\ref{four} 
these Bond Order Potential (BOP) methods are fairly complex and difficult to implement. 
The basic idea in the BOP method is to calculate the off diagonal elements of the 
density matrix as the derivative of diagonal elements of a density matrix defined with 
respect to a transformed basis. Even though it is now possible to calculate 
forces within the BOP method, they are unfortunately not consistent 
with the total energy. In another version of the recursion method the GDOS 
method (Horsfield {\it et al.,} 1996b, Horsfield {\it et al.,} 1996c) the exact forces can be calculated.
It is however necessary to calculate 
some generalized moments $H^k$ called interference terms from the recursion coefficients. 
This inversion is ill conditioned and becomes unstable if one tries to calculate more than 
20 moments (Bowler {\it et al.,} 1997).  With such a low number of moments it is however not 
possible to describe many realistic systems (Kress and Voter 1995) and the 
error one reaches when the instability sets in is frequently still 
much too large to be acceptable (Bowler {\it et al.,} 1997).
So recursion based methods seem not to be a general purpose tool for 
electronic structure calculations where accurate energies and forces are 
required. BOP methods can however give insight into basic bonding properties of 
crystalline solids (Pettifor). Since in these methods related to the recursion algorithm 
all the diagonal elements of the density matrix have to be calculated, they are 
obviously not very efficient if a very large number of basis functions per atom is used 
and they were indeed primarily proposed for Tight Binding or other schemes with 
a small number of basis function. The only exception is a version proposed by 
Baroni {\it et al.} (Baroni and Giannozzi 1995) who suggest to use delta functions 
as a basis in a Density Functional type calculation. With his basis set the diagonal 
elements of the density matrix 
are enough to determine the charge density, whereas for more general basis functions 
the off diagonal elements are needed as well (Equation~(\ref{rhotrace})). Because the number of 
delta functions has to be very large even in the most favorable case of silicon, 
the crossover point was estimated to be around 1000 silicon atoms. This method is therefore 
clearly not competitive with most other methods where the cross over point is much lower.

Another approach to improve the scaling properties is based on so called 
pseudo-diagonalization (Stewart {\it et al.,} 1982). The method is closely 
related to the well known Jacobi method for matrix diagonalization.
Whereas in the original Jacobi method rotation transformations are 
applied until all off-diagonal elements vanish, only the entries 
which couple occupied and unoccupied states are annihilated in the 
pseudo-diagonalization method. One obtains thus not the occupied eigenvectors 
of the Hamiltonian but an arbitrary set of vectors which span the same occupied 
space. In its original formulation (Stewart and Pulay 1982) this method still 
had cubic scaling however with a smaller prefactor than complete matrix 
diagonalization. Nearly linear scaling has been reported with this method 
(Stewart 1996) if the Hamiltonian matrix is constructed with respect to 
a set of well localized orbitals. In this way most of the elements in the block 
coupling occupied and unoccupied states are zero at the start of the transformations.
The annihilation of certain matrix elements during the rotation steps causes 
only a controlled spread of other rows and columns in the matrix. So at the 
end each column and row extends over a region which is comparable to the 
localization region in other O(N) methods.

A method proposed by Galli and Parrinello (1992) can be considered as a predecessor of the 
OM method. The total energy is minimized with respect to a set of localized 
Wannier functions. In contrast to the OM method one has however the 
old conventional functional (Equation~(\ref{convfunc})) 
which has to be minimized under the constraint of orthogonality, whereas in the 
OM method the orthogonalization constraint is contained in the 
modified functional (Equation~(\ref{omfunc})). In this scheme it is necessary to calculate 
the inverse of the overlap matrix between the Wannier functions. From timing 
considerations this 
is not a serious drawback since this part is only a small fraction of the 
total workload even for large systems and even if it is done with a 
scheme which scales cubically. There are, however, problems of numerical stability. 
As pointed out by Pandey {\it et al.} (1995) 
the overlap matrix becomes close to singular and the introduction of localization 
constraints can under these circumstances falsify the results. 

Also vaguely related to the OM functional are methods where certain constrains 
are included by a penalty function. Wang and Teter (1992) included the orthogonality 
constraint in this way, Kohn (1996) the idempotency condition of the 
zero temperature density matrix.

O(N) implementations of electronic structure methods based on the multiple scattering 
theory have also been reported. In the simplest version (Wang {\it et al.,} 1995) it is 
essentially a DC method with the difference that within each localization region 
the calculation is done with the multiple scattering method. A further development was 
to replace the buffer region by an effective medium 
(Abrikosov {\it et al.,} 1995, Abrikosov {\it et al.,} 1996). This can considerably 
reduce the prefactor of the calculation, especially in metallic systems where large 
localization regions are needed. For this class of methods no force formulas have 
however been reported, a deficiency restricting their applicability.

A scheme which possibly leads to a reduced scaling behavior has also be 
proposed by Alavi {\it et al.,} (1994) It is based on a direct representation under the form of a 
sparse matrix of the Mermin finite temperature functional (Mermin 1965). 
So it allows for a finite temperature as does the FOE method.

As was already mentioned the polynomial FOE method becomes inefficient 
in cases where one has a large basis set causing the highest eigenvalue to 
grow very large. This would necessitate a Chebychev polynomial of very high degree.
A elegant method to overcome this bottleneck within a polynomial 
method has been proposed by Baer and Head-Gordon (1998). 
They write the density matrix at a low temperature $T$ as 
a telescopic expansion of differences of density matrices at higher 
temperatures $Tq^j$.
\begin{equation} 
 F_{T} = F_{Tq^n} + \left( F_{Tq^{n-1}} - F_{Tq^n} \right)
                  + \left( F_{Tq^{n-2}} - F_{Tq^{n-1}} \right)
                  +  ...   
                  + \left( F_{Tq^{0}} - F_{Tq^{1}} \right)
\end{equation}
The factor $q$ determining the geometric sequence of temperatures 
is chosen from considerations of numerical convenience.
As the temperature is lowered the numerical rank of each difference term 
becomes smaller and smaller 
since the difference of two Fermi distributions of similar temperature is 
vanishing to within numerical precision over most of the spectrum.
Hence it is necessary to find Chebychev expansions only over successively smaller 
regions of the spectrum and it is also possible to calculate the 
traces (which in turn determine all physically 
relevant quantities) within spaces of smaller and smaller dimension.

Ordejon {\it et al.} (1995) proposed a method based on the OM method to calculate 
phonons from the electronic structure with linear scaling. 

This article concentrated on O(N) methods which are able to calculate the total 
energy of a system as well as its derived quantities such as the forces 
acting on the atoms. Let us finally still point out that there are 
also several O(N) methods which primarily 
calculate the density of states and thus give information about the eigenvalue 
spectrum of a system. These methods (Drabold and Sankey 1993, 
Wang 1994, Silver and Roeder 1997) are not described in this article. 
In principle, it would also be possible to derive band structure and total 
energies from the spectrum by an integration up to the chemical potential.
Attempts in this direction have however not been very successful up to now 
with the exception of the smeared density of states Kernel Polynomial method 
(Voter {\it et al.,} 1996) which is closely related to the FOE method.

\section{Some further aspects of O(N) methods}

Hierse and Stechel (1994, 1996) 
examined whether non-orthogonal Wannier like orbitals are transferable from one 
chemical environment to another similar one. If this was the case one could reuse 
precalculated Wannier orbitals as a very good initial starting guess. 
Such a property would thus reduce the 
prefactor of any method regardless of its scaling behavior. Unfortunately, they 
could find reasonable transferability only under rather restrictive conditions. 
When they added $C H_2$ units to a $C_n H_{2n+2}$ polymer they found good 
transferability of the orbitals associated with this building block 
within a Density Functional scheme (Hierse and Stechel 1994). 
As soon as they started bending the polymer (Hierse and Stechel 1996), 
the transferability was however lost in the Density Functional scheme.
Only in a non-self-consistent Harris functional scheme some efficiency gains 
were still possible.

Hernandez {\it et al.} (1997) suggest a solution to the basis set problem 
in O(N) methods.  As was mentioned in section~\ref{four} 
O(N) techniques are difficult to reconcile with extended basis sets such 
as plane waves. Plane waves have however several important advantages and are 
therefore widely used in conventional (i.e. not O(N)) electronic 
structure calculations. One of their main advantages is that the accuracy 
can easily be improved by increasing one single parameter, namely the 
minimal wavelength which corresponds to the resolution in real space. 
Hernandez {\it et al.} proposed now a basis set  of ``blip" functions which 
combines both advantages. 
It is localized and its resolution can systematically be improved.
As an alternative to the ``blip" functions one could also use finite 
difference techniques (Chelikowsky {\it et al.,} 1994) or wavelets 
(Lippert {\it et al.,} 1998, Goedecker and Ivanov 1998b, Arias 1998, Goedecker 1998b) 
since they share the same advantages. As shown by Goedecker and Ivanov (1998c) 
wavelets allow for highly compact representation of both the 
density matrix and the Wannier functions, since they are localized 
both in real and Fourier space. 

Much of the work of Ordejon (1996) 
is also based on a new set of basis functions 
proposed by Sankey and Niklewski (1989). 
This basis set consists of atomic orbitals which are modified 
in such a way that they are strictly zero outside a certain
spherical support region. This then gives rise to a 
Hamiltonian matrix which is strictly sparse. 
By tabulating these matrix elements it is possible to do Density Functional 
calculations whose computational requirements are in between the requirements of 
traditional Density Functional calculations and Tight Binding calculations.
Obviously one has to find a compromise between accuracy and speed. If the basis 
functions extend about a larger support region, one has a more accurate basis, but the 
the numerical effort increases because the Hamiltonian is less sparse.

Horsfield and Bratkovsky (1996d) have incorporated entropy terms 
in O(N) methods within the FOE algorithm. As soon as one has systems at 
nonzero temperature, these terms should in principle be added, however 
in most systems their effects are very small 
at room temperature. For computational convenience temperatures much 
larger than room temperature can however be used.
Wentzcovitch {\it et al.} (1992) and Weinert and Davenport (1992) 
showed that the inclusion of entropy gives 
simplified force formulas, since only the Hellmann-Feynman term survives.
The free energy $A$ is given by
\begin{equation}
 A = Tr [F \: H - k_B T \: S(F)] \: ,
\end{equation}
where the entropy $S$ is a matrix function of $F$
\begin{equation}
 S = - \left( F \: \ln (F)  + (1-F) \: \ln (1-F) \right) \: .
\end{equation}
Writing $S$ as a Chebychev polynomial in $H$ and analyzing everything in terms of the 
eigenfunctions of $H$ they find that one has to do a Chebychev fit to 
a distribution very similar to the Fermi distribution just with 
some additional features close to the chemical potential. 
Using a formalism by Gillan (1998) they then extrapolate their 
results to zero temperature obtaining faster convergence  to the 
zero temperature limit in this way. Let us stress again, that with 
the FOE method it is possible to build up density matrices corresponding to 
several temperatures without significant extra cost.
A set of generalized Fermi distributions which allow an efficient
extrapolation to the zero temperature limit by eliminating 
arbitrarily high powers of $T$ has also been proposed 
by Methfessel and Paxton (1989) 

As was mentioned in the introduction the fundamental cubic term in an electronic structure 
calculation based on orbitals comes from the orthogonalization requirement.
In traditional pseudopotential calculations based on a Fourier space 
formulation there is however a second very large cubic term, arising from 
the nonlocal part of the pseudopotential. This second cubic term can 
be eliminated by using pseudopotentials which are separable in real 
space (King-Smith {\it et al.,} 1991, Goedecker {\it et al.,} 1996, Hartwigsen {\it et al.,} 1998)

\section{Non-orthogonal basis sets}
\label{nonorthog}
Up to now we have always implicitly assumed that we are dealing with 
orthogonal basis sets, i.e. that 
\begin{equation} \label{ortho}
  \int \phi^*_i({\bf r}) \phi_j({\bf r}) d{\bf r} = \delta_{i,j} \: .
\end{equation}
Non-orthogonal basis sets give rise to an overlap matrix $S$, 
\begin{equation}
S_{i,j}  =  \int \phi^*_i({\bf r}) \phi_j({\bf r}) d{\bf r}  \: .
\end{equation}
An orthogonality relation similar to Equation~(\ref{ortho}) can be obtained 
by introducing the dual basis functions $\tilde{\phi}_i({\bf r})$ given by 
\begin{equation}
\tilde{\phi}_i({\bf r}) = \sum_j S^{-1}_{i,j} \phi_j({\bf r}) \: ,
\end{equation}
where $S^{-1}$ is the inverse of the overlap matrix $S$. It is then easy to verify that
\begin{equation} 
  \int \tilde{\phi}^*_i({\bf r}) \phi_j({\bf r}) d{\bf r} = \delta_{i,j} \: .
\end{equation}
As has been mentioned in section~\ref{four} all realistic atom centered localized 
basis sets are non-orthogonal. Within the Tight Binding context, there are 
also many non-orthogonal schemes on the market. There is thus 
certainly a need to apply O(N) techniques also for these schemes. 
Most of the basic O(N) algorithms presented previously have therefore 
been generalized to the non-orthogonal case and we will present these 
generalizations in the following. 
In the context of a non-orthogonal scheme one has to distinguish carefully 
between the eigenfunctions $\Psi_n$ and the associated eigenvector ${\bf c}^n$ which 
contain the expansion coefficients $c_i^n$ such that 
$\Psi_n({\bf r}) = \sum_i c_i^n \phi_i({\bf r})$.
The eigenvector ${\bf c}^n$ satisfies the generalized eigenvalue equation
\begin{equation}
 H {\bf c}^n  =  \epsilon_n S {\bf c}^n   \label{gevcs}  \: .
\end{equation}
In the same way one has to distinguish carefully between the density matrix operator 
$F({\bf r},{\bf r}')$ and the density matrix $F_{i,j}$ itself.
While the expression ~(\ref{fdens}) for the density matrix operator remains the same,
\begin{eqnarray} \label{frrno}
 F({\bf r},{\bf r}') = \sum_n f(\epsilon_n) \Psi^*_n({\bf r}) \Psi_n({\bf r}') 
     =  \sum_n f(\epsilon_n) \sum_{i,j} c^{n *}_i  c_j^n  \phi^*_i({\bf r}) \phi_j({\bf r}') 
\end{eqnarray}
the expression for the number of electrons (Equation~(\ref{nel})) is modified to
\begin{equation}
N_{el} = Tr[F \: S] \: .
\end{equation}
The expression for the band structure energy ~(\ref{ebstrace}) however remains valid.
In the definition of the density matrix $F_{i,j}$ (Equation~(\ref{fijdef})) 
one has to use now the dual basis functions instead of the ordinary,
\begin{equation} \label{fijdefn}
F_{i,j}  =  \int \int \tilde{\phi}^*_i({\bf r}) F({\bf r},{\bf r}') \tilde{\phi}_j({\bf r}') d{\bf r} d{\bf r}'  = \sum_n c^{n *}_i  c_j^n f(\epsilon_n)
\end{equation}
This replacement can have important consequences for the locality of the density 
matrix $F_{i,j}$.  If we have a set of localized orthogonal basis functions 
(the only known set of basis functions with this property are the Daubechies 
scaling functions), whose extension is less than the oscillation period of the 
density matrix operator then Equation~(\ref{fijdef}) ensures that 
$F_{i,j}$ will have the same decay properties as $F({\bf r},{\bf r}')$. This does 
not necessarily hold true for Equation~(\ref{fijdefn}). Even if the basis functions 
$\phi_i({\bf r})$ are well localized this is in general not true for 
their duals $\tilde{\phi}_i({\bf r})$. If the duals have a very slow decay then 
this slow decay will be inherited by $F_{i,j}$ and it might not be possible to 
use O(N) algorithms for the calculation of $F_{i,j}$. 
Problems might for instance arise if high quality 
Gaussian basis sets containing diffuse functions are used. Preliminary experience 
indicates that for small basis sets of moderate quality the duals are not 
so delocalized as to destroy the localization of $F_{i,j}$. 

In the case of the DC method the 
generalization to the non-orthogonal case is trivial. Since it is based on diagonalization 
within each subvolume one has to solve just the generalized eigenvalue 
problem (Equation~(\ref{gevcs})) instead of an ordinary one. In the Density Functional context, the 
DC method has actually only been used with non-orthogonal orbitals.

The FOE method using a Chebychev representation of the density matrix 
has been generalized by Stephan {\it et al.}(1998). It is easy to see that all the 
central equations of the FOE method remain correct if the 
Hamiltonian $H$ is replaced by a modified Hamiltonian $\bar{H}$ 
(that is not any more hermitian) given by 
\begin{equation} \label{modham}
\bar{H} = S^{-1} H \: .
\end{equation}
In particular, it remains true that the density matrix 
is given within arbitrary precision by 
\begin{equation}
F \approx \frac{c_0}{2} I + \sum_{j=1}^{n_{pl}} c_j T_j(\bar{H})
\end{equation}
if a sufficiently large $n_{pl}$ is used. 
The problem is how to find $\bar{H}$ efficiently. Even if $S$ is a sparse matrix 
the inverse $S^{-1}$ is not sparse in general and $\bar{H}$ would be a full 
matrix as well, destroying immediately 
the linear scaling. However, it turns out that the matrix elements of $\bar{H}$
decay faster than the corresponding matrix elements of $H$ (Stephan 1998, Gibson 1993). 
One can therefore cut off the Tight Binding Hamiltonian $\bar{H}$ at the same 
distance where one would usually cut off $H$. In this way all the matrices involved 
are reduced to sparse matrices and $\bar{H}$ can be constructed by 
solving the set of linear systems
\begin{equation} 
S \bar{H} = H \: .
\end{equation}
Since both the right hand sides in $H$ and the solution vectors 
making up $\bar{H}$ are sparse different systems of equations are 
only coupled by subblocks of $S$. So the big matrix inversion problem 
is decoupled into many small systems of equations and the scaling 
is therefore strictly linear.

If the FOP method is used in connection with a rational approximation 
the generalization to the non-orthogonal case can be done straightforwardly 
and without any approximation (Goedecker 1995)
\begin{equation} 
 F = \sum_{\nu} \frac{w_{\nu} }{H-z_{\nu} S }  \: .
\end{equation}
One has then to solve 
a systems of linear equations which is a generalization of Equation~(\ref{padeeq})
\begin{eqnarray}
 (H-z_{\nu} S ) F_{\nu} & = & I  \label{green} \\
  F & = & \sum_{\nu} w_{\nu} F_{\nu}   \: .
\end{eqnarray}
A similar approach was adopted by Jayanthis {\it et al.} (1998).
They formulated their method in terms of the Green function. However, 
$F_{\nu}$ in Equation~(\ref{green}) is a Green function for a complex 
energy $z_{\nu}$ and the methods are essentially equivalent.

If the FOE method is used to calculate Wannier functions the required 
projection operator $F_p$ is slightly different from the density matrix and it is given by
\begin{equation} 
 F_p = \sum_{\nu} \frac{w_{\nu} S }{H-z_{\nu} S }  \: .
\end{equation}

The DMM method has also been generalized to the non-orthogonal case (Nunes and Vanderbilt 1994). 
Introducing a modified density matrix $\bar{F}$ defined as 
\begin{equation} 
 \bar{F} = S^{-1} F S^{-1}
\end{equation}
the DMM functional~(\ref{dmmfunc}) becomes
\begin{equation} 
\Omega = Tr[ (3 \bar{F} S \bar{F} - 2 \bar{F} S \bar{F} S \bar{F}) (H - \mu I)] \: .
\end{equation} 
This has the advantage that one does not have to invert $S$ if one minimizes 
directly with respect to $\bar{F}$.
The calculation of the gradient of the DMM functional in the non-orthogonal case 
is a tricky point. The definition of the gradient is not as absolute as one 
might think. It is the direction of steepest descent per unit change of the variables, 
and one must therefore define a norm for the multidimensional space of variables 
before defining the gradient (D. Vanderbilt private communication).
Two different gradient expression have been proposed by Nunes and Vanderbilt (1994) 
and by White {\it et al.} (1997) which correspond to two different choices of metric. The gradient of 
White {\it et al.} (1997) requires less minimization steps (Gillan {\it et al.,} 1998), 
but each minimization step is more expensive since it requires the calculation of the 
inverse of the overlap matrix. From the point of view of overall numerical 
efficiency it is therefore not clear which gradient expression is more efficient.

Another generalization (Millam and Scuseria 1997, Daniels {\it et al.,} 1997) 
of the DMM method which is similar in spirit to 
Stephan's generalization of the FOE method consists in first performing 
a transformation to an orthogonal basis set by finding the LU decomposition 
of the overlap matrix
\begin{equation} \label{ludec}
 S  = U^T U \: ,
\end{equation} 
where U is an upper triangular matrix. If in addition $U$ is approximated as 
a sparse matrix with $m$ off-diagonal elements then Equation~(\ref{ludec}) can be solved 
with a scaling proportional to $n \: m^2$ where $n$ is the dimension of the 
matrices involved. Thus the scaling with respect to the size of the system is linear 
as it should be.

The OM method can easily be generalized to the non-orthogonal case 
(Ordejon {\it et al.,} 1996). The orbital overlap $\sum_l c^n_l c^m_l$ in the OM 
functional (Equation~(\ref{omfunc})) has to be generalized to 
$\sum_{l,k} c^n_l S_{l,k} c^m_k$.

\section{The calculation of the selfconsistent potential} 
We will now discuss an issue which is relevant only in selfconsistent 
electronic structure calculations, namely the calculation of the potential arising 
from the electronic charge. This potential consists essentially of two 
parts, the electrostatic or Hartree potential and the exchange correlation 
potential.

\subsection{The electrostatic potential} 
\label{coulomb}
The solution of Poisson's equation to find the electrostatic potential 
arising from a charge distribution $\rho$ is a basic problem found 
in many branches of physics. Solution techniques are 
described in a wide variety of books and articles. 
We will therefore only point out a few aspects which are important in the 
special context of O(N) electronic structure calculations.

If one is dealing with an electronic charge density which has only one length 
scale Poisson's equation can be solved efficiently and with a scaling which 
is close to linear. Charge densities of this type are encountered in the 
context of pseudopotential calculations where one has eliminated the core 
electrons and where the characteristic length scale is the typical extension 
of an atomic valence electron. One can, for instance, use plane wave
or multigrid techniques (Briggs) which both have a scaling proportional to 
$n \: \log (n)$ where $n$ is the number of grid points.

The situation becomes problematic when core electrons are included. In this case 
one could in principle still use the above mentioned techniques with 
a increased resolution. One would still have linear scaling, but 
the prefactor would be so large to make it completely impractical both from 
timing and memory considerations. Exactly the same arguments apply 
to the representation of the wave functions themselves and for this reason 
ordinary plane waves are not used for all electron calculations. 

A widely used basis set for all electron calculations are Gaussian type basis sets 
(Boys 1960). 
By varying the width of the Gaussians one can describe both core and valence electrons.
The popularity of Gaussian type basis functions comes from the fact that 
many important operations can be done analytically (S. Obara and Saika, 1986)
One property which we will use is that the product of two Gaussians is again a 
Gaussian centered in between the two original Gaussian type functions.
The matrix elements of the electrostatic potential part of the  Hamiltonian with 
respect to a set of Gaussian orbitals  $g_i({\bf r}), i = 1, ... , M_b$  are given by
\begin{equation} \label{gtohij}
 H_{i,j} = \int d{\bf r} g_i({\bf r}) 
           \left( \int d{\bf r}' \sum_{k,l} \frac{ F_{k,l} \: g_k({\bf r}') \: g_l({\bf r}') }{|{\bf r}-{\bf r}'|} \right) g_j({\bf r}) \: .
\end{equation} 
The elementary integral
\begin{equation} 
  \int d{\bf r}  \int d{\bf r}' \frac{ g_i({\bf r}) g_j({\bf r}) g_k({\bf r}') g_l({\bf r}') }{|{\bf r}-{\bf r}'|} 
\end{equation} 
can also be calculated analytically (S. Obara and Saika, 1986).
A straightforward evaluation of Equation~(\ref{gtohij}) would then result in 
a quartic scaling. There are, however, many 
well known techniques (Challacombe {\it et al.,} 1995) to reduce this scaling. 
The most obvious trick 
starts from the observation that there is a negligible contribution to 
the charge density $\rho$ if the Gaussians $g_l$ and $g_k$ 
are centered far apart. Consequently the charge density is not a sum 
over $M_b^2$ product Gaussians $G_m$, ($G_m = g_i g_j$), 
but only over $M_a$ such Gaussians 
\begin{equation} \label{rhoprod}
\rho({\bf r}') = \sum_{k=1}^{M_b} \sum_{k=1}^{M_b} F_{k,l} \: g_k({\bf r}') \: g_l({\bf r}') 
            \approx  \sum_{m=1}^{M_a} c_m G_m({\bf r}') \: .
\end{equation} 
The size of the auxiliary basis set $M_a$ is proportional to $M_b$  with a large 
prefactor which depends on the ratio of the largest to the smallest extension of 
the Gaussians as well as on the accuracy target.
In a similar way matrix elements $H_{i,j}$ become negligible if 
the basis functions $g_i$ and $g_j$ are centered very far apart.
Using these two approximations one obtains a quadratic scheme with a very big prefactor. 
An approximate quadratic scaling is also found in numerical tests
(Strout and Scuseria 1995). 

Another widely used method consists in fitting the 
charge density by a set of $M_a$ auxiliary Gaussians $G_i$. 
Even though we use the same symbols ($M_a$, $G_i$) as above they denote now somewhat different 
quantities. The allowed number of auxiliary Gaussians $M_a$ is now much smaller, 
namely a few times $M_b$. The auxiliary functions themselves are therefore not anymore 
taken to be all the possible product functions, but determined by empirical rules in such a way as 
to give the best possible fit in Equation~(\ref{rhoprod}).
The fitting involves the solution 
of an ill conditioned system of linear equations and has therefore cubic scaling, however
with a small prefactor. The evaluation of the matrix elements using this representation 
of the charge density has then quadratic scaling if one again neglects small elements.

To obtain an even better scaling behavior requires the introduction of completely 
new concepts. One possibility is to build upon algorithms which solve the classical 
Coulomb problem for point particles. The classical Coulomb problem requires the 
calculation of the electrostatic potential arising from all the $N$ particles with 
charge $Z_j$ at all the positions ${\bf r}_i$
\begin{equation} \label{fmm}
U( {\bf r}_i ) = \sum_j \frac{Z_j}{|{\bf r}_i-{\bf r}_j|} \: .
\end{equation} 
By grouping particles into hierarchical groups and by describing their 
potential far away from such groups in an controlled approximate way 
by multipoles these fast algorithms allow to evaluate Equation~(\ref{fmm}) 
with linear scaling instead of the expected quadratic one.
There are now several proposals (Strain {\it et al.,} 1996, White {\it et al.,} 1994, 
Perez-Jorda 1997), how one can modify one of these 
fast algorithms, the Fast Multipole Method (Greengard 1994) in 
such a way that it can handle also the continuous charge distributions
arising in the context of electronic structure calculations.
The basic idea is fairly straightforward. 
As we saw the charge distributions is always given as a weighted sum 
of auxiliary Gaussians (Equation ~(\ref{rhoprod})). Now the electrostatic potential 
of such a Gaussian particle looks the same from a distance as the potential 
of a point particle. Concerning the far field, one can thus essentially 
take over the existing algorithms. To account for the non point like 
nature of the Gaussian particles one has however to correct the near field. 
Since the calculation of these local corrections have 
linear scaling the whole procedure has linear scaling as well.
There are two problems with this kind of approach. First, one has only linear scaling 
with respect to the size of the basis set if the volume covered by the 
basis set grows at least as fast as the size of the basis set. 
If one adds for example basis functions for a molecule of fixed size, to improve 
the accuracy of the basis set one does not any more have linear scaling. 
This is related to the fact that one can apply the fast far field treatment now 
to a smaller number of Gaussian particle interactions. This behavior can be 
easily understood by considering the extreme case where all Gaussian particles 
are centered very close together within a radius which is smaller than their width.
In this case one evidently cannot use any more any far field techniques. 
The second problem is closely akin to the first. If one adds extended 
Gaussians to the system the efficiency deteriorates.

A method which scales strictly linear with respect to the size of the basis set,  
independently of whether the volume is increased at the same time or not 
and which can be applied within the context of any basis set, 
is based on wavelets (Goedecker and Ivanov 1998a). As the input to this method the charge 
density is needed on a real space grid which can have varying resolution. 
THus near the core region of the atoms in a molecule the resolution can 
be arbitrarily increased. Using interpolating wavelets this charge density can 
uniquely be mapped to a wavelet expansion, since a wavelet expansion 
can compactly describe nonuniform functions. In the wavelet basis 
one can then iteratively solve Poissons equation 
\begin{equation}  \label{poiss}
 \nabla^2 V = -4 \pi \rho \: .
\end{equation} 
The matrix representing the Laplace operator $\nabla^2$ is sparse and the
matrix times vector multiplications needed for the iterative solution of 
Equation~(\ref{poiss}) scale linearly. Using a preconditioning scheme in 
a basis of lifted wavelets the condition number is independent of the size 
and of the maximal resolution of the wavelet expansion and the number 
of iterations is therefore constant. Thus one obtains an overall linear scaling.

\subsection{The exchange correlation potential} 
Within the most popular versions of Density Functional theory the exchange 
correlation potential is a purely local function. In the case of the 
Local Density Approximation (Parr and Yang) the exchange correlation potential 
at a certain point depends only on the 
density at that point, in the case of Generalized Gradient Approximations (Perdew {\it et al.,} 1996,
Becke 1988, Lee {\it et al.,} 1988) it depends in addition still on the gradient of the 
density at that point. If one uses real space methods such as finite differences 
or finite elements as well as plane wave methods where the calculation of the 
exchange correlation potential is done on a real space grid as well, it is 
obvious that the numerical effort is linear with respect to the system size. 
If one uses more extended basis functions such as Gaussian type orbitals it becomes 
more difficult to achieve linear scaling (Stratmann {\it et al.,} 1996). 

In the case of Hartree Fock calculations the exchange energy 
\begin{equation}  \label{hfex}
  \sum_{i,j} \int \int d{\bf r} d{\bf r}' 
  \frac{ \Psi_i({\bf r}) \Psi_j({\bf r}) \Psi_i({\bf r}') \Psi_j({\bf r}') }
  {|{\bf r}-{\bf r}'|} 
\end{equation} 
seems to be as nonlocal as the Coulomb potential. Using Equation~(\ref{denst0}) 
one can however rewrite the expression ~(\ref{hfex}) to obtain
\begin{equation}  
  \int \int d{\bf r} d{\bf r}' 
  \frac{ F({\bf r},{\bf r}') F({\bf r},{\bf r}')} {|{\bf r}-{\bf r}'|} 
\end{equation} 
showing that the exchange energy in an insulator is indeed a local 
quantity whose locality is determined by the decay properties of the density matrix.
A linear method to evaluate exchange terms within Gaussian type electronic structure 
calculations based on the aforementioned locality properties 
has been devised by Schwegler and Challacombe (1996).
An alternative method based on the Fast Multipole Method has been developed by
Burant {\it et al.,} (1996).

\section{Obtaining self-consistency}
To do a self-consistent electronic structure calculation, two ingredients have 
to be blended. The first is the calculation of the density matrix in 
a fixed external potential, a topic which is the main focus of this article. The 
second is the calculation of the potential from a given electronic charge density 
which was discussed in the 
preceeding section~(\ref{coulomb}). Even if both of these basic parts exhibit linear 
scaling, it is however not yet granted, that one has overall linear scaling.
It might happen that the number of times one has to repeat these two basic 
parts increases with the size of the system.

The easiest scheme to combine the calculation of the density matrix and the calculation 
of the potential is the so called scalar mixing scheme. Given an input charge 
density $\rho_{in}$ which determines the potential one obtains after the 
calculation of the density matrix for this potential via 
Equation~(\ref{rhotrace}) a new output density $\rho_{out}$. The new input 
density $\rho_{in}^{new}$ is 
now not the output density $\rho_{out}$, but rather a linear 
combination of the old input density and the output density
\begin{equation}
\rho_{in}^{new}({\bf r}) =  
  \rho_{in}({\bf r}) + \alpha ( \rho_{out}({\bf r}) - \rho_{in}({\bf r}))  \: .
\end{equation}
Here $\alpha$ is the mixing parameter. Overall linear scaling is endangered if 
one has to decrease $\alpha$ for reasons of numerical stability 
as the system becomes bigger and if one consequently 
needs a larger number of iterations. 

The standard theory of mixing (Dederichs and Zeller 1983, Ho {\it et al.,} 1982) 
is based on the dielectric response function in ${\bf k}$ space. Within this theory 
numerical instabilities arise if the the dielectric response functions tends to infinity 
as $k$ tends to zero. This happens in a metal but not in an insulator where the 
dielectric response function always remains finite. Following the general 
philosophy of this paper to remain within a real space formalism, we 
will elucidate mixing under this 
perspective. The final conclusions are of course the same as the one based on the 
Fourier space theory.

Let us first consider a metal. We assume that we 
are doing a calculation for a one-dimensional metallic structure of length $L$. Let us also 
assume that due to deviations from the converged self-consistent charge density 
we transfer an incremental charge $\Delta Q_{in}$ from one end of the sample to the 
other. Using the basic formula for the potential in a capacitor we get 
a constant electric field in the sample giving rise to a potential difference 
of $\Delta U = L \: \Delta Q_{in}$ between the two ends. 
In a metal this potential difference will most likely be 
larger than the HOMO LUMO separation (which vanishes for large systems) 
and we get a large charge transfer $\Delta Q_{out}$.
This charge transfer is related to the density of states 
at the Fermi level, $D(\mu)$, which in our one dimensional case is the 
number of states per length unit and per energy unit.
So the total charge transfer $\Delta Q_{out}$ is given by
\begin{equation}
\Delta Q_{out} \approx D(\mu) \: L \: \Delta U = D(\mu) \: L^2  \: \Delta Q_{in} \: .
\end{equation}
If this induced charge transfer $\Delta Q_{out}$ is larger than the initial transfer 
$\Delta Q_{in}$ then the charge transfer will exponentially increase in 
subsequent iterations and we have the numerical instability called 
``charge sloshing''. To avoid it the mixing factor $\alpha$ has to be proportional to 
$\frac{1}{D(\mu) \: L^2}$. Doing the same analysis in a three-dimensional 
structure all the lateral dimensions cancel and we get the same result concerning 
$\alpha$. Denoting the volume of our sample by $v$ we find that 
$\alpha$ is proportional to $v^{-2/3}$. So $\alpha$ has to be decreased with 
increasing volume and the number of iterations in the mixing schemes 
increases with increasing system size. Fortunately and contrary to the 
implications of Annett (1995), this charge sloshing can be 
eliminated by state-of-the-art techniques (Kresse, 1996). One possibility (Kerker 1981)
is just to do the mixing in Fourier space and to have a $k$ 
dependent mixing parameter $\alpha(k)=\alpha_0 \frac{k^2}{k^2+k_0^2}$
\begin{equation}
\rho_{in}^{new}({\bf k}) =  \rho_{in}({\bf k}) + \alpha_0 \frac{k^2}{k^2+k_0^2} ( \rho_{out}({\bf k}) - \rho_{in}({\bf k}))  \: .
\end{equation}
As we see long wavelength components (corresponding to small $k$ values) 
are now strongly damped by 
\begin{equation}
\alpha_0 k^2 = \alpha_0 \left( \frac{2 \pi}{\lambda} \right)^2  
\end{equation}
and the dampening has the correct dimensional behavior with respect to 
the wavelength $\lambda$ which corresponds to the length $L$ in our above
dimensional analysis.
Short wavelength contributions are just damped by $\alpha_0$ 
and this constant dampening sets in when $k$ becomes comparable in magnitude to $k_0$. 
We know, that for wavelengths of the order of the interatomic spacing a mixing parameter 
somewhat smaller than 1 works well and so we can determine by these conditions the values 
of $\alpha_0$ and $k_0$.

Let us next examine whether we can have charge sloshing in an insulator.
We will assume that the potential difference across the sample is not larger 
than the gap, in which case the discussion for the metallic case would rather apply.
Again we consider a sample of length $L$.
According to the modern theory of polarization in solids (King-Smith and Vanderbilt 1989)
a polarization arises because the centers of the Wannier functions are 
displaced under the action of an electric field. Since the Wannier functions 
are exponentially localized, the charge which will build up at the two surfaces 
of our sample is mainly due to the displacements of the Wannier function in the 
elementary cells of the crystal right on the surface and the charge $\Delta Q_{out}$ 
is thus practically independent of the length of the sample. So the optimal mixing constant 
$\alpha$ is nearly independent of the size of the system and the number of iterations 
as well. 

In conclusion, we see that linear scaling can also be obtained in the 
selfconsistent case and that even in a metal charge sloshing problems 
can be overcome. 

Mixing is the natural choice if the DC or the FOE methods are used in 
a self-consistent calculation. If methods based on minimization 
(DMM and OM) are used one can alternatively also obtain the ground state 
by a single minimization loop (R. Car and M. Parrinello 1985, I. \v{S}tich {\it et al.,} 1989,
M. P. Teter {\it et al.,} 1989, M. Payne, {\it et al.,} 1992 ) without distinguishing 
between density matrix optimization cycles and potential mixing cycles.
As is not surprising after our discussion of mixing one finds (Annett 1995) that 
in an insulator the number of iterations does not depend upon whether 
one has a self-consistent type of calculation where the 
potential is varying during each minimization step or whether one 
has a fixed potential. In other words there is no charge sloshing.
In metallic systems it is essential to have finite electronic 
temperature (Wentzcovitch {\it et al.,} 1992, Weinert and Davenport 1992, Kresse 1996) 
and therefore the minimization schemes cannot be 
applied straightforwardly in any case. Insofar Annett's analysis (Annett 1995) 
showing that in this case the scaling is at least proportional to 
$N_{at}^{4/3}$ is irrelevant.

A completely different approach to the mixing problem has recently been proposed 
by Gonze (1996). He calculates the gradient of the total energy with 
respect to the potential. His gradient expression does not depend on 
the wavefunctions and could thus well be combined with O(N) schemes.

\section{Applications of O(N) methods}
This chapter is not intended to be a comprehensive or even complete review 
of all the applications which were done using O(N) methods. It is rather intended 
as an illustration of the wide range of areas where O(N) methods allowed to 
study systems which were too big to be studied with conventional methods.
In general one can say that most large-scale Tight Binding studies are nowadays done 
in the connection with O(N) methods. In those cases systems comprising from a few hundred up to 
many thousand atoms are typically studied. Treating such a large number of atoms 
with O(N) Density Functional methods is much more difficult. In the case of 
Density Functional calculations the benchmarking and verification aspect 
is usually dominating whereas in tight binding calculations the 
focus was in most cases on how to solve challenging physical problems. 

\begin{figure}[ht]
     \begin{center}
      \setlength{\unitlength}{1cm}
       \begin{picture}( 8.,9.5)           % figure dimensions
        \put(-0.,-1.0){\includegraphics{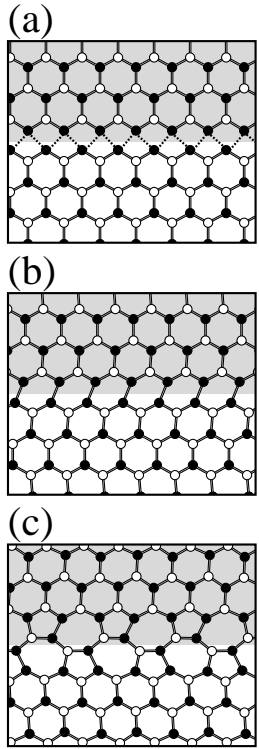}}   % VAX
       \end{picture}
       \caption{\label{nunes} \it  (a) Symmetric reconstruction of the 
               $90^0$ partial dislocation in silicon. Shaded area 
               indicates stacking fault. (b) the single-period 
               symmetry breaking reconstruction. (c) The double-period 
               symmetry breaking reconstruction. The figure is reproduced with 
               the kind permission of the authors from Nunes {\it et al.} (1998). }
      \end{center}
     \end{figure}

Questions concerning extended 
defects in crystalline materials were one of the main focus of these Tight Binding 
studies. Because several good Tight Binding parameters are available for silicon, 
most studies were done for this material.
The $90^0$ partial dislocation in silicon was at the focus of interest of a series of tight 
binding studies. The three structures that were 
examined are shown in Figure~\ref{nunes}. 

The energy difference between the structure (a) and (b) of Figure~\ref{nunes} 
was studied both by Nunes {\it et al.} (1996) and by Hansen {\it et al.} (1995). 
Even though they used different TB parameters and different O(N) algorithms 
(DMM and FOE) the both obtained exactly the energy difference of .18eV/$\AA$ 
in favor of structure (b). Later Benetto and al. (1997) discovered 
a new structure (c) that is even lower in energy. To validate their tight 
binding results they did conventional density functional calculations 
for smaller subsystems finding perfect agreement with the O(N) tight binding 
results. This new structure is 
experimentally difficult to distinguish form structure (b) and so this result 
is a convincing illustration of the power of these new O(N) algorithms 
in materials science. All these tight binding calculations necessitated electronic structure 
calculations involving a few hundred atoms and would therefore have been 
unfeasible with standard algorithms. 

Extended \{311\} defects in silicon systems containing more than 1000 atoms 
and their relation to point defects 
were studied by Kim {\it et al.} (1997) using the OM method in the improved version 
of Kim {\it et al.} (1995). An understanding of these processes is 
important for the fabrication of semiconductor devices, since defects have 
a strong influence of the diffusion properties of semiconductors. 
Unfortunately, the more realistic questions involving boron dopant atoms in 
addition to the bulk silicon atoms can not be treated with current tight binding models. 

Ismail-Beigi and Arias (1998) examined the surface reconstruction properties 
of silicon nanobars, finding that the influence of edges in these small 
structures is strong enough to lead to surface reconstructions that are 
different form the ones in bulk silicon. They also both employed traditional 
density functional calculations and O(N) FOE tight binding calculations 
and also found good agreement between both for small subsystem which are 
accessible to both approaches. 

Roberts and Clancy (1998) simulated vacancy and interstitial 
diffusion processes in silicon using the FOE 
tight binding method. The diffusion constants they obtain are in good agreement with 
similar calculations based on classical force fields and density functional 
calculations. Compared to the density functional calculations they could 
also significantly enlarge both the number of atoms (216) and the simulation times. 
The diffusion constant predicted by all these simulations is 
however orders of magnitude larger than the experimental one, a fact for which 
no explanation is known up to now.

Besides silicon there is another material for which several good Tight Binding schemes 
are available, namely carbon. Fullerene systems are therefore another focus of Tight 
Binding studies. 

Galli  and Mauri (1994) did molecular crash test of $C_{60}$ fullerenes 
colliding with a diamond surface using the OM method. They found three different impact energy 
regimes where the impinging fullerenes either survive the collision undamaged, 
slightly damaged or get completely destroyed. Even though the interaction region between 
the impinging fullerene and the surface does not comprise a very large number of atoms,
their computational box contained more than 1000 atoms. The reason why 
the box has to be so large is that the phonons emitted during the collision 
may not be reflected back from the walls of the box during the time scale of the collision.
This reflection of phonons is also a serious problem in classical force field 
simulations of crack propagation where for this reason systems comprising several 
million atoms are sometimes necessary (Zhou {\it et al.,} 1997). 
In the case of this molecular crash test most of the 
carbon atoms are propagating the phonons. Phonons are well described by 
classical force fields and one could use this scheme for the majority of the atoms,  
while it would be necessary to use the more expensive 
tight binding scheme only for the atoms in the collision region. 
Unfortunately such schemes combining molecular methods of different speed and accuracy 
have not yet been developed and thus it is therfore a feature of many O(N) calculations 
that one is doing an overkill in a certain sense, treating a large number of 
essentially inactive atoms with highly accurate methods. 
Canning {\it et al.} (1997) examined thin films of $C_{28}$ fullerenes with the same method, 
finding that thin superatom films can be formed. 

\begin{figure}[ht]
     \begin{center}
      \setlength{\unitlength}{1cm}
       \begin{picture}( 8.,9.5)           % figure dimensions
        \put(0.,0.7){\includegraphics{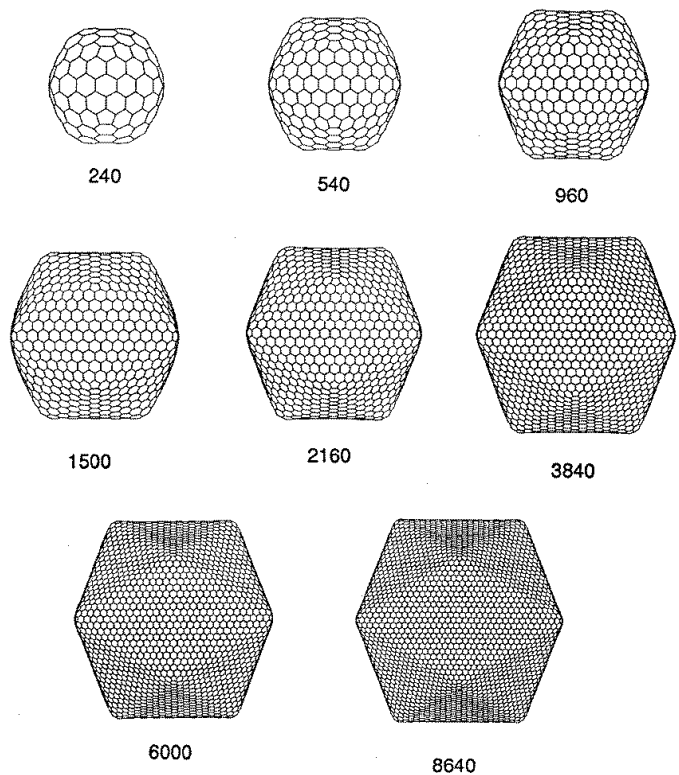}}   % VAX
       \end{picture}
       \caption{\label{scuseria} \it  Tight binding equilibrium geometries of 
                selected icosahedral fullerenes viewed along the $C_2$ symmetry axis. 
               This figure is reproduced with the kind permission of the authors 
               from Xu and Scuseria (1996). } 
      \end{center}
     \end{figure}

The equilibrium geometries of large fullerenes such as $C_{240}$ was also studied 
by several groups with O(N) techniques. The central question here is whether 
such large fullerenes have a spherical form or a polyhedrally faceted shape, 
where nearly flat polyhedral regions are alternating with edges where the curvature 
is concentrated. 
York {\it et al.} (1994) used the original 
formulation of the DC method in terms of densities to do density functional calculations 
of $C_{240}$ and found spherical shapes. Itoh and al. using both empirical and 
ab initio tight binding in the context of the OM method found however polyhedral 
shapes. This result is also supported by Xu and Scuseria (1996) who 
investigated fullerenes up to $C_{8640}$ using the DM method. 
The optimized geometries they found for various large fullerenes are shown 
in Figure~\ref{scuseria}. 

Ajayan {\it et al.} (1998) used tight binding FOE molecular dynamics to simulate irradiation 
mediated knock-out of carbon atoms out of carbon nanotubes. In agreement with 
experimental observations they found that this atom removal leads to a shrinking 
of the diameter of the nanotube, but leaves the nanotube essentially intact 
until the diameter is practically zero. 
They could identify in their virtual 400 atom sample processes on the atomic level 
that are responsible for the rapid healing of the 
defects created by the removal of atoms. 

Recently developed Tight Binding parameters (Horsfield 1996a) 
made it possible to study also a composite system, namely hydrocarbons. 
This set of tight binding parameters includes some kind of self-consistency 
by imposing a local charge neutrality requirement. Local charge neutrality 
is essential if different phases are studied because it prevents any 
unphysically large charge transfer between different phases arising 
from different chemical potentials in different phases.
Using this new tight binding scheme, Kress {\it et al.} (1998) studied the 
dissociation of $CH_4$ under high pressure and at high pressure 
using the FOE molecular dynamics. 
Previous Density functional based molecular dynamics studies by Ancilotto {\it et al.} (1997) 
were limited to very small system sizes of 16 $CH_4$ molecules and short simulation times of 
2 ps. At variance with experimental findings these density functional simulations 
could not find a phase separation 
of methane into hydrogen and carbon. FOE molecular dynamics allowed to treat much larger 
systems of 128 molecules and also much longer simulation times of 8 ps. After 4 ps 
a phase separation was indeed observed. 

Sanchez-Portal {\it et al.} (1997) compared the experimental X-ray structure 
of a large DNA molecule  comprising 650 atoms with the geometric structure 
obtained from a density functional based OM relaxation. They obtained 
a root mean square deviation from the experimental geometry of 0.23 $\AA$. 
Their method relies 
however on fairly drastic approximations resulting in errors that are 
by far larger than the error one generally expects from a density functional 
calculation. 

A similar study of a large biomolecule is reported by Lewis {\it et al.} (1997). 

York {\it et al.} used the DC method in the context of the semi-empirical 
AM1 method (Dewar {\it et al.,} 1985) to calculate heats of formation, 
solvation free energies and densities 
of states for protein and DNA systems containing up to 2700 atoms.

Daniels and Scuseria reported AM1 semi-empirical benchmark calculations for 
up to 20000 atoms using DMM, FOE and pseudo-diagonalization methods.

Applications of O(N) methods within Density Functional theory, that 
use basis sets large enough such the basis set errors are not dominating 
the Density Functional error at present do not exist. If calculations 
of this type were done, such as the calculations of a cell containing 
6000 silicon atoms by Goringe {\it et al.} (1997), then the performance evaluation 
aspect was always the dominating one. With the 
advance of faster computers and improved algorithms this situation will,  
however, certainly soon change. It is also interesting to note in this 
context that the 1998 version of the very popular Gaussian software 
package will contain O(N) algorithms.

Let us finally come back to a point briefly mentioned in the introduction.
The development of O(N) methods has also deepened our understanding 
of locality in quantum mechanical systems and has thereby also fostered 
the development of theories based on a local picture rather than the 
conventional nonlocal Bloch function picture. An example is 
the theory of polarization in crystalline 
materials by King-Smith and Vanderbilt (1989). Their theory is based on 
a local picture in terms of Wannier functions and allows for an 
intuitive understanding of these phenomena that were difficult to understand 
before.

\section{Conclusions}
O(N) methods have become an essential part of most large scale atomistic 
simulations based either on Tight Binding or semiempirical methods.
The physical foundations of O(N) methods are well understood. They are related 
to the decay properties of the density matrix. The use of O(N) methods 
within Density Functional methods is still waiting for their widespread appearance.
All the algorithms that would allow us to treat very large basis sets 
within density functional theory have certain shortcomings. 
The OM and OBDMM method suffer from ill-conditioning problems, 
and in both the OM and FOP method detailed knowledge about the 
bonding properties is required to form the input guess. 
Thus there is probably still some algorithmic progress 
necessary before these obstacles can be overcome. It is also not quite 
clear what the localization properties of very large complicated molecules 
are and whether perhaps a quadratic scaling rather than a linear scaling 
is the optimum one can obtain in certain cases.
It is clear that the elimination of the cubic scaling bottleneck 
is a very important achievement and that it will pave the way for 
calculations of unprecedented size in the future. Such calculations 
will not only be beneficial to physics, but they will also nourish progress 
in many related fields such as chemistry, materials science and biology.
Even with O(N) algorithms it will not be possible in the foreseeable future 
to treat systems containing millions of atoms 
at a highly accurate Density functional level using large basis sets, 
as would be necessary for certain materials science applications.
Such problems can only be approached if one succeeds 
in combining methods of different accuracy such as density functional 
methods with classical force fields, applying the high accuracy 
method only to regions where the low accuracy method is expected to fail.
Hybrid methods of this type will certainly be based on the same notions 
of locality and similar techniques as O(N) methods. 

\section{Acknowledgments}
I would like to express my deep gratitude to all the people who helped me to discover 
errors in the manuscript and who made valuable suggestions how to 
improve the paper. They are in alphabetical order
David Bowler, 
David Drabold, 
Olle Gunnarson,
Eduardo Hernandez, 
Andrew Horsfield,
Jurg Hutter,
Ove Jepsen, 
Joel Kress,
Klaus Maschke, 
Richard Martin, 
Chris Mundy
Ricardo Nunes, 
Michele Parrinello,
Anna Puttrino,
Gustavo Scuseria,
Uwe Stephan, 
and
David Vanderbilt
.

\pagebreak
\section{Appendix: Decay properties of Fourier transforms}
The density matrix is defined in terms of a Fourier transform given by 
Equation~(\ref{ftrans}). The decay properties of the density matrix are thereby 
closely related to the decay properties of Fourier transforms.
All the properties described in this paragraph are well known, 
for completeness we will briefly outline them.

For simplicity let us just consider a one-dimensional Fourier transform
\begin{equation} \label{frdef} 
g(r) = \int_{-\infty}^{\infty} e^{i k r} g(k) dk \: ,
\end{equation} 
where $g(k)$ is an integrable piecewise continuous function 
tending rapidly to $0$ for $k$ tending to $\pm \infty$.

For any function $g(k)$ of this type, $g(r)$ will obviously tend to zero 
when $r$ tends to infinity. In this case $e^{i k r}$ 
is a very rapidly oscillating function and the product $ e^{i k r} g(k)$ 
will therefore change sign very rapidly and thus the integral will tend to zero.
The exact decay properties depend on how many derivatives are continuous.
Let us consider first a function which is piecewise constant  and has only 
a finite number of discontinuities. A function which falls into this class 
is the function $g(k)$ which is $1$ in the interval [-1:1] and zero everywhere else.
Calculating the Fourier transform one finds $g(r) = 2 \frac{\sin (r)}{r}$.
Since any piecewise function $g(k)$ can be written as a linear combination of the 
above prototype function, its transform $g(r)$ will always decay like $1/r$.
Using integration by parts we see that
\begin{equation}
r^l g(r) = \int_{-\infty}^{\infty} g(k) \: i^{-l} \left( \frac{\partial}{\partial k} \right)^l e^{ikr} dk
         = (-i)^{-l} \int_{-\infty}^{\infty} e^{ikr} \left( \frac{\partial}{\partial k} \right)^l g(k) dk \: .
\end{equation}
If the $l$-th derivative is integrable then the integral will vanish for the reasons 
discussed above.
So if we can do $l$ integrations by parts each transformation 
will accelerate the decay by one inverse power of $r$ and we can do such a transformation 
whenever our function has at least continuous first derivatives. Hence we arrive at the 
rule that if $l$ derivatives of $g(k)$ are continuous, $g(r)$ will decay 
like $r^{-(l+1)}$. 

If we have a function $g(k)$ which is analytic, i.e. one for which 
an infinite number of derivatives 
exists, then the transform will decay faster than any power of $r$. One then says that 
it decays exponentially instead of algebraically. This notion of exponential decay does 
not necessarily mean that it decays strictly like an exponential function. As an example we 
could for instance take $g(k)= \exp(-k^2)$, where we know that the transform is again a Gaussian 
and decays thus faster than an ordinary exponential function.
The rate of decay will be related to the smallest length scale of $g(k)$. If the 
smallest length scale of $g(k)$ is $k_{min}$ then $g(r)$ will roughly decay 
like $\exp (- c |r| k_{min} )$, where $c$ is a constant of order of 1.
This follows from the fact that one will have an important cancellation of terms 
of opposite sign in the integral in Equation~(\ref{frdef}) only if 
several oscillations occur within the interval $k_{min}$.

Another qualitative feature of the Fourier transform is that it 
will have oscillations whenever $g(k)$ is shifted off center. The oscillation period 
is determined by this shift. As an example let us look at the Fourier transform of 
a shifted Gaussian $g(k)=\exp (-\frac{1}{2}(k-a)^2)$. 
The result is $g(r)=\sqrt{2 \pi} \: \exp (i a r) \: \exp (-\frac{1}{2} r^2)$, which 
is the transform of the unshifted Gaussian times an oscillatory term.

\pagebreak
\section{References}

 Abrikosov, I., A. Niklasson, S. Simak, B. Johansson, A. Ruban and H. Skriver, Phys. Rev. Lett. {\bf 76}, 4203 (1996) 

 Abrikosov, I., S. Simak, B. Johansson, A. Ruban and H. Skriver, Phys. Rev. B {\bf 56}, 9319 (1997) 

 Ajayan, P., V. Ravikumar and C. Charlier, Phys. Rev. Lett. {\bf 81}, 1437 (1998) 

 Alavi, A., J. Kohanoff, M. Parrinello and D. Frenkel, Phys. Rev. Lett. {\bf 73}, 2599 (1994) 

 Ancilotto, F., G. Chiarotti, S, Scandolo and E. Tosatti, Science {\bf 275}, 1288 (1997)

 Annett, J., Comp. Mater. Science {\bf 4}, 23 (1995)

 Aoki, M., Phys. Rev. Lett. {\bf 71}, 3842 (1993) 

 Arias, T., Rev. of Mod. Phys. {\bf ??}, ?? (1998) 

 Ashcroft, N., and N. D. Mermin {\it ``Solid State Physics"} Saunders College, Philadelphia, 1976
  
 Ayala, P. and G. Scuseria,  J. Chem. Phys., in press
 
 Baer, R., and M. Head-Gordon, Phys. Rev. Lett. {\bf 79}, 3962 (1997)

 Baer, R., and M. Head-Gordon, J. Phys. Chem. {\bf 107}, 10003 (1997)

 Baer, R., and M. Head-Gordon, submitted to J. Phys. Chem. 

 Baroni, S., and P. Giannozzi, Europhys. Lett,  {\bf 17}, 547 (1992)

 Bates, K. R., A. D. Daniels, and G. E. Scuseria, J. Chem. Phys. {\bf 109}, 3308 (1998a).

 Becke, A., Phys. Rev. A {\bf 38}, 3098  (1988)

 Bennetto, J., R.W. Nunes, and David Vanderbilt, Phys. Rev. Lett. {\bf 79}, 245 (1997)

 Blount, E., in {\it ``Solid State Physics''} vol. 13, page 305, 
                 edited by F. Seitz and C. Turnbull, Academic Press, New York, 1962

 Bowler, D., M. Aoki, C. Goringe, A. Horsfield and D. Pettifor, Modelling Simul. Mater. Sci. Eng.  {\bf 5}, 199 (1997) 

 Bowler, D., M. Gillan, Comp. Phys. Comm. to be published (1998)

 Boys, S.,Proc. R. Soc. London, Ser. A {\bf 200}, 542 (1950)

 Burant, J., G. Scuseria and M. Frisch, J. Phys. Chem. {\bf 105}, 8969 (1996)

 Briggs, W., {\it ``A Multigrid Tutorial''}, SIAM, Philadelphia, 1987

 Canning, A., G. Galli, F. Mauri, A. de Vita and R. Car, Comp. Phys. Comm. {\bf 94}, 89 (1996) 

 Canning, A., G. Galli and J. Kim, Phys. Rev. Lett. {\bf 78}, 4442 (1997)

 Car, R., and M. Parrinello, Phys. Rev. Lett. {\bf 55}, 2471 (1985); 

 Challacombe, M., E. Schwegler and J. Alml\"{o}f, University of Minnesota Supercomputer Institut Research Report UMSI 95/186

 Chalvet, O., Daudel, S. Diner, and J.-P. Malrieu (eds.), 
             {\it ``Localization and Delocalization in Quantum Chemistry, vols. I - IV''}, 
              Reidel, Dordrecht, 1976

 Chelikowsky, J., N. Troullier and Y. Saad, Phys. Rev. Lett. {\bf 72}, 1240 (1994) 

 Cloizeaux, J. des, Phys. Rev. {\bf 135}, A685 and A698 (1964)

 Daniels, A., J. Millam and G. E. Scuseria, J. Chem. Phys. {\bf 107}, 425 (1997)

 Daniels, A., and G. E. Scuseria,  J. Chem. Phys. (1998b), in press

 Dederichs, P., and R. Zeller, Phys. Rev. B {\bf 28}, 5462 (1983)
 
 Devar, M., E. Zoebisch. E. Healy and J. Stewart, J. Am. Chem. Soc. {\bf 107}, 3902 (1985)

 Drabold, D., and O. Sankey, Phys. Rev. Lett. {\bf 70}, 3631 (1993)

 Daw, M., Phys. Rev. B {\bf 47}, 10895  (1993)

 Dickson, R., and A. Becke, J. Chem. Phys. {\bf 99}, 3898 (1993)

 Erdelyi, A., editor,  {\it ``Tables of Integral Transforms"}, McGraw-Hill, New York, 1954

 Feynman, R., Phys. Rev. {\bf 56}, 340  (1939)

 Fernandez, P., A. Dal Corso , A. Baldereschi and F. Mauri, Phys. Rev. B {\bf 55}, R1909 (1997) 

 Fulde, P. , {\it `` Electron Correlations in Molecules and Solids"},
             Springer Series in Solid-State Sciences, Vol 100, 1995

 Gagel, F., J. of Comp. Phys. {\bf 139}, 399 (1998)

 Galli, G., and M. Parrinello, Phys. Rev. Lett. {\bf 69}, 3547 (1992) 

 Galli, G., and F. Mauri, Phys. Rev. Lett. {\bf 73}, 3471 (1994)

 Galli, G., Current Opinion in Solid State and Materials Science {\bf 1}, 864 (1996)

 Gibson, A., R. Haydock and J. P. LaFemina, Phys. Rev. B {\bf 47}, 9229 (1993) 

 Gillan, M., J. Phys. Condens. Matter, {\bf 1}, 689 (1989) 

 Gillan, M., D. Bowler, C .Goringe and E.Hern\'andez, Proceedings 
             of the Symposium on Complex Liquids, 10-12 November 1997, Nagoya,
             Japan, ed. T.Fujiwara and M.Doi (World Scientific, 1998)

 Goedecker, S., Phys. Rev. B {\bf 48}, 17573 (1993) 

 Goedecker, S., L. Colombo, Phys. Rev. Lett. {\bf 73}, 122 (1994) 

 Goedecker, S., L. Colombo, Proc. Supercomputing 94, Washington D.C., November (1994)

 Goedecker, S., M. Teter, Phys. Rev. B {\bf 51}, 9455 (1995) 

 Goedecker, S., J. of Comp. Phys. {\bf 118}, 261 (1995)

 Goedecker, S., M. Teter, J. Hutter, Phys. Rev. B {\bf 54}, 1703 (1996) 

 Goedecker, S., and A. Hoisie, Los Alamos National laboratory, unclassified report, LA-UR-97-1504 (1997)

 Goedecker, S., O. Ivanov, Sol. State Comm., {\bf 105}, 665 (1998a)

 Goedecker, S., and I. Ivanov , Comp. in Phys. {\bf ??}, ?? (1998b) 

 Goedecker, S., and I. Ivanov , Phys. Rev. B {\bf ??}, ?? (1998c) 

 Goedecker, S., Phys. Rev. B {\bf 58}, 3501 (1998a)

 Goedecker, S., {\it ``Wavelets and their application for the solution of 
            differential equations"}, Presses Polytechniques Universitaires et Romandes, 
             Lausanne, Switzerland (ISBN 2-88074-398-2), (1998b)

 Gonze, X., Phys. Rev. B {\bf 54}, 4383  (1996)

 Goodwin, L., J. Phys.: Condens. Matter {\bf 3}, 3869 (1991) 

 Goringe, C., D. Bowler and E. Hernandez, Rep. Prog. Phys. {\bf 60}, 1447 (1997a) 

 Goringe, C., E. Hernandez, M. Gillan and I. Bush, Comp. Phys. Comm. {\bf 102}, 1 (1997b) 

 Greengard, L., Science,  {\bf 265}, 909 (1994)

 Hammond, B.L.,  W.A. Lester and P.J. Reynolds,
 {\it ``Monte Carlo Methods in Ab Initio Quantum Chemistry"}, World Scientific 1994

 Hansen, L., K. Stokbro, B. Lundquist, K. Jacobsen and D. Deaven, Phys. Rev. Let., {\bf 75}, 4444 (1995)

 Harrison, W., {\it ``Electronic Structure and the Properties of Solids"} Dover, New York, 1980

 Hartwigsen, C., S. Goedecker and J. Hutter, Phys. Rev. B {\bf 58}, 3641  (1998)

 Haydock, R., Solid State Phys. {\bf 35}, 215 (1980)

 Hehre, W. J., L. Radom, J. A. Pople and P. V. R. Schleyer, 
         {\it ``Ab Initio Molecular Orbital Theory"}, John Wiley and Sons, New York, 1996.

 Heine, V., Solid State Phys. {\bf 35}, 1 (1980)

 Hernandez, E., and M. Gillan, Phys. Rev. B {\bf 51}, 10157  (1995)

 Hernandez, E., M. Gillan, and C. Goringe, Phys. Rev. B {\bf 53}, 7147 (1996)

 Hernandez, E., M. Gillan, and C. Goringe, Phys. Rev. B {\bf 55}, 13485  (1997)

 Hierse, W., and E. Stechel, Phys. Rev. B {\bf 50}, 17811  (1994)

 Hierse, W., and E. Stechel, Phys. Rev. B {\bf 54}, 16515  (1996)

 Ho, K-M., J. Ihm and J. Joannopoulos, Phys. Rev. B {\bf 25}, 4260 (1982)

 Horsfield, A., P. D. Godwin, D. G. Pettifor and A. P. Sutton, Phys. Rev. B {\bf 54}, 15773 (1996a)

 Horsfield, A., A. Bratkovsky, D. Pettifor and M Aoki, Phys. Rev. B {\bf 53}, 1656 (1996b) 

 Horsfield, A., A. Bratkovsky, M. Fearn, D. Pettifor and M Aoki, Phys. Rev. B {\bf 53}, 12694 (1996c) 

 Horsfield, A., A. Bratkovsky, Phys. Rev. B {\bf 53}, 15381 (1996d) 

 Ismail-Beigi, S., and T. Arias, Phys. Rev. B {\bf 57}, 11923 (1998)

 Itoh, S., P. Ordejon and R. M. Martin, Comp. Phys. Comm. {\bf 88}, 173 (1995) 

 Itoh, S., P. Ordejon, D. Drabold and R. Martin, Phys. Rev. B {\bf 53}, 1 (1996)

 Jayanthis, C., S. Wu, J. Cocks, N. Luo, Z. Xie, M. Menon and G. Yang, Phys. Rev. B {\bf 57}, 3799 (1998)

 Johnson, B., P. Gill and J. Pople. J. Chem. Phys. {\bf 98}, 5612  (1993)

 Kerker, G., Phys. Rev. B {\bf 23}, 3082 (1981)

 Kim, J., F. Mauri and G. Galli, Phys. Rev. B {\bf 52}, 1640   (1995)

 Kim, J., and J. Wilkins, F. Khan and A. Canning, Phys. Rev. B {\bf 55}, 16186 (1997)

 King-Smith, R., and D. Vanderbilt, Phys. Rev. B {\bf 47}, 1651 (1993)

 King-Smith, R., M. Payne and J. Lin, Phys. Rev. B {\bf 44}, 13063 (1991)

 Kittel, C., {\it ``Quantum Theory of Solids" } John Wiley and Sons, New York, 1963

 Kohn, W., Phys. Rev. {\bf 115}, 809 (1959)

 Kohn, W., Phys. Rev. {\bf 133}, A 171 (1964)

 Kohn, W., Phys. Rev. B {\bf 7}, 4388 (1973)

 Kohn, W., Chem. Phys. Lett. {\bf 208},  (1993)

 Kohn, W., Int. J. Quant. Chem.  {\bf 56}, 229 (1995)

 Kohn, W., Phys. Rev. Lett. {\bf 76}, 3168  (1996)

 Kress, J., and A. Voter, Phys. Rev. B {\bf 52}, 8766 (1995) 

 Kress, J., S. Goedecker, A. Hoisie, H. Wassermann, O. Lubeck, L. Collins and B. Holian, 
                    Journal of Computer-Aided Materials Design, in press

 Kresse, G., Phys. Rev. B {\bf 54}, 11169 (1996)

 Landau, E. M. and L. P. Lifshitz, {\it ``Statistical Physics, Part 1"}, Third edition,
         Pergamon Press, New York, 1980

 Lee, C., W. Yang and R. Parr, Phys. Rev. B {\bf 37}, 785  (1988)

 Lewis, J., P. Ordejon and O. Sankey, Phys. Rev. B {\bf 55}, 6880 (1997)

 Li, X.-P., W. Nunes and D. Vanderbilt, Phys. Rev. B{\bf 47}, 10891  (1993)

 Lippert, R., T. Arias and A. Edelman, to appear in J. Comp. Physics

 Majewski, J., and P. Vogl in ``{\it The Structure of Binary Compounds}",
             North Holland, Amsterdam (1989), p. 287

 March, N., W. Young and S. Sampanthar, {\it ``The Many-Body Problem in Quantum Mechanics"} 
                                     Cambridge University Press, Cambridge, England, 1967

 Marzari, N., and D. Vanderbilt, Phys. Rev. B {\bf 56}, 12847 (1997) 

 Maslen, P., C. Ochsenfeld, C. White, M. Lee and M. Head-Gordon, J. Phys. Chem. {\bf 102}, 2215 (1998)

 Mauri, F., G. Galli and R. Car, Phys. Rev. B {\bf 47}, 9973  (1993)

 Mauri, F., and G. Galli, Phys. Rev. B {\bf 50}, 4316  (1994)

 McWeeny, R., Rev. Mod. Phys, {\bf 32}, 335  (1960)

 McWeeny, R., {\it ``Methods of Molecular Quantum Mechanics"}, Academic Press, 
          New York 1989

 Mermin, N., Phys. Rev. {\bf 137 A}, 1441 (1965)

 Methfessel, M., and A. Paxton, Phys. Rev. B {\bf 40}, 3616 (1989)

 Millam, J., and G. Scuseria, J. Chem. Phys. {\bf 106}, 5569 (1997) 

 Nicholson, D., and X.-G. Zhang, Phys. Rev. B {\bf 56}, 12805 (1997)

 Nightingale, M. P. and C. J. Umrigar editors
 {\it ``Quantum Monte Carlo Methods in Physics and Chemistry"}, 
                NATO ASI Series, Series C, Mathematical and Physical
                Sciences, Vol. XXX, Kluwer Academic Publishers, 1998

 Nunes, R., and D. Vanderbilt, Phys. Rev. B {\bf 50}, 17611 (1994) 

 Nunes, R., J. Bennetto, and David Vanderbilt, Phys. Rev. Lett. {\bf 77}, 1516 (1996)

 Nunes, R., J. Bennetto, and David Vanderbilt, submitted to Physical Review B

 Obara, S., and A. Saika, J. Chem. Phys. {\bf 84}, 3963 (1986)

 Ordejon, P., D. Drabold, M. Grumbach and R. Martin, Phys. Rev. B {\bf 48}, 14646  (1993)

 Ordejon, P., D. A. Drabold, R. M. Martin and M. P. Grumbach, Phys. Rev. B {\bf 51}, 1456  (1995)

 Ordejon, P., E. Artacho and J. Soler, Phys. Rev. B {\bf 53}, R10441 (1996)

 Pandey, K., A. Williams and J. Janak, Phys. Rev. B {\bf 52}, 14415 (1995) 

 Parr, R., and W. Yang, ``Density-Functional Theory of Atoms and Molecules", Oxford University Press, 1989

 Payne, M., M. Teter, D. Allan, T. Arias and J. Joannopoulos, Rev. of Mod. Phys. {\bf 64}, 1045, (1992) 

 Perdew, J., K. Burke, and M. Ernzerhof, Phys. Rev. Lett. {\bf 77} 3865 (1996)

 Perez-Jorda, J., and W. Yang, J. Chem. Phys. {\bf 107}, 1218 (1997)
 
 Pettifor, D. ``Bonding and structure of molecules and solids", Claredon Press, Oxford, 1995

 Press, W., B. P. Flannery, S. A. Teukolsky
                and W. T. Vetterling, {\it ``Numerical Recipes, The Art of Scientific
                Computing''} Cambridge University Press, Cambridge, England, 1986

 Pulay, P., in {\it ``Modern Theoretical Chemistry''} , H. F. Schaefer editor, 
                 Plenum Press, New York, 1977

 Qui, S-Y., C.  Wang, K. Ho and C. Chan, J. Phys, Condens. Matter {\bf 6}, 9153 (1994) 

 Roberts, B., and P. Clancy, submitted to Phys. Rev. B

 Saad, Y., {\it ``Iterative methods for sparse linear systems ''}
	        PWS Publishing Company, Boston 1996

 Sankey, O., and D. Niklewski, Phys. Rev. B {\bf 40}, 3979 (1989)

 Sankey, O., D. Drabold and A. Gibson, Phys. Rev. B {\bf 50}, 1376 (1994)

 Schwegler, E., and M. Challacombe, J. Chem. Phys. {\bf 105}, 2726 (1996)

 Silver, R., H. Roeder, A. Voter and J. Kress, J. of Comp. Phys. {\bf 124}, 115 (1996)

 Silver, R., and H. Roeder, Phys. Rev. E {\bf 56}, 4822 (1997)

 Sloan, I., and S. Joe, {\it ``Lattice methods for multiple integration"} 
           Oxford Science Publications, New York 1994

 Stephan, U., and D. Drabold, Phys. Rev. B {\bf 57}, 6391 (1998) 

 Stewart, J., P. Csaszar and P. Pulay, J. of Comp. Chem. {\bf 39}, 4997 (1982)

 Stewart, J., Int. J. of Quant. Chem. {\bf 58}, 133 (1996)

 \v{S}tich, I., R. Car, M. Parrinello, and S. Baroni, Phys. Rev. B {\bf 39}, 4997 (1989)

 Strain, M., G. Scuseria and M. Frisch, Science {\bf 271}, 51 (1996)

 Stratmann, R., G. Scuseria and M. Frisch, Chem. Phys. Lett. {\bf 257}, 213 (1996)

 Strout, D.,  and G. Scuseria, J. Chem. Phys. {\bf 102}, 8448 (1995)

 Szabo, A. and N. Ostlund {\it ``Modern Quantum Chemistry"} McGraw Hill, New York, 1982

 Teter, M., M. Payne, and D. Allan, Phys. Rev. B {\bf 40}, 12255 (1989) 

 Voter, A., J. Kress and R. Silver, Phys. Rev. B {\bf 53}, 12733 (1996)

 Wang, L.-W., and M. Teter, Phys. Rev. B {\bf 46}, 12798 (1992)

 Wang, L.-W., Phys. Rev. B {\bf 49}, 10154 (1994)

 Wang, Y., G. Stocks, W. Shelton D. Nicholson, Z. Szotek and W. Temmerman, Phys. Rev. Lett. {\bf 75}, 2867 (1995) 

 Weinert, M., and J. Davenport, Phys. Rev. B {\bf 45}, 11709 (1992)

 Wentzcovitch, R., J. L. Martins and P. Allen, Phys. Rev. B {\bf 45}, 11372 (1992)

 White, S., J. Wilkins and M. Teter, Phys. Rev. B {\bf 39}, 5819 (1989)

 White, C., B. Johnson, P. Gill and M. Head-Gordon, Chem. Phys. Lett. {\bf 230}, 8 (1994)

 White, C., P. Maslen, M. Lee and M. Head-Gordon, Chem. Phys. Lett.  {\bf 276}, 133 (1997) 
 
 Wimmer, E., Mater. Sci. Eng. B {\bf 37}, 72 (1996)

 Xu, C., and G. Scuseria, Chem. Phys. Lett. {\bf 262}, 219  (1996)

 Yang, W., Phys. Rev. Lett. {\bf 66}, 1438  (1991)

 Yang, W., J. Chem Phys.  {\bf 94}, 1208 (1991)

 Yang, W., and T-S. Lee, J. Chem. Phys. Rev. B {\bf 103}, 5674   (1995)

 Yang, W., Phys. Rev. B {\bf 56}, 9294  (1997)

 York, D., T-S. Lee and W. Yang, J. Am. Chem. Soc. {\bf 118}, 10940 (1996)

 Zhao, Q., and W. Yang, J. Chem. Phys. {\bf 102}, 9598  (1995)

 Zhou, S., D. Beazley, P. Lomdahl and B. Holian, Phys. Rev. Lett, {\bf 78}, 479  (1997)

\end{document}